\pdfoutput=1
\documentclass[11pt,a4paper]{article}
\usepackage{jheppub}

\usepackage{preprintcover}
\PreprintCoverPaperTitle{Measurement of the $Z/\gamma^*$ boson transverse momentum distribution 
in $pp$ collisions at  $\sqrt{s} =7$~TeV with the ATLAS detector} 
\PreprintIdNumber{CERN-PH-EP-2014-075 }  
\PreprintCoverAbstract{
This paper describes a measurement of the $Z/\gamma^*$ boson transverse momentum spectrum
using ATLAS proton--proton collision data at a centre-of-mass energy of 
$\sqrt{s} = 7$~TeV at the LHC. The measurement is performed in the
$Z/\gamma^* \rightarrow e^+e^-$ and $Z/\gamma^* \rightarrow \mu^+\mu^-$ channels, using data 
corresponding to an integrated
luminosity of 4.7~fb$^{-1}$. Normalized differential cross sections as a function of the 
$Z/\gamma^*$ boson transverse momentum  are
measured for transverse momenta up to 800~GeV. The measurement
is performed inclusively for $Z/\gamma^*$ rapidities up to 2.4, as
well as in three rapidity bins. The channel results are combined,
compared to perturbative and resummed QCD calculations and used to 
constrain the parton shower parameters of Monte Carlo generators.
}  
\PreprintJournalName{JHEP}  

\usepackage{atlasphysics}
\usepackage{zptdefs}
\usepackage{multirow}
\usepackage{rotating}
\usepackage{booktabs}  
\usepackage{graphicx}
\usepackage{overpic}
\usepackage{amssymb}
\usepackage{lineno}
\usepackage{lscape}
\usepackage{subfigure}
\usepackage{rotating}
\usepackage{overpic}
\usepackage{parskip}

\title{\boldmath Measurement of the $Z/\gamma^*$ boson transverse momentum distribution 
in $pp$ collisions at  $\sqrt{s} =7$~TeV with the ATLAS detector}

\author{The ATLAS Collaboration}

\abstract{ This paper describes a measurement of the $Z/\gamma^*$ boson transverse momentum spectrum
using ATLAS proton--proton collision data at a centre-of-mass energy of 
$\sqrt{s} = 7$~TeV at the LHC. The measurement is performed in the
$Z/\gamma^* \rightarrow e^+e^-$ and $Z/\gamma^* \rightarrow \mu^+\mu^-$ channels, using data 
corresponding to an integrated
luminosity of 4.7~fb$^{-1}$. Normalized differential cross sections as a function of the 
$Z/\gamma^*$ boson transverse momentum  are
measured for transverse momenta up to 800~GeV. The measurement
is performed inclusively for $Z/\gamma^*$ rapidities up to 2.4, as
well as in three rapidity bins. The channel results are combined,
compared to perturbative and resummed QCD calculations and used to 
constrain the parton shower parameters of Monte Carlo generators. }


\begin{document}
\maketitle
\flushbottom
\clearpage

\section{Introduction}
\label{sec:input}

The transverse momentum distribution of $W$ and $Z$ bosons produced in
hadronic collisions is a traditional probe of strong interaction
dynamics. The low transverse momentum (\pt) range is governed by
initial-state parton radiation (ISR) and the intrinsic transverse
momentum of the initial-state partons inside the proton, and modeled
using soft-gluon resummation~\cite{Balazs:1997xd} or parton shower models~\cite{pythia,herwig}. Quark-gluon
scattering dominates at high \pt\ and is described by perturbative QCD~\cite{Melnikov:2006kv,Gavin:2010az,Li:2012wna}. 
The correct modelling of the vector boson $\pt$ distribution is
important in many physics analyses at the LHC for which the production
of $W$ or $Z$ bosons constitutes a significant background.
Moreover, it is crucial for a precise measurement of the W boson mass.
The transverse momentum distribution also probes the gluon density of
the proton~\cite{Brandt:2013hoa}. Vector boson \pt\ distribution
measurements were published by ATLAS~\cite{Aad:2011gj,Aad:2011fp} and
CMS~\cite{PhysRevD.85.032002} based on 35--40 $\ipb$ of proton--proton
collisions at a centre-of-mass energy of $\rts=7$~TeV. The typical
precision of these measurements is 4\% to 10\%. 

This paper presents a measurement of the normalized $Z$ boson transverse momentum distribution (\ptz)
with the ATLAS detector, in the \Zgee\ and  \Zgmm\ channels, using LHC proton--proton
collision data taken in 2011 at a centre-of-mass energy of $\rts =
7$~TeV and corresponding to an integrated  luminosity of
4.7~\ifb~\cite{Aad:2013ucp}. The large integrated luminosity allows
the measurement to be performed in three  different $Z$ boson rapidity
($\yz$) bins, probing the transverse momentum  dynamics over a wide
range of the initial-state parton momentum fraction. With respect to
previous results, the present analysis aims at reduced uncertainties,
finer binning and extended measurement range. 

Reconstructed from the final-state lepton kinematics, \ptz\ is affected by lepton energy and momentum measurement uncertainties. 
To minimize the impact of these uncertainties, the $\phi^\star_\eta$
observable\footnote{$\phi^\star_\eta$ is defined as $\tan(\phi_{\rm{}acop}/2)\sin\theta_\eta^\star$,
with $\phi_{\rm{}acop}=\pi-\Delta\phi$ and
$\theta_\eta^\star=\tanh[\Delta\eta/2]$, $\Delta\phi$ the opening
angle between the $Z$ boson decay leptons in the transverse plane, and
$\Delta\eta=\eta^--\eta^+$ the difference in pseudorapidity between
the negatively and positively charged lepton.} was introduced as an 
alternative probe of \ptz~\cite{Banfi:2010cf}, pioneered at the
Tevatron~\cite{Abazov:2007ac,Abazov:2010kn,Aaltonen:2012fi}, and
studied by ATLAS using the present data set~\cite{Aad:2013ps}
and LHCb~\cite{Aaij:1501937}.
The correlation between $\phi^\star_\eta$ and \ptz\ is, however, only partial and the good 
experimental resolution on $\phi^\star_\eta$ is counterbalanced by a reduced sensitivity 
to the underlying transverse momentum distribution; in addition,
interpreting $\phi^\star_\eta$ as a probe of \ptz\ assumes that the 
final-state lepton angular correlations are correctly modeled. The
measurement presented in this paper allows the effects of the $Z$
boson transverse momentum and the lepton angular correlations to be
disentangled unambiguously.

QCD predictions for the \ptz\ distribution are described in the next section. After a
brief description of the experiment in section~\ref{sec:atlas}, the
measurement is presented in
sections~\ref{sec:evt_sim}-\ref{sec:results}. The results are compared
to available QCD predictions in section~\ref{sec:compaqcd} and used to constrain phenomenological
models describing the low-\ptz\ region in section~\ref{sec:tuning}; the compatibility of the
$\phi^\star_\eta$ measurement with the \ptz-constrained models is
also tested. Section~\ref{conclusion} concludes the paper.

\section{QCD predictions}
\label{sec:qcdpred}

The measurements are compared to a representative set of theoretical
predictions. They rely on perturbative QCD (pQCD) only, or include
resummation of soft-gluon emissions. Resummation is treated either
analytically, or using Monte Carlo methods.

Fully differential inclusive boson-production
cross sections can be obtained to second order in the
strong coupling constant \alphas\ (NNLO) using the {\sc
Fewz}3.1~\cite{Melnikov:2006kv,Gavin:2010az,Li:2012wna} and  {\sc
Dynnlo}1.3~\cite{Catani:2009sm,Catani:2007vq} programs. The
$\mathcal{O}(\alphas^2)$ cross-section predictions are valid at
large \ptz, where the cross section is dominated by the radiation of
high-$\pt$ gluons. At low \ptz, multiple soft-gluon emissions
predominate and fixed-order pQCD predictions are not
appropriate. 

The {\sc ResBos} calculation relies on soft-gluon resummation at low \ptz\ and matches the
$\mathcal{O}(\alphas^2)$ cross section at high \ptz. 
It simulates the vector boson decays but does not include
a description of the hadronic activity in the event. Two versions
are used here, which differ in the non-perturbative parameterization
used to perform the resummation. The original
parameterization~\cite{Balazs:1997xd} and a recent
development~\cite{Guzzi:2013aja} are referred to as {\sc
ResBos-BLNY} (NLO+NNLL) and {\sc ResBos-GNW} (NNLO+NNLL), respectively, in this paper. Further predictions at
$\mathcal{O}(\alphas^2)$ and including resummation terms at
next-to-next-to-leading-logarithmic accuracy (NNLO+NNLL) were also obtained~\cite{Banfi:2012du},
primarily focusing on the $\phi_\eta^\star$
observable.

The {\sc Pythia}~\cite{pythia} and {\sc Herwig}~\cite{herwig} generators use the parton shower approach to describe
the low-$\ptz$ region and include an $\mathcal{O}(\alphas)$ matrix element for the emission of one hard parton.
The NLO Monte Carlo generators
\mcatnlo~\cite{mcatnlo} and {\sc Powheg}~\cite{Alioli:2008gx} 
consistently incorporate NLO QCD matrix elements into the parton shower
frameworks of {\sc Herwig} or {\sc Pythia}.
The {\sc Alpgen}~\cite{Mangano:2002ea} and {\sc Sherpa}~\cite{Gleisberg:2008ta} generators
implement tree-level matrix elements for the generation of multiple hard partons
in association with the boson for various parton multiplicities.
The generators listed above are used in performing the measurement, as described in section~\ref{sec:evt_sim}.

The generators contain phenomenological parameters which are not 
constrained by the theory but can be adjusted to improve their
description of the measured distributions. The ATLAS measurement is
thus compared to the current state-of-the-art models. In
section~\ref{sec:tuning}, the low-\ptz\ region is used to adjust the
parton shower parameters in {\sc Pythia}, used as full event generator
or interfaced to {\sc Powheg}.

\section{The ATLAS detector}
\label{sec:atlas}

ATLAS~\cite{DetectorPaper:2008} is a multipurpose detector\footnote{ATLAS uses a right-handed coordinate system with its origin at the 
nominal interaction point (IP) in the centre of the detector and the $z$-axis 
along the beam pipe. The $x$-axis points from the IP to the centre of the LHC 
ring, and the $y$-axis points upward. Cylindrical coordinates $(r,\phi)$ are 
used in the transverse plane, $\phi$ being the azimuthal angle around the 
beam pipe. The pseudorapidity is defined in terms of the polar angle 
$\theta$ as $\eta=-\ln\tan(\theta/2)$.} consisting of an inner tracking 
system (ID) inside a  2~T superconducting solenoid, electromagnetic and hadronic 
calorimeters and, outermost, a toroidal large acceptance muon spectrometer (MS), 
surrounding the interaction point with almost full coverage.

The ID allows precision tracking of charged particles for 
$|\eta|<2.5$. 
The three innermost layers constitute the pixel detector. The semiconductor 
tracker, at intermediate radii,  consists of four double-sided silicon strip 
layers allowing reconstruction of three-dimensional space points.
The outer layers, made of straw tubes sensitive to transition
radiation, complete the momentum measurement for $|\eta|<2$ and provide
ability to distinguish electrons from pions.

The calorimeters between the ID and the MS measure the energy of particles in the range
$|\eta| < 4.9$. The high-granularity electromagnetic (EM) 
calorimeter is made of lead absorbers immersed in a liquid-argon 
active medium, and is divided into barrel ($|\eta| < 1.5$) and end-cap 
($1.4 < |\eta| < 3.2$) regions. For $|\eta| < 2.5$, it is finely segmented in
$\eta$ and $\phi$ for position measurement and particle
identification purposes, and has three layers in depth to enable longitudinal  
EM-shower reconstruction. The hadronic calorimeter surrounding the EM 
calorimeter is divided into a central part covering $|\eta| < 1.7$, made of 
alternating steel and plastic scintillator tiles, and end-cap ($1.5<|\eta|<3.2$) and forward ($|\eta|<4.9$)  sections
included in the liquid argon end-cap cryostats, and using copper and tungsten as
absorbing material, respectively.

The MS, covering a range of $|\eta|<2.7$, consists of three stations of drift tubes and 
cathode-strip chambers, which allow precise muon track measurements and 
of resistive-plate and thin-gap chambers for muon triggers and additional
measurements of the $\phi$ coordinate.

\section{Event simulation}
\label{sec:evt_sim}

The response of the ATLAS detector to generated Monte Carlo (MC) events is simulated~\cite{Aad:2010ah}
using {\sc Geant4}~\cite{geant4} for the description
of the ATLAS detector geometry, and the interaction 
of particles with the material defined by that geometry. These samples are used to model 
the signal, estimate the backgrounds and to correct the observed \ptz\
spectrum for detector effects back to the particle level, a procedure
hereafter referred to as unfolding. 

The MC signal samples used as baseline for the measurement are
obtained using the {\sc Powheg} generator version r1556 interfaced
with {\sc Pythia}6.425 to model the parton shower, hadronization and underlying event with
parameters set according to tune AUET2B~\cite{ATLAS:2011zja}. {\sc Powheg} events are generated using the CT10
parton distribution function (PDF) set~\cite{Lai:2010vv}.  
The predicted \ptz\ distribution is then modified to match that of
\pythia 6.425 with the AMBT1 tune~\cite{ATLAS-CONF-2010-031}, denoted by {\sc Pythia}6-AMBT1, 
which agrees with the data within 5\% accuracy~\cite{Aad:2011gj}.
These samples are referred to as {\sc Powheg+Pythia6}. 

Additional signal samples, used for comparison, are based on {\sc
Pythia}6.425 with tune {\sc AUET2B} and PDF set
MRSTMCal~\cite{Sherstnev:2007nd} (referred to as {\sc Pythia6-AUET2B});  
\mcatnlo4.01 with the CT10 PDF set, interfaced to {\sc
Herwig}6.520 to model the parton shower and hadronization, and to {\sc
Jimmy}4.31~\cite{Butterworth:1996zw} for the simulation of multiple interactions, with parameters set according to tune
AUET2~\cite{ATLAS:2011gmi}; and finally \sherpa1.4.0 with the CT10 PDFs. 
The MC generators used in tuning studies described in section~\ref{sec:tuning} are {\sc Pythia}
version~8.176~\cite{Corke:2010yf,Sjostrand:2007gs} and {\sc Powheg} version~r2314.

Background processes include $W^\pm\to\ell^\pm\nu$, $Z\to\tau^+\tau^-$
and $\bbbar,\ccbar\to\ell^\pm+X$ and are generated with {\sc Pythia6-AUET2B}. The $t\bar{t}$ background
sample is based on \mcatnlo\ interfaced to {\sc Herwig+Jimmy}. Backgrounds from weak boson pair production are simulated
using {\sc Herwig+Jimmy}, tuned with AUET2. All
generators are interfaced to {\sc Photos}2.154~\cite{photos} and {\sc
Tauola}2.4~\cite{Jadach:1993hs} to simulate QED final-state radiation (FSR) and
$\tau$-lepton decays, except \sherpa\ and {\sc Pythia8}, which rely on their internal
treatment. Photon-induced dilepton production, i.e. the double
dissociative process $q\bar{q} \to \ell^+\ell^-$ and inelastic photon-induced $pp \to \ell^+\ell^-$, is simulated using {\sc
Horace}~\cite{CarloniCalame:2007cd} and {\sc
Herwig++}~\cite{Bahr:2008pv}, interfaced to the MRST2004qed PDFs~\cite{Martin:2004dh}. 

The MC events are simulated with additional interactions in the same or neighbouring bunch 
crossings to match the pile-up conditions during LHC operation, and are weighted to reproduce 
the distribution of the average number of interactions per bunch crossing in data.

\section{Event reconstruction and selection}
\label{sec:reco}

Electrons are reconstructed from energy deposits measured in the EM
calorimeter and matched to ID tracks. They are required to have
$\pt > 20$~GeV and $|\eta| < 2.47$ excluding 
$1.37 < |\eta| < 1.52$, which corresponds to the transition region
between the barrel and end-cap EM calorimeters. The electrons
are identified using shower shape, track--cluster matching and transition radiation 
criteria~\cite{Aad:2014fxa}. The \Zgee\ event trigger
requires two such electrons with $\pt > 12$~\GeV. 
Muons are reconstructed from high-quality MS segments matched to ID tracks. 
They are required to have $\pt>20\GeV$, $|\eta|<2.4$ and to be isolated 
to suppress background from heavy-flavour decays. The isolation
requires the sum of transverse momenta of additional tracks with
$\pt>1\,\GeV$ and within a cone of size $\Delta R \equiv \sqrt{(\Delta\eta)^2+(\Delta\phi)^2} = 0.2$ around the muon
to be less than 10\% of the muon $\pt$. The \Zgmm\ event trigger requires one muon with $\pt > 18$~\GeV. 
  
Events are required to have at least one primary vertex 
reconstructed from at least three tracks with $\pt > 500$~MeV, and to contain exactly two
oppositely charged same flavour leptons, selected as described above, with invariant mass satisfying $66\GeV
< \mass{\ell} < 116\GeV$ ($\ell=e,\mu$). This broad interval is chosen to
minimize the impact of QED FSR on the signal acceptance. The total
selected sample consists of 1228863 \Zgee\ and 1816784 \Zgmm\ candidate events.  

Monte Carlo events are corrected to take into account differences with
data in lepton reconstruction, identification and trigger
efficiencies, as well as energy and momentum scale and resolution. The
efficiencies are determined using a tag-and-probe method based on
reconstructed $Z$ and $W$ events~\cite{Aad:2014fxa}. The isolation
requirement used in the muon channel induces significant \ptz\
dependence in the muon selection efficiency, and the efficiency
determination is repeated in each \ptz\ bin. The energy resolution and
scale corrections are obtained comparing the lepton pair invariant mass
distribution in data and simulation~\cite{Aad:2011mk,Aad:2014zya}.

\section{Background estimation}
\label{sec:bg}

The background to the observed $Z$ signal includes contributions from  \Zgtt,
$W \rightarrow \ell\nu$, gauge boson pair production, single top quark and
$t\bar{t}$ production, and multijet production. The electroweak and top quark
background contributions are estimated from simulation and normalized
using theoretical cross sections calculated at NNLO accuracy. For the
multijet background, which dominates at low $\ptz$,  the leptons
originate from semileptonic decays or from hadrons or photons
misidentified as electrons, which cannot be simulated accurately and
are determined using data-driven methods. 

In the electron channel, the multijet background fraction is determined
from the electron isolation distribution observed in data. The
isolation variable, $x$, is defined as the transverse energy contained
in a cone of size $\Delta R=0.3$ 
around the electron energy cluster (excluding the electron itself), divided by the electron transverse
energy. On average, isolated electrons from $Z/\gamma^*\rightarrow e^+ e^-$ decays
are expected at lower values of $x$ than multijet background
events. The signal distribution, $S(x)$, is given by the simulation
and shifted to match the data in the signal-dominated low-$x$ region. A jet-enriched
sample is extracted from data by requiring electron 
candidates to fail the track--cluster matching or shower
shape criteria in the first EM calorimeter layer, but otherwise pass the analysis selections, giving
$B(x)$. This distribution is corrected for the residual contribution
from electroweak and top quark
backgrounds, which are estimated using simulation. The multijet
background normalization is then given by a fit of
$D(x)=qB(x)+(1-q)S(x)$, where $D(x)$ is the isolation distribution 
observed in data and $q$ is the fitted background fraction. The above
procedure is repeated, separating events with same charge sign (SS) 
and opposite charge sign (OS) leptons
in the background-enriched sample, and varying $\Delta R$
between 0.2 and 0.4. The average of the results and their envelope
define the multijet background fraction and its uncertainty, yielding $q =
(0.14^{+0.10}_{-0.05})\%$. The \ptz\ shape of the background is
assumed to follow that of the background-enriched sample; this
assumption is verified by repeating the procedure in three
coarse \ptz\ bins. The uncertainty on the shape is defined from the
difference between the SS and OS samples.

In the muon channel, the multijet background is estimated using muon isolation
information in signal- and background-dominated invariant-mass
regions. Four two-dimensional regions are defined, characterized by
a mass window and according to whether both muons pass or fail the isolation
cut described in section~\ref{sec:reco}. The signal region (region A), the
two control regions (regions B and C) and the multijet region (region D)
are defined as follows:  

\begin{center}
 \begin{tabular}{lll}
        Region A (signal region): & $66\,\GeV < m_{\mu\mu} < 116\,\GeV$, & isolated     \\ 
        Region B:     & $47\,\GeV < m_{\mu\mu} < 60\,\GeV$, & isolated     \\ 
        Region C:     & $66\,\GeV < m_{\mu\mu} < 116\,\GeV$, & non-isolated \\ 
        Region D (multijet region):     & $47\,\GeV < m_{\mu\mu} < 60\,\GeV$,  & non-isolated \\
 \end{tabular}
\end{center}
Assuming the $m_{\mu\mu}$ and isolation distributions are not
correlated, the number of multijet events in the signal region is
determined from the number of events observed in regions B, C and D, as $n_\mathrm{A} =
n_\mathrm{B} \times n_\mathrm{C} / n_\mathrm{D}$, where $n_\mathrm{B}$, $n_\mathrm{C}$ and $n_\mathrm{D}$  are corrected for the
residual contribution from electroweak and top processes. In an alternative method, the multijet
background is assumed to be dominated by heavy-flavour decays, and its
normalization is derived from the number of observed SS muon pairs, corrected
by the expected OS/SS ratio in heavy-flavour jet events, as predicted
by {\sc Pythia}. 
Since the results of the two methods differ by more than their estimated
uncertainty, the background normalization used for this channel is
defined as the average of the two computations, and its uncertainty as their half difference,
giving an expected fraction of $(0.11\pm 0.06)\%$.
The \ptz\ shape of the multijet background is defined from the
control sample with the inverted isolation cut (region D); using that obtained from
the SS sample instead has negligible impact on the measurement result.

Figure~\ref{fig:control_pt} shows the \ptz\ distributions for data and
Monte Carlo samples including the experimental corrections discussed
in section~\ref{sec:reco} as well as the background estimates, in the
electron and muon channels.

\begin{figure}
  \centering
  \includegraphics[width=0.7\textwidth]{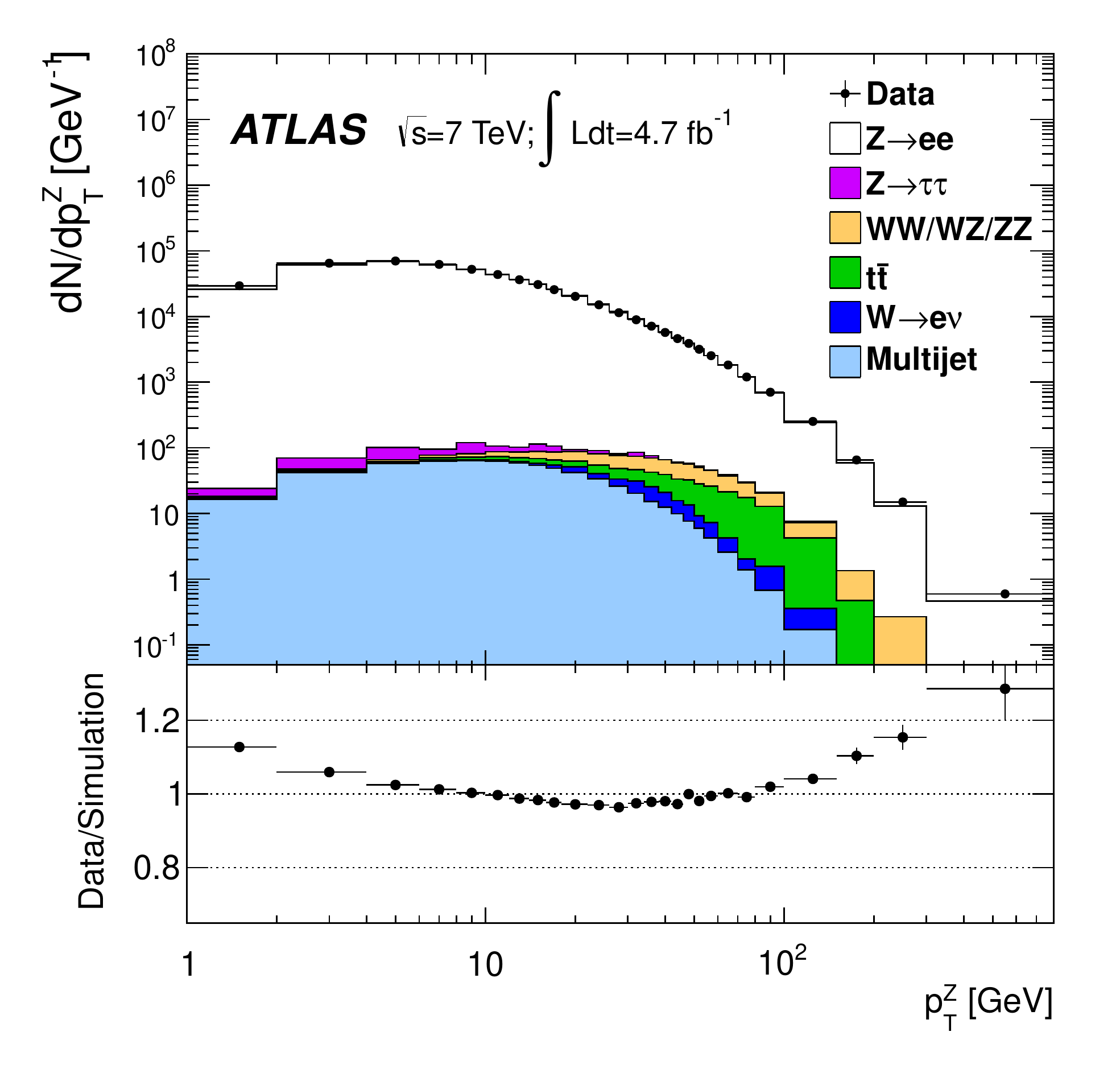}
  \includegraphics[width=0.7\textwidth]{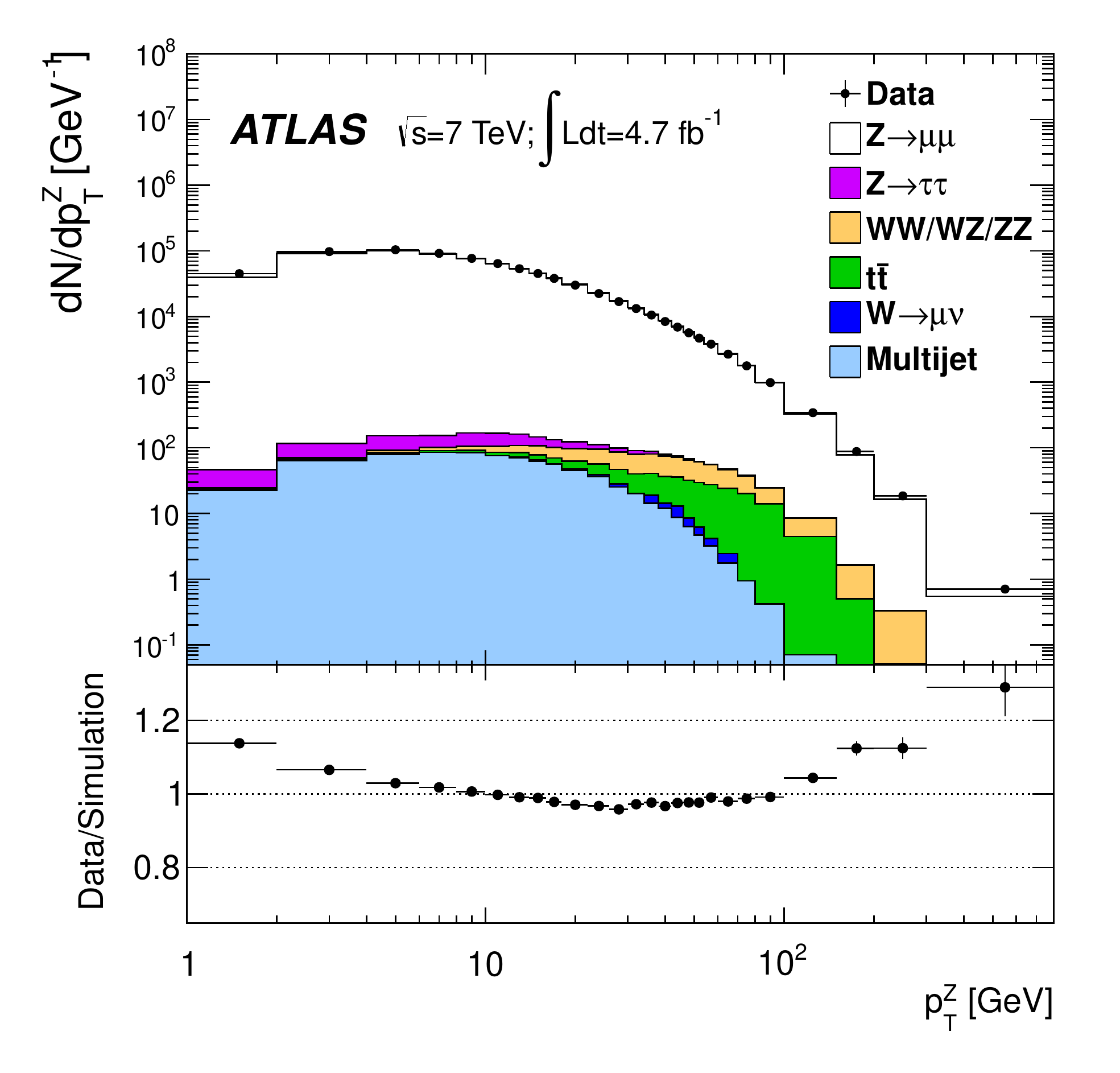}
  \caption{Distributions of \ptz\ for data and simulation, and their ratios, in the
  electron channel (top) and muon channel (bottom).
  The plots show statistical uncertainties only.}
  \label{fig:control_pt}
\end{figure}

\section{Unfolding and systematic uncertainties}
\label{sec:uncer}

The normalized differential cross section as a function of \ptz\ is defined as 
$(1/\sigma^{\mathrm{fid}})(\der\sigma^{\mathrm{fid}}/\der\ptz)$,
where $\sigma^{\mathrm{fid}}$ is the inclusive $pp\ra\Zg$
cross section measured within the fiducial acceptance
defined by requiring $\pt>20$~GeV 
and $|\eta|<2.4$ for the decay leptons; the invariant mass of the
pair must satisfy $66<m_{\ell\ell}<116$~GeV. In addition to the
rapidity-inclusive measurement, the measurement is performed for
$0\leq|\yz|<1$, $1\leq|\yz|<2$ and $2\leq|\yz|<2.4$.

The measurement is performed for three definitions of the
particle-level final-state kinematics. The Born and bare kinematics are defined from
the decay lepton kinematics before and after FSR, respectively. The
dressed kinematics are defined by combining the bare momentum of each lepton
with that of photons radiated within a distance smaller than $\Delta
R=0.1$. Conversion factors from the Born to the bare and dressed
levels are defined from the ratio of the corresponding particle-level \ptz\
distributions and denoted by $k_{\rm bare}(\ptz)$ and $k_{\rm dressed}(\ptz)$,
respectively. 

The \Zg\ transverse momentum is reconstructed from the measured lepton four-momenta. 
The \ptz\ range is divided into 26 bins of varying width between $0\,\GeV$ and 
$800\,\GeV$, with finer granularity in the low-$\ptz$ range, as shown in 
tables~\ref{tab:results}--\ref{tab:combined2}. The
bin purity, defined as the fraction of reconstructed events for
which $\ptz$ falls in the same bin at reconstruction and particle
level, is everywhere above 50\%. 

The total background is subtracted from the observed \ptz\
distribution. The electroweak background cross sections are assigned a 
5\% uncertainty derived by varying the PDFs within their
uncertainties and from QCD renormalization and factorization scale
variations; in addition, a relative uncertainty of 1.8\% on the total
integrated luminosity is  taken into account. The normalization of the
top background was verified comparing data and simulation at high
missing transverse energy ($E_{\rm T}^{\rm miss}$), defined  for each event as the
total transverse momentum imbalance of the reconstructed objects. An
uncertainty of 12\% is assigned comparing data and simulation for
$E_{\rm T}^{\rm miss}>100$~GeV and $20<\ptz<120$~GeV, where this
background contribution dominates. The multijet background uncertainty is discussed in
section~\ref{sec:bg}.

The \ptz\ distribution is
subsequently corrected for resolution effects and QED final-state
radiation back to the Born level, as well as for the differences
between the reconstruction- and particle-level fiducial acceptance,
with an iterative Bayesian unfolding
method~\cite{DAgostini:1994zf,DAgostini:2010,Adye:2011gm}; three
iterations are used. The response matrix used for the unfolding is
defined as a two-dimensional histogram correlating the Born-level and
reconstructed \ptz\ distributions. The prior probability distribution for the Born-level
$\ptz$ distribution is defined from the modified {\sc Powheg+Pythia6}
prediction described in section~\ref{sec:evt_sim}, and matches that of
{\sc Pythia}6-AMBT1. 

The statistical uncertainty on the unfolded spectrum is obtained by
generating random replicas of the reconstruction-level \ptz\
distribution. For each trial, Poisson-distributed fluctuations are
applied to the number of entries in each bin, and the measurement procedure is
repeated. The obtained ensemble of fluctuated measurement results is
used to fill a covariance matrix, including correlations between the
bins introduced by the unfolding and normalization procedure. The
relative statistical uncertainty remains below 0.6\% for $\ptz<30$~GeV in both channels,
and below 1.1\% up to 150~\GeV. The uncertainty induced by the 
size of the MC samples is determined by applying the same method to
the response matrix, and stays below 0.4\% and 0.5\% up
to $\ptz=150$~\GeV\ in the muon and electron channel, respectively,
reaching 2\% for the bin $300<\ptz<800$~\GeV.

Systematic uncertainties from experimental sources such as trigger,
reconstruction and identification efficiency corrections, energy
scale and resolution corrections, and the background normalization
and \pt\ distribution are evaluated by repeating the analysis varying the
corresponding parameters within their uncertainties and comparing
to the nominal result. For each channel, the impact of a given source of
uncertainty is evaluated preserving correlations across the
measurement range. The uncertainty on the normalization of the
electroweak and top quark backgrounds is treated as fully correlated between
the two channels. The electron- and muon-specific uncertainties are uncorrelated
between channels.  

In the electron channel, the uncertainties on the trigger, reconstruction and
identification efficiency corrections are propagated preserving
their correlations across lepton $\eta$ and \pt. These sources 
contribute a relative uncertainty of the order of $10^{-4}$ up
to $\ptz=100$~\GeV\ and less than 0.2\% over the full measurement
range. The uncertainty induced by the background subtraction is
typically 0.1\%, except around $\ptz=100$~\GeV\ where it reaches 0.3\%  
because of the top quark background contribution.
The uncertainty induced by charge misidentification,
estimated from the difference between the results obtained with and
without an opposite-sign requirement on the leptons, amounts to less than
0.2\% over the whole \ptz\ range. The dominant experimental
uncertainties in the electron channel arise from the electron
energy scale, resolution, mis-modelling of the electron
energy tails caused by uncertainties in the treatment of electron
multiple scattering in {\sc Geant4} and in passive detector material.  
The combined contribution from energy scale and resolution uncertainties to the total
systematic uncertainty is typically 0.3\% per bin between 4~GeV and
70~\GeV, and reaches about 2\% at the end of the spectrum. The uncertainty from
the energy tails amounts to 0.8\% at most, contributing mainly at very
low \ptz\ and at very high \ptz\ where the statistical uncertainty dominates.

In the muon channel, the trigger, reconstruction and isolation
efficiency corrections contribute an uncertainty of 0.6\% on average,
spanning 0.2\% to 1.7\% across the measurement range. The momentum scale
and resolution uncertainties amount to 0.2\%, except in the last
three \ptz\ bins where they stay below 1.5\%. The uncertainty
contributed by the background subtraction is below 0.1\% over the
whole \ptz\ range except around $\ptz=100$~\GeV\ where it reaches 0.13\% 
because of the top quark background contribution.

The dominant contribution to the systematic uncertainties for both channels comes from the
unfolding method. Two effects are addressed: the bias of the result towards the
prior, and the dependence of the result on the theoretical calculation used to
determine the response matrix. The first item is evaluated by repeating the measurement using the 
nominal result as the prior. The difference between the nominal result and
this iteration is less than 0.1\% up to 100~\GeV, and less than 1.3\%
for the rest of the distribution. The second effect is evaluated by
unfolding the \ptz\ distribution using an alternative response matrix,
constructed from a $\Zgll$ sample obtained with \mcatnlo\ instead of {\sc Powheg}, and
modified to match the \pythia 6-AMBT1 spectrum as it was done for {\sc
Powheg}. A systematic uncertainty of about 0.3\% over the whole \ptz\ range is assigned from the
difference between the two results. The PDF uncertainties are
estimated by reweighting the baseline sample to each of the CT10 PDF
error sets~\cite{Lai:2010vv} and repeating the unfolding. In each bin,
the sum in quadrature of deviations with respect to the nominal result is
used to define the associated uncertainty, which is below 0.1\% up to
60~\GeV\ and below 0.3\% over the remaining \ptz\ range. The unfolding
systematic uncertainties are assumed to be fully correlated between the
electron and muon channels.

The uncertainty arising from the accuracy of the theoretical description
of QED FSR is obtained by comparing $k_{\rm bare}(\ptz)$ and $k_{\rm
dressed}(\ptz)$ as predicted by \photos\ and \sherpa. The differences
obtained for $k_{\rm bare}(\ptz)$ are representative of the QED uncertainty
in the muon channel, and amount to 0.3\% across the \ptz\
distribution. From the differences obtained for $k_{\rm dressed}(\ptz)$, a
0.1\% uncertainty is assigned to the electron channel. Photon-induced
dilepton production is significant only in the lowest \ptz\ bin (0--2~\GeV), where
it contributes 0.4\%. The cross sections obtained for this process when
evaluating the MRST2004qed PDFs in the current and constituent quark
mass schemes differ by 30\%, and contribute an uncertainty of 0.1\% to
the measurement in this bin.

Figure~\ref{fig:plot_systematics_inclusive} presents the contributions from 
the different uncertainties to the inclusive
\ptz\ measurement  integrated over the $Z$ rapidity.  

\begin{figure}  
\centering 
\includegraphics[width=0.65\textwidth]{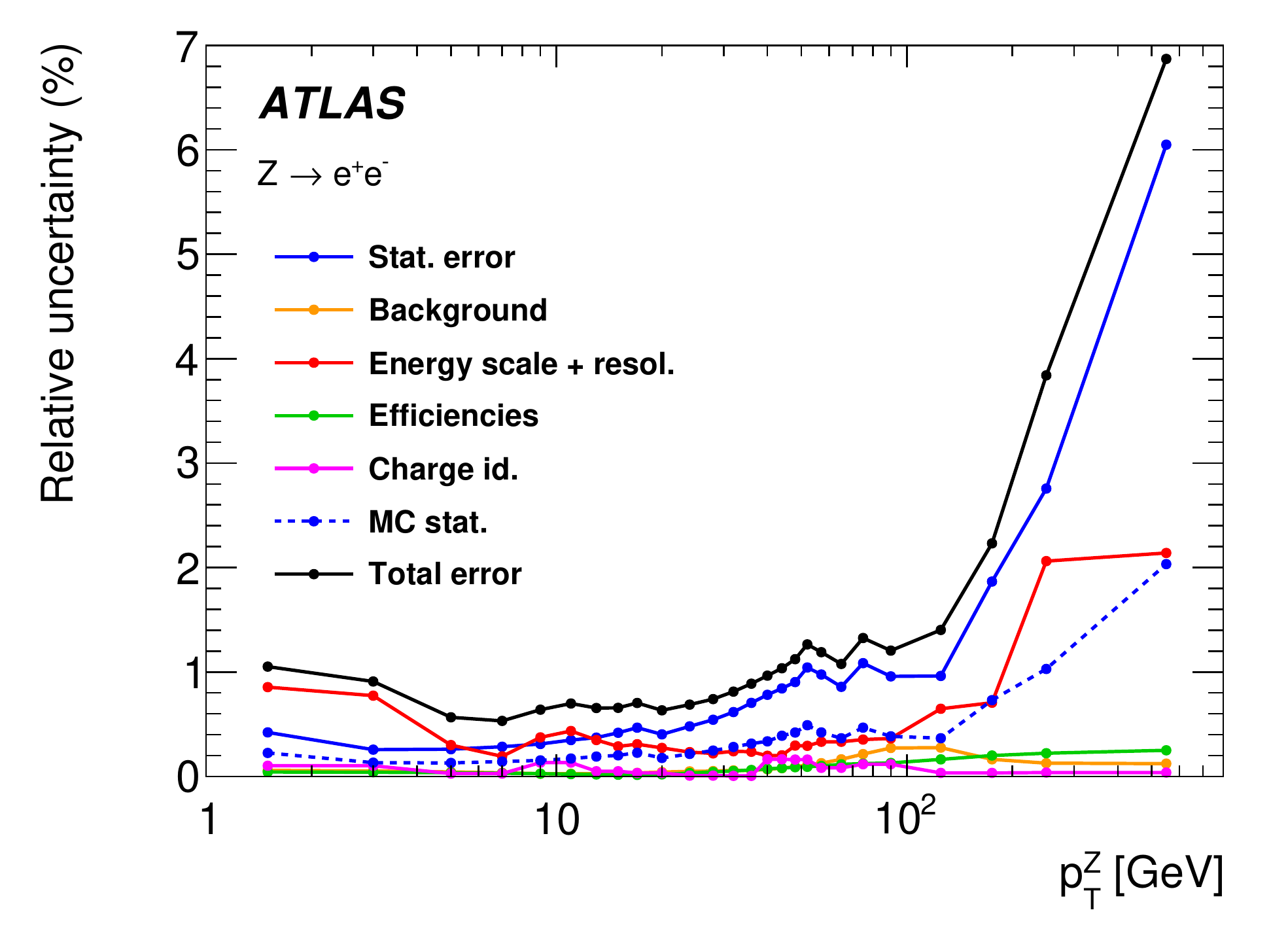}
\includegraphics[width=0.65\textwidth]{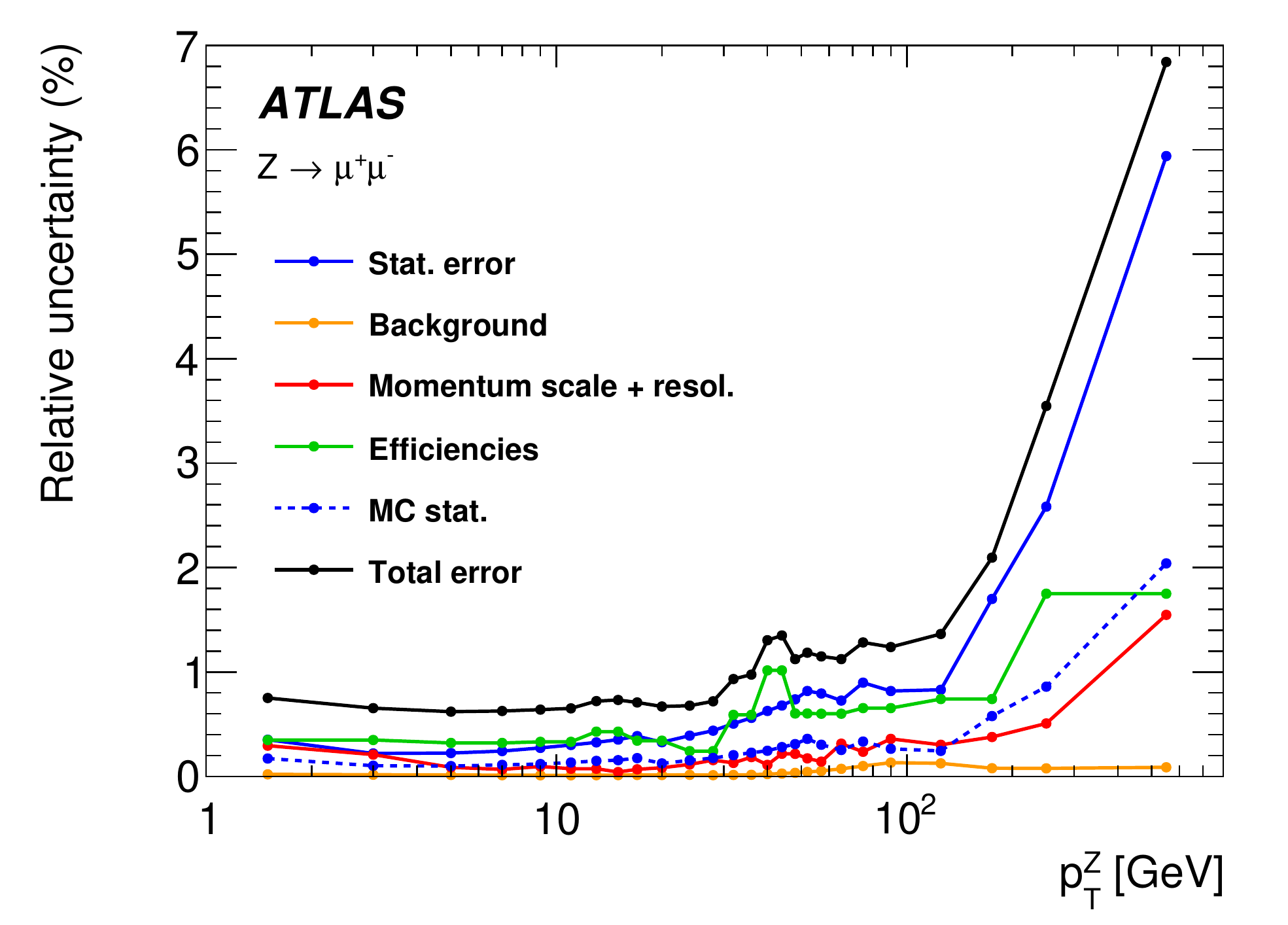}
\includegraphics[width=0.65\textwidth]{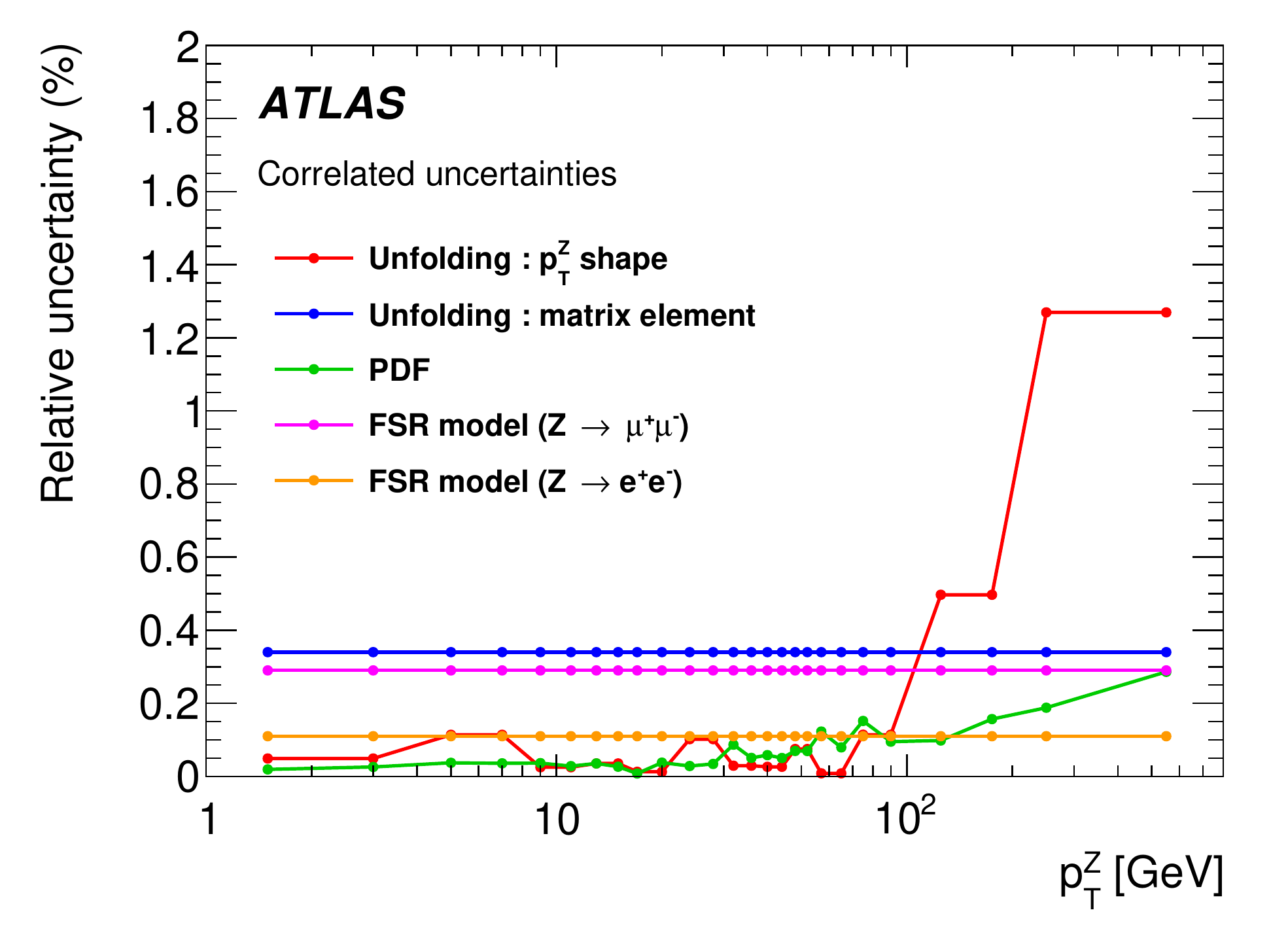}
\caption{Summary of uncertainties for the \yz-integrated measurement, 
given as a percentage of the central value of the bin. Electron channel (top),
muon channel (middle), correlated uncertainties (bottom).}
\label{fig:plot_systematics_inclusive}
\end{figure}

\section{Results}
\label{sec:results}

The inclusive normalized cross sections 
$(1/\sigma^{\mathrm{fid}})(\der\sigma^{\mathrm{fid}}/\der\ptz)$
measured in the \Zgee\ and \Zgmm\
channels are presented in table~\ref{tab:results} including
statistical, uncorrelated and correlated systematic uncertainties. 
The sizes of the correlated uncertainties depend on the channel because of different
resolutions and background levels. The measurement results are reported at
Born level and factors \kbare\ and \kdressed\ are given to translate
to the bare and dressed levels. In each channel, the total
uncertainty is between 0.5\% and 1\% for $\ptz<30$~\GeV, below 1.5\%
per bin up to $\ptz=150$~\GeV\, and rises to ~7\% at the end of the spectrum. 

\begin{sidewaystable}
\centering
\small
\begin{tabular}{lllllllllllllll}
\hline
                  & \multicolumn{5}{c}{\Zgee}  &  & \multicolumn{5}{c}{\Zgmm} &  & \multicolumn{2}{c}{Common}\\
                  \cline{2-6} \cline{8-12} \cline{14-15}\\
$p_{\rm T}$ range       & \multicolumn{3}{l}{$\frac{1}{\sigma^{\rm fid}}\frac{\der\sigma^{\rm fid}}{\der\ptz}$ $[1/\GeV]$}
		  & $\delta_{\rm {Stat}}$ & $\delta^{\rm {uncor}}_{\rm {Syst}}$  & 
		  & \multicolumn{3}{l}{$\frac{1}{\sigma^{\rm fid}}\frac{\der\sigma^{\rm fid}}{\der\ptz} $ $[1/\GeV]$} 
		  & $\delta_{\rm {Stat}}$ & $\delta^{\rm {uncor}}_{\rm {Syst}}$  &  & $\delta^{\rm {cor}}_{\rm {Syst}}$ & $\delta^{\rm {cor}}_{\rm {Syst}}$\\
		 \cline{2-4}  \cline{8-10} \\
   $[\GeV]$          & Born & $k_{\rm bare}$ & $k_{\rm dressed}$ & $[\%]$ & $[\%]$ &  & Born & $k_{\rm bare}$ & $k_{\rm dressed}$ &
   $[\%]$ & $[\%]$ &  & $ee$ $[\%]$ & $\mu\mu$ $[\%]$ \\ 
		\cline{1-6}  \cline{8-12}\cline{14-15}\\

0--2     & 2.811 $10^{-2}$ & 0.916 & 0.974 & 0.42 & 0.85 & & 2.836 $10^{-2}$ & 0.953 & 0.974 & 0.35 & 0.50 &  & 0.36 & 0.36 \\ 
2--4     & 5.840 $10^{-2}$ & 0.935 & 0.980 & 0.26 & 0.76 & & 5.833 $10^{-2}$ & 0.964 & 0.980 & 0.22 & 0.43 &  & 0.35 & 0.34 \\ 
4--6     & 5.806 $10^{-2}$ & 0.969 & 0.990 & 0.26 & 0.39 & & 5.800 $10^{-2}$ & 0.982 & 0.990 & 0.22 & 0.35 &  & 0.36 & 0.36 \\ 
6--8     & 4.908 $10^{-2}$ & 1.002 & 1.000 & 0.28 & 0.31 & & 4.929 $10^{-2}$ & 1.002 & 1.000 & 0.24 & 0.35 &  & 0.36 & 0.36 \\ 
8--10    & 4.074 $10^{-2}$ & 1.025 & 1.007 & 0.31 & 0.43 & & 4.082 $10^{-2}$ & 1.014 & 1.007 & 0.27 & 0.44 &  & 0.34 & 0.34 \\ 
10--12   & 3.381 $10^{-2}$ & 1.040 & 1.012 & 0.35 & 0.49 & & 3.375 $10^{-2}$ & 1.023 & 1.012 & 0.30 & 0.45 &  & 0.34 & 0.34 \\ 
12--14   & 2.815 $10^{-2}$ & 1.055 & 1.016 & 0.37 & 0.42 & & 2.814 $10^{-2}$ & 1.031 & 1.016 & 0.33 & 0.46 &  & 0.34 & 0.34 \\ 
14--16   & 2.374 $10^{-2}$ & 1.060 & 1.017 & 0.42 & 0.38 & & 2.376 $10^{-2}$ & 1.032 & 1.017 & 0.35 & 0.46 &  & 0.34 & 0.34 \\ 
16--18   & 2.014 $10^{-2}$ & 1.060 & 1.017 & 0.47 & 0.38 & & 2.011 $10^{-2}$ & 1.032 & 1.016 & 0.39 & 0.48 &  & 0.34 & 0.34 \\ 
18--22   & 1.598 $10^{-2}$ & 1.052 & 1.016 & 0.40 & 0.32 & & 1.593 $10^{-2}$ & 1.029 & 1.016 & 0.33 & 0.47 &  & 0.34 & 0.34 \\ 
22--26   & 1.199 $10^{-2}$ & 1.033 & 1.010 & 0.48 & 0.31 & & 1.201 $10^{-2}$ & 1.018 & 1.010 & 0.39 & 0.50 &  & 0.36 & 0.36 \\ 
26--30   & 9.164 $10^{-3}$ & 1.021 & 1.006 & 0.54 & 0.33 & & 9.172 $10^{-3}$ & 1.010 & 1.006 & 0.44 & 0.53 &  & 0.36 & 0.36 \\ 
30--34   & 7.236 $10^{-3}$ & 1.007 & 1.003 & 0.62 & 0.38 & & 7.256 $10^{-3}$ & 1.006 & 1.003 & 0.50 & 0.54 &  & 0.35 & 0.35 \\ 
34--38   & 5.806 $10^{-3}$ & 0.997 & 1.000 & 0.70 & 0.40 & & 5.800 $10^{-3}$ & 0.999 & 1.000 & 0.56 & 0.58 &  & 0.35 & 0.35 \\ 
38--42   & 4.666 $10^{-3}$ & 0.992 & 0.999 & 0.78 & 0.45 & & 4.619 $10^{-3}$ & 0.997 & 0.999 & 0.63 & 0.63 &  & 0.35 & 0.35 \\ 
42--46   & 3.760 $10^{-3}$ & 0.990 & 0.998 & 0.84 & 0.49 & & 3.795 $10^{-3}$ & 0.992 & 0.998 & 0.68 & 0.68 &  & 0.35 & 0.34 \\ 
46--50   & 3.216 $10^{-3}$ & 0.977 & 0.995 & 0.90 & 0.53 & & 3.137 $10^{-3}$ & 0.990 & 0.995 & 0.73 & 0.66 &  & 0.37 & 0.37 \\ 
50--54   & 2.604 $10^{-3}$ & 0.982 & 0.996 & 1.04 & 0.59 & & 2.586 $10^{-3}$ & 0.987 & 0.996 & 0.82 & 0.68 &  & 0.37 & 0.36 \\ 
54--60   & 2.097 $10^{-3}$ & 0.972 & 0.994 & 0.98 & 0.55 & & 2.113 $10^{-3}$ & 0.986 & 0.994 & 0.79 & 0.65 &  & 0.38 & 0.36 \\ 
60--70   & 1.501 $10^{-3}$ & 0.966 & 0.992 & 0.86 & 0.52 & & 1.484 $10^{-3}$ & 0.982 & 0.992 & 0.72 & 0.71 &  & 0.39 & 0.36 \\ 
70--80   & 9.820 $10^{-4}$ & 0.959 & 0.989 & 1.08 & 0.56 & & 9.886 $10^{-4}$ & 0.976 & 0.989 & 0.89 & 0.78 &  & 0.44 & 0.39 \\ 
80--100  & 5.599 $10^{-4}$ & 0.955 & 0.991 & 0.96 & 0.50 & & 5.449 $10^{-4}$ & 0.979 & 0.991 & 0.81 & 0.83 &  & 0.46 & 0.39 \\ 
100--150 & 1.920 $10^{-4}$ & 0.957 & 0.991 & 0.96 & 0.74 & & 1.917 $10^{-4}$ & 0.976 & 0.991 & 0.83 & 0.83 &  & 0.67 & 0.62 \\ 
150--200 & 4.809 $10^{-5}$ & 0.953 & 0.994 & 1.86 & 1.02 & & 4.982 $10^{-5}$ & 0.975 & 0.994 & 1.70 & 1.11 &  & 0.64 & 0.60 \\ 
200--300 & 1.085 $10^{-5}$ & 0.950 & 0.995 & 2.76 & 2.51 & & 1.074 $10^{-5}$ & 0.974 & 0.995 & 2.58 & 1.99 &  & 1.33 & 1.34 \\ 
300--800 & 3.910 $10^{-7}$ & 0.949 & 0.995 & 6.05 & 3.12 & & 4.047 $10^{-7}$ & 0.958 & 0.995 & 5.84 & 3.20 &  & 1.35 & 1.30 \\ \hline 
\end{tabular}
        \caption{The measured normalized cross section
		  $(1/\sigma^{\mathrm{fid}})(\der\sigma^{\mathrm{fid}}/\der\ptz)$
                  in bins of \ptz\ for the \Zgee\ and \Zgmm\ channels,
                  and correction factors to the bare- and dressed-level cross sections. 
		  The relative statistical and total uncorrelated systematic uncertainties 
		  are given for each channel as well as the correlated systematic uncertainties.}
\label{tab:results}
\end{sidewaystable}

The electron- and muon-channel cross sections are
combined using $\chi^2$ minimization, following the best linear
unbiased estimator prescription (BLUE)~\cite{Lyons:1988rp,Valassi:2003mu}.  
The combination is performed for the Born-level and dressed-level distributions.
When building the $\chi^2$, the measurement uncertainties are categorized into
uncorrelated and correlated sources. 
Table~\ref{tab:combined1} presents the combined results  for the
inclusive measurement for Born level and dressed lepton
kinematics. The combined precision is between
0.5\% and 1.1\% for $\ptz<150$ \GeV, rising to ~5.5\% towards the
end of the spectrum. The combination has $\chi^2/{\rm dof}=12.3/25$ ($\chi^2$ per degree of freedom). 
The individual channels are compared to the combined result in
figure~\ref{fig:combinationInc}. 

\begin{table}
\centering
\small
\begin{tabular}{llllll}
\hline
	      & Born & Dressed & & &   \\
\pt\ range   & $\frac{1}{\sigma^{\rm fid}}\frac{\der\sigma^{\rm fid}}{\der\ptz}$ &
$\frac{1}{\sigma^{\rm fid}}\frac{\der\sigma^{\rm fid}}{\der\ptz}$ 
& $\delta_{\rm {Stat}}$ & $\delta^{\rm {uncor}}_{\rm {Syst}}$ &  $\delta^{\rm {cor}}_{\rm {Syst}}$ \\
 $[\GeV]$ & $[1/\GeV]$ & $[1/\GeV]$& $[\%]$  &  $[\%]$  & $[\%]$  \\
                  \cline{2-6}\\
0--2     & 2.822 $10^{-2}$ & 2.750 $10^{-2}$ & 0.27 & 0.37 & 0.36 \\ 
2--4     & 5.840 $10^{-2}$ & 5.723 $10^{-2}$ & 0.17 & 0.32 & 0.35 \\ 
4--6     & 5.805 $10^{-2}$ & 5.749 $10^{-2}$ & 0.17 & 0.23 & 0.36 \\ 
6--8     & 4.917 $10^{-2}$ & 4.920 $10^{-2}$ & 0.18 & 0.22 & 0.36 \\ 
8--10    & 4.076 $10^{-2}$ & 4.103 $10^{-2}$ & 0.20 & 0.24 & 0.34 \\ 
10--12   & 3.380 $10^{-2}$ & 3.420 $10^{-2}$ & 0.23 & 0.26 & 0.34 \\ 
12--14   & 2.815 $10^{-2}$ & 2.860 $10^{-2}$ & 0.25 & 0.26 & 0.34 \\ 
14--16   & 2.375 $10^{-2}$ & 2.415 $10^{-2}$ & 0.27 & 0.26 & 0.34 \\ 
16--18   & 2.012 $10^{-2}$ & 2.046 $10^{-2}$ & 0.30 & 0.27 & 0.34 \\ 
18--22   & 1.595 $10^{-2}$ & 1.621 $10^{-2}$ & 0.25 & 0.25 & 0.34 \\ 
22--26   & 1.200 $10^{-2}$ & 1.212 $10^{-2}$ & 0.30 & 0.28 & 0.36 \\ 
26--30   & 9.166 $10^{-3}$ & 9.223 $10^{-3}$ & 0.34 & 0.31 & 0.36 \\ 
30--34   & 7.242 $10^{-3}$ & 7.267 $10^{-3}$ & 0.39 & 0.33 & 0.35 \\ 
34--38   & 5.802 $10^{-3}$ & 5.803 $10^{-3}$ & 0.44 & 0.35 & 0.35 \\ 
38--42   & 4.641 $10^{-3}$ & 4.636 $10^{-3}$ & 0.49 & 0.39 & 0.35 \\ 
42--46   & 3.777 $10^{-3}$ & 3.769 $10^{-3}$ & 0.53 & 0.43 & 0.35 \\ 
46--50   & 3.172 $10^{-3}$ & 3.157 $10^{-3}$ & 0.57 & 0.43 & 0.37 \\ 
50--54   & 2.593 $10^{-3}$ & 2.582 $10^{-3}$ & 0.64 & 0.46 & 0.37 \\ 
54--60   & 2.104 $10^{-3}$ & 2.091 $10^{-3}$ & 0.61 & 0.43 & 0.37 \\ 
60--70   & 1.492 $10^{-3}$ & 1.480 $10^{-3}$ & 0.55 & 0.44 & 0.38 \\ 
70--80   & 9.851 $10^{-4}$ & 9.738 $10^{-4}$ & 0.69 & 0.49 & 0.43 \\ 
80--100  & 5.525 $10^{-4}$ & 5.474 $10^{-4}$ & 0.62 & 0.49 & 0.44 \\ 
100--150 & 1.918 $10^{-4}$ & 1.901 $10^{-4}$ & 0.63 & 0.53 & 0.65 \\ 
150--200 & 4.891 $10^{-5}$ & 4.860 $10^{-5}$ & 1.26 & 0.72 & 0.63 \\ 
200--300 & 1.081 $10^{-5}$ & 1.075 $10^{-5}$ & 1.88 & 1.40 & 1.33 \\ 
300--800 & 3.985 $10^{-7}$ & 3.966 $10^{-7}$ & 4.20 & 2.04 & 1.32 \\ \hline
\end{tabular}
\caption{The measured normalized combined (electron and muon channels) cross section
$(1/\sigma^{\mathrm{fid}})(\der\sigma^{\mathrm{fid}}/\der\ptz)$, inclusive in
rapidity. The cross sections at Born and dressed levels are given as
well as the relative statistical ($\delta_{\rm {Stat}}$) and total systematic
($\delta_{\rm {Syst}}$) for uncorrelated and correlated sources.} 
\label{tab:combined1}
\end{table}

\begin{figure}
  \centering
  \includegraphics[width=0.8\textwidth]{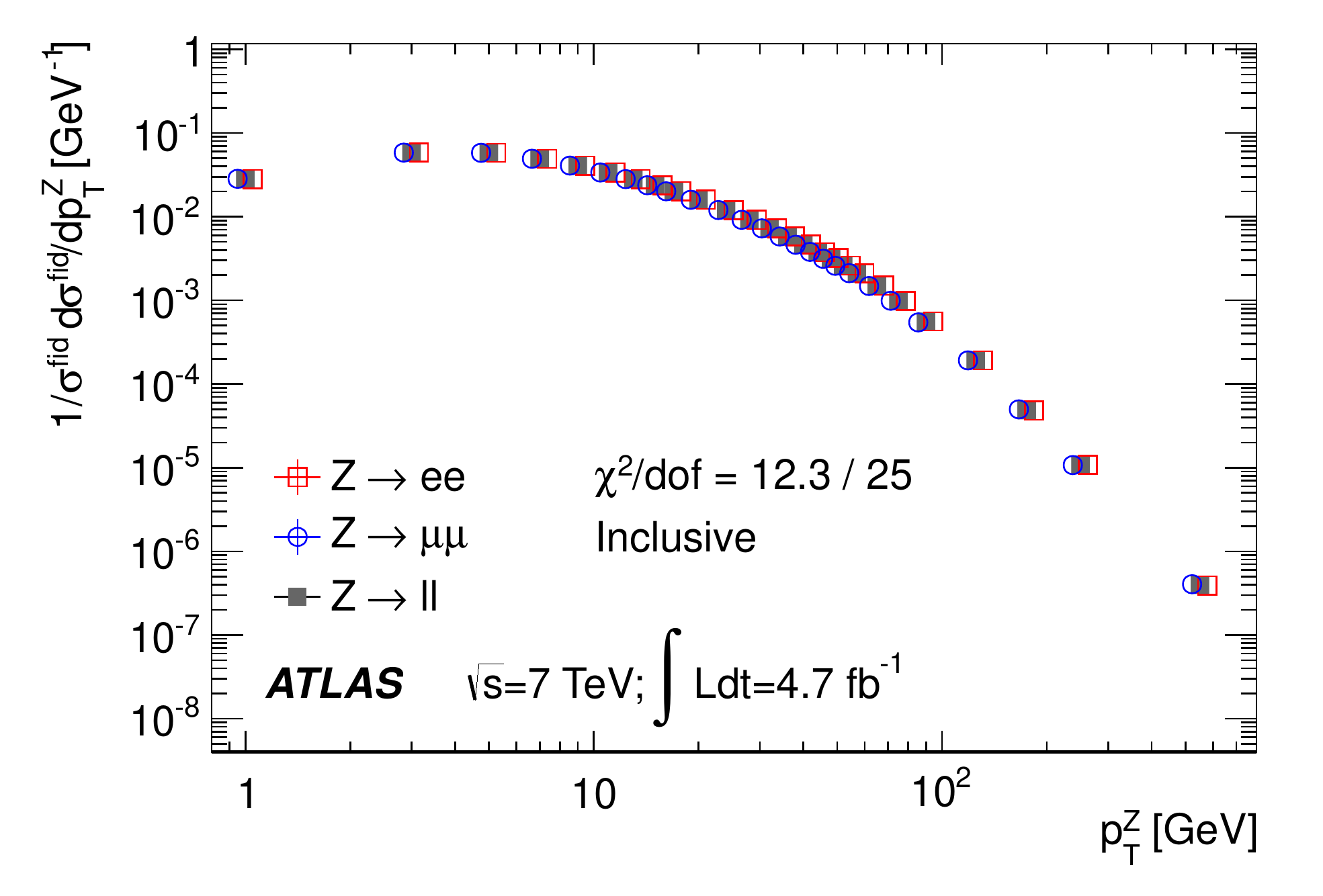}
  \includegraphics[width=0.8\textwidth]{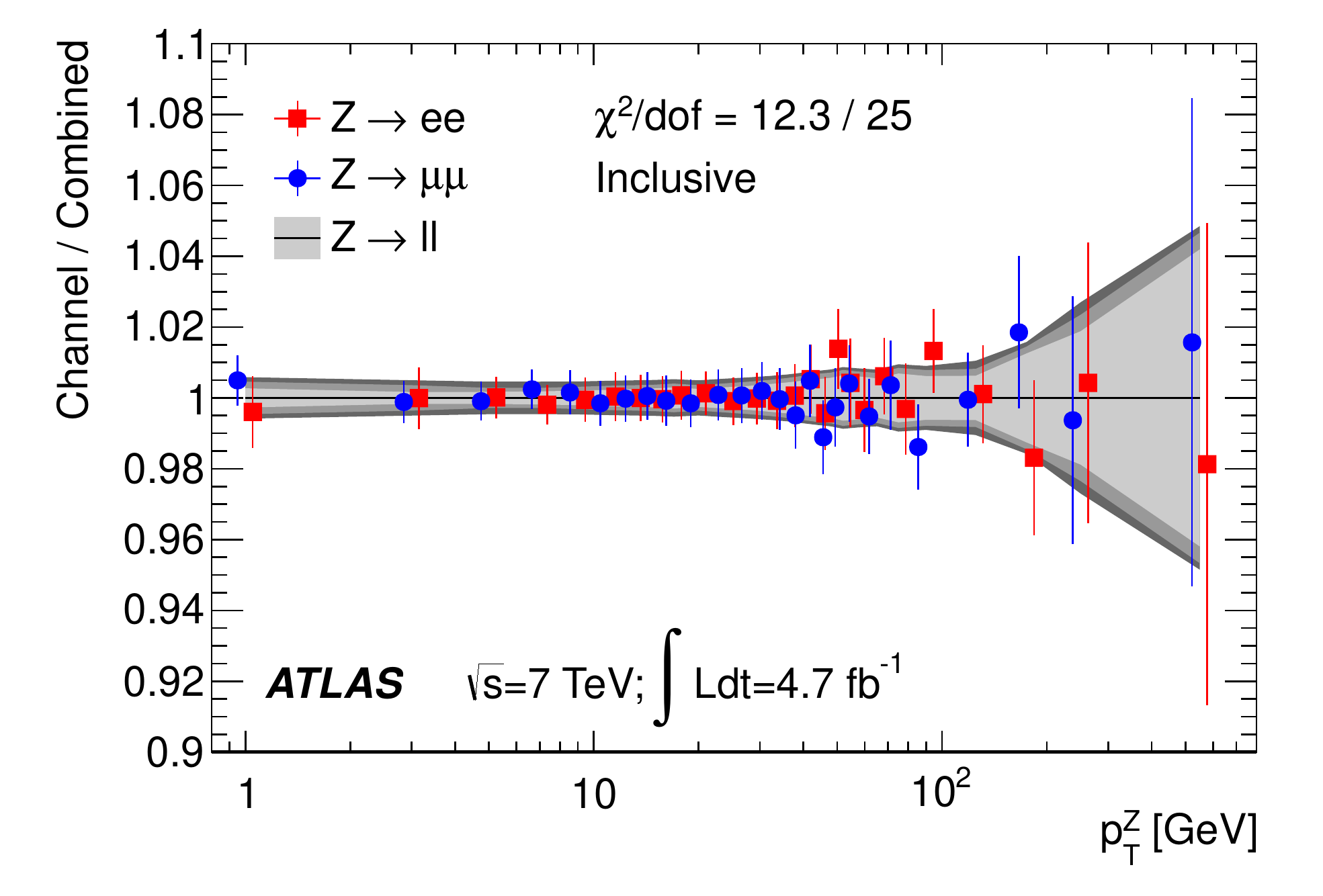}
  \caption{The measured inclusive  normalized cross section
  $(1/\sigma^{\mathrm{fid}})(\der\sigma^{\mathrm{fid}}/\der\ptz)$
  as a function of \ptz\ for the electron and muon channels and the
  combined result (top). Ratio of the electron and muon channels to the
  combined result (bottom). The uncertainty bands represent the statistical, total
  uncorrelated and total uncertainties, from light gray to dark gray respectively.}
  \label{fig:combinationInc}
\end{figure}

The measurements are repeated in three exclusive boson rapidity bins,
namely $0\leq|\yz|<1$, $1\leq|\yz|<2$ and $2\leq|\yz|<2.4$.
The combined results, corrected to the Born level, are given in
table~\ref{tab:combined2} with statistical, correlated and
uncorrelated systematic uncertainties for the three rapidity bins. The
measurement results in each channel and their combination are illustrated in
figures~\ref{fig:combinationRap1}-\ref{fig:combinationRap3}. 

\begin{sidewaystable}
\centering
\small
\resizebox{\textwidth}{!}{
\begin{tabular}{lrrrrr|rrrrr|rrrrr}
\hline
    &  \multicolumn{5}{c|}{$0\leq|\yz|<1$} & \multicolumn{5}{c|}{$1\leq|\yz|<2$} &  \multicolumn{5}{c}{$2\leq|\yz|<2.4$} \\
              \cline{2-16}
    & Born & Dressed & & & & Born & Dressed & & & & Born & Dressed & & & \\
\pt\ range   & $\frac{1}{\sigma^{\rm fid}}\frac{\der\sigma^{\rm fid}}{\der\ptz}$ &
$\frac{1}{\sigma^{\rm fid}}\frac{\der\sigma^{\rm fid}}{\der\ptz}$ 
& $\delta_{\rm {Stat}}$ & $\delta^{\rm {uncor}}_{\rm {Syst}}$ &  $\delta^{\rm {cor}}_{\rm {Syst}}$ &
 $\frac{1}{\sigma^{\rm fid}}\frac{\der\sigma^{\rm fid}}{\der\ptz}$ & 
 $\frac{1}{\sigma^{\rm fid}}\frac{\der\sigma^{\rm fid}}{\der\ptz}$ & $\delta_{\rm {Stat}}$ & 
 $\delta^{\rm {uncor}}_{\rm {Syst}}$ &  $\delta^{\rm {cor}}_{\rm {Syst}}$ &
 $\frac{1}{\sigma^{\rm fid}}\frac{\der\sigma^{\rm fid}}{\der\ptz}$ & 
 $\frac{1}{\sigma^{\rm fid}}\frac{\der\sigma^{\rm fid}}{\der\ptz}$ & $\delta_{\rm {Stat}}$ & 
 $\delta^{\rm {uncor}}_{\rm {Syst}}$ &  $\delta^{\rm {cor}}_{\rm {Syst}}$\\
 $[\GeV]$ & $[1/\GeV]$ & $[1/\GeV]$& $[\%]$  &  $[\%]$  & $[\%]$ & $[1/\GeV]$ & $[1/\GeV]$&$[\%]$  & $[\%]$  & $[\%]$ & $[1/\GeV]$ & $[1/\GeV]$& $[\%]$  &  $[\%]$  & $[\%]$ \\
                  \cline{2-16}\\
0--2     & 2.861 $10^{-2}$ & 2.792 $10^{-2}$ & 0.37 & 0.34 & 0.36 & 2.781 $10^{-2}$ & 2.704 $10^{-2}$ & 0.42 & 0.48 & 0.37 & 2.71 $10^{-2}$ & 2.63 $10^{-2}$ & 1.3 & 1.0 & 0.5 \\ 
2--4     & 5.874 $10^{-2}$ & 5.763 $10^{-2}$ & 0.23 & 0.31 & 0.34 & 5.802 $10^{-2}$ & 5.680 $10^{-2}$ & 0.26 & 0.39 & 0.35 & 5.68 $10^{-2}$ & 5.53 $10^{-2}$ & 0.8 & 0.7 & 0.4 \\ 
4--6     & 5.834 $10^{-2}$ & 5.784 $10^{-2}$ & 0.23 & 0.23 & 0.35 & 5.782 $10^{-2}$ & 5.720 $10^{-2}$ & 0.25 & 0.27 & 0.39 & 5.64 $10^{-2}$ & 5.56 $10^{-2}$ & 0.8 & 0.5 & 0.5 \\ 
6--8     & 4.972 $10^{-2}$ & 4.974 $10^{-2}$ & 0.26 & 0.22 & 0.34 & 4.868 $10^{-2}$ & 4.872 $10^{-2}$ & 0.28 & 0.27 & 0.38 & 4.71 $10^{-2}$ & 4.70 $10^{-2}$ & 0.8 & 0.6 & 0.5 \\ 
8--10    & 4.106 $10^{-2}$ & 4.134 $10^{-2}$ & 0.28 & 0.24 & 0.34 & 4.047 $10^{-2}$ & 4.074 $10^{-2}$ & 0.31 & 0.30 & 0.34 & 3.95 $10^{-2}$ & 3.97 $10^{-2}$ & 0.9 & 0.6 & 0.4 \\ 
10--12   & 3.385 $10^{-2}$ & 3.424 $10^{-2}$ & 0.31 & 0.26 & 0.35 & 3.381 $10^{-2}$ & 3.423 $10^{-2}$ & 0.34 & 0.32 & 0.34 & 3.22 $10^{-2}$ & 3.27 $10^{-2}$ & 1.0 & 0.7 & 0.4 \\ 
12--14   & 2.819 $10^{-2}$ & 2.859 $10^{-2}$ & 0.35 & 0.27 & 0.34 & 2.823 $10^{-2}$ & 2.876 $10^{-2}$ & 0.38 & 0.32 & 0.35 & 2.66 $10^{-2}$ & 2.71 $10^{-2}$ & 1.1 & 0.7 & 0.4 \\ 
14--16   & 2.375 $10^{-2}$ & 2.412 $10^{-2}$ & 0.37 & 0.27 & 0.35 & 2.385 $10^{-2}$ & 2.427 $10^{-2}$ & 0.40 & 0.32 & 0.34 & 2.27 $10^{-2}$ & 2.33 $10^{-2}$ & 1.3 & 0.7 & 0.5 \\ 
16--18   & 1.997 $10^{-2}$ & 2.028 $10^{-2}$ & 0.42 & 0.29 & 0.35 & 2.034 $10^{-2}$ & 2.070 $10^{-2}$ & 0.44 & 0.35 & 0.35 & 1.99 $10^{-2}$ & 2.03 $10^{-2}$ & 1.4 & 0.8 & 0.5 \\ 
18--22   & 1.587 $10^{-2}$ & 1.609 $10^{-2}$ & 0.35 & 0.27 & 0.34 & 1.606 $10^{-2}$ & 1.634 $10^{-2}$ & 0.39 & 0.32 & 0.35 & 1.60 $10^{-2}$ & 1.64 $10^{-2}$ & 1.2 & 0.6 & 0.5 \\ 
22--26   & 1.187 $10^{-2}$ & 1.199 $10^{-2}$ & 0.41 & 0.29 & 0.35 & 1.217 $10^{-2}$ & 1.228 $10^{-2}$ & 0.47 & 0.36 & 0.36 & 1.23 $10^{-2}$ & 1.24 $10^{-2}$ & 1.4 & 0.8 & 0.6 \\ 
26--30   & 9.065 $10^{-3}$ & 9.113 $10^{-3}$ & 0.46 & 0.31 & 0.35 & 9.275 $10^{-3}$ & 9.340 $10^{-3}$ & 0.52 & 0.41 & 0.35 & 9.68 $10^{-3}$ & 9.81 $10^{-3}$ & 1.7 & 0.8 & 0.6 \\ 
30--34   & 7.143 $10^{-3}$ & 7.165 $10^{-3}$ & 0.53 & 0.35 & 0.35 & 7.339 $10^{-3}$ & 7.363 $10^{-3}$ & 0.59 & 0.46 & 0.35 & 7.82 $10^{-3}$ & 7.90 $10^{-3}$ & 1.8 & 0.9 & 0.5 \\ 
34--38   & 5.707 $10^{-3}$ & 5.707 $10^{-3}$ & 0.59 & 0.38 & 0.34 & 5.880 $10^{-3}$ & 5.883 $10^{-3}$ & 0.66 & 0.49 & 0.35 & 6.34 $10^{-3}$ & 6.34 $10^{-3}$ & 2.0 & 1.0 & 0.5 \\ 
38--42   & 4.559 $10^{-3}$ & 4.554 $10^{-3}$ & 0.66 & 0.44 & 0.35 & 4.709 $10^{-3}$ & 4.704 $10^{-3}$ & 0.74 & 0.51 & 0.35 & 5.09 $10^{-3}$ & 5.09 $10^{-3}$ & 2.2 & 1.2 & 0.4 \\ 
42--46   & 3.757 $10^{-3}$ & 3.747 $10^{-3}$ & 0.73 & 0.47 & 0.35 & 3.745 $10^{-3}$ & 3.739 $10^{-3}$ & 0.82 & 0.57 & 0.38 & 4.38 $10^{-3}$ & 4.40 $10^{-3}$ & 2.4 & 1.2 & 0.4 \\ 
46--50   & 3.150 $10^{-3}$ & 3.140 $10^{-3}$ & 0.79 & 0.48 & 0.38 & 3.156 $10^{-3}$ & 3.134 $10^{-3}$ & 0.86 & 0.62 & 0.37 & 3.57 $10^{-3}$ & 3.55 $10^{-3}$ & 2.6 & 1.4 & 0.4 \\ 
50--54   & 2.584 $10^{-3}$ & 2.575 $10^{-3}$ & 0.88 & 0.52 & 0.36 & 2.568 $10^{-3}$ & 2.556 $10^{-3}$ & 0.99 & 0.67 & 0.36 & 3.00 $10^{-3}$ & 2.99 $10^{-3}$ & 2.9 & 1.5 & 0.6 \\ 
54--60   & 2.052 $10^{-3}$ & 2.040 $10^{-3}$ & 0.81 & 0.48 & 0.37 & 2.125 $10^{-3}$ & 2.110 $10^{-3}$ & 0.92 & 0.59 & 0.35 & 2.66 $10^{-3}$ & 2.65 $10^{-3}$ & 2.7 & 1.3 & 0.4 \\ 
60--70   & 1.466 $10^{-3}$ & 1.457 $10^{-3}$ & 0.73 & 0.46 & 0.39 & 1.494 $10^{-3}$ & 1.481 $10^{-3}$ & 0.87 & 0.64 & 0.39 & 1.82 $10^{-3}$ & 1.80 $10^{-3}$ & 2.5 & 1.3 & 0.4 \\ 
70--80   & 9.646 $10^{-4}$ & 9.557 $10^{-4}$ & 0.92 & 0.55 & 0.43 & 9.979 $10^{-4}$ & 9.845 $10^{-4}$ & 1.08 & 0.71 & 0.40 & 1.14 $10^{-3}$ & 1.12 $10^{-3}$ & 3.3 & 1.6 & 0.7 \\ 
80--100  & 5.458 $10^{-4}$ & 5.413 $10^{-4}$ & 0.83 & 0.53 & 0.47 & 5.566 $10^{-4}$ & 5.509 $10^{-4}$ & 0.99 & 0.69 & 0.48 & 5.96 $10^{-4}$ & 5.89 $10^{-4}$ & 3.1 & 1.4 & 0.8 \\ 
100--150 & 1.874 $10^{-4}$ & 1.859 $10^{-4}$ & 0.83 & 0.54 & 0.57 & 1.974 $10^{-4}$ & 1.954 $10^{-4}$ & 1.00 & 0.70 & 0.71 & 1.98 $10^{-4}$ & 1.96 $10^{-4}$ & 3.3 & 1.5 & 2.1 \\ 
150--200 & 4.826 $10^{-5}$ & 4.794 $10^{-5}$ & 1.67 & 0.74 & 0.51 & 4.990 $10^{-5}$ & 4.959 $10^{-5}$ & 2.03 & 0.99 & 0.69 & 5.08 $10^{-5}$ & 5.05 $10^{-5}$ & 6.7 & 2.8 & 2.2 \\ 
200--300 & 1.126 $10^{-5}$ & 1.124 $10^{-5}$ & 2.38 & 1.40 & 1.43 & 1.018 $10^{-5}$ & 1.011 $10^{-5}$ & 3.17 & 2.05 & 1.20 & 9.09 $10^{-6}$ & 9.12 $10^{-6}$ & 10.9 & 4.4 & 0.8 \\ 
300--800 & 4.783 $10^{-7}$ & 4.768 $10^{-7}$ & 5.02 & 2.00 & 1.50 & 3.048 $10^{-7}$ & 3.028 $10^{-7}$ & 8.02 & 3.67 & 1.03 & 1.47 $10^{-7}$ & 1.45 $10^{-7}$ & 34.0 & 15.8 & 0.9 \\ \hline
\end{tabular}}
\caption{The measured normalized combined (electron and muon channels) cross section
$(1/\sigma^{\mathrm{fid}})(\der\sigma^{\mathrm{fid}}/\der\ptz)$, for
$0\leq|\yz|<1$, $1\leq|\yz|<2$ and $2\leq|\yz|<2.4$. The
cross sections at Born and dressed levels are given as well as the
relative statistical ($\delta_{\rm {Stat}}$) and systematic  ($\delta_{\rm {Syst}}$)
uncertainties for uncorrelated and correlated sources.}
\label{tab:combined2}
\end{sidewaystable}

\begin{figure}
  \centering
  \includegraphics[width=0.8\textwidth]{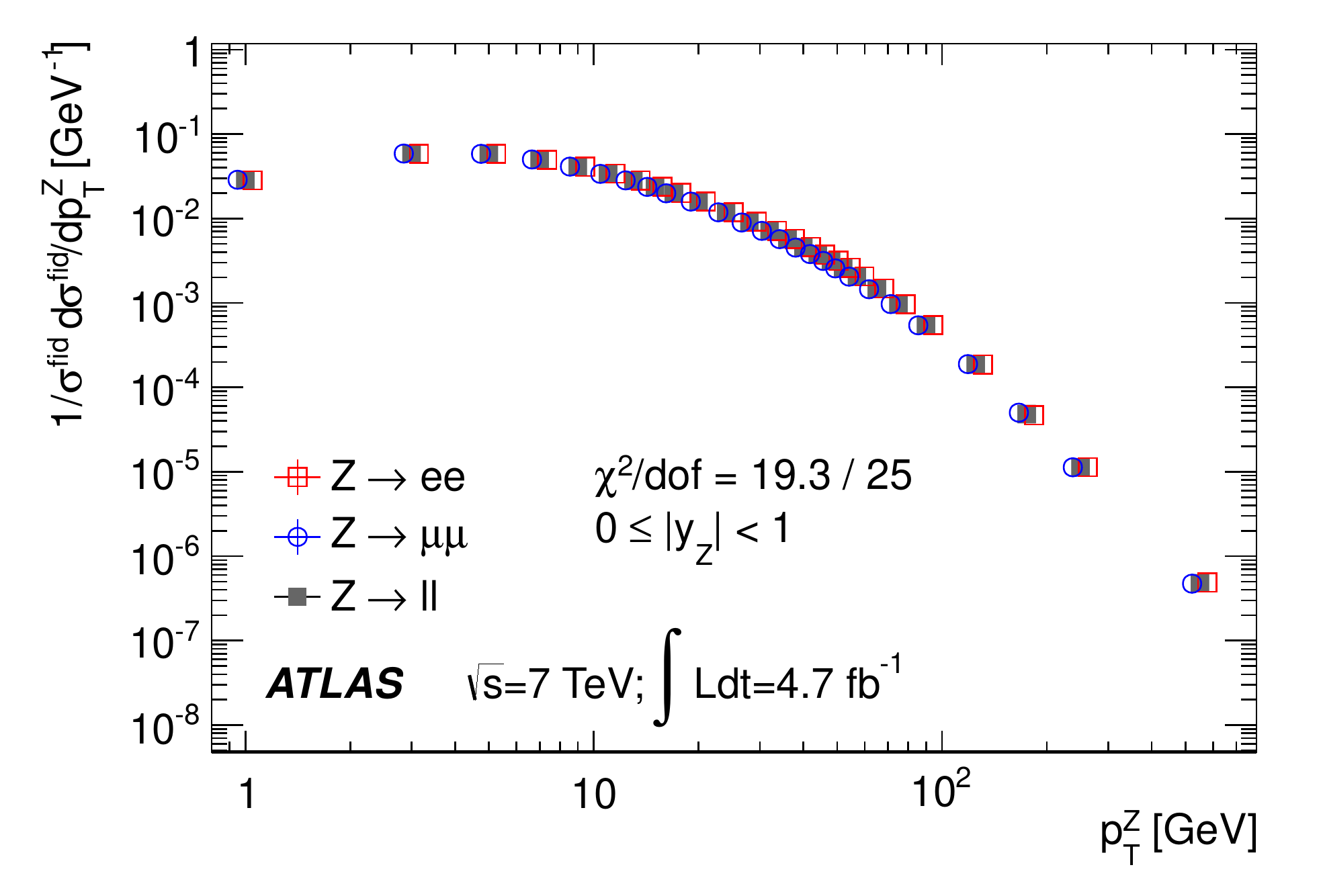}
  \includegraphics[width=0.8\textwidth]{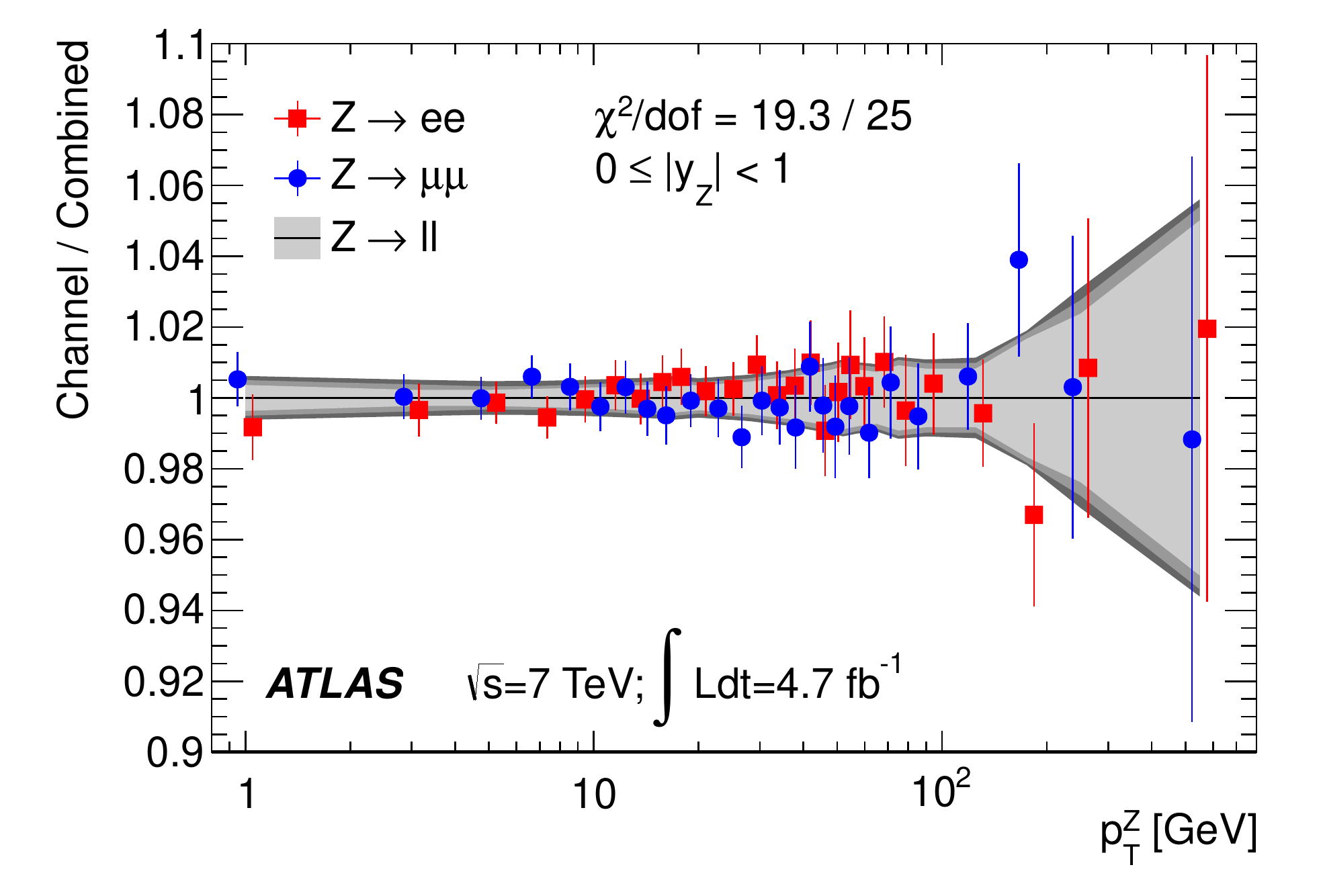}
  \caption{The measured normalized cross section
  $(1/\sigma^{\mathrm{fid}})(\der\sigma^{\mathrm{fid}}/\der\ptz)$
  for $0\leq|\yz|<1$, as a function of \ptz\ for the electron and muon channels and the
  combined result (top). Ratio of the electron and muon channels to the
  combined result (bottom). The uncertainty bands represent the statistical, total
  uncorrelated and total uncertainties, from light gray to dark gray respectively.}
  \label{fig:combinationRap1}
\end{figure}

\begin{figure}
  \centering
  \includegraphics[width=0.8\textwidth]{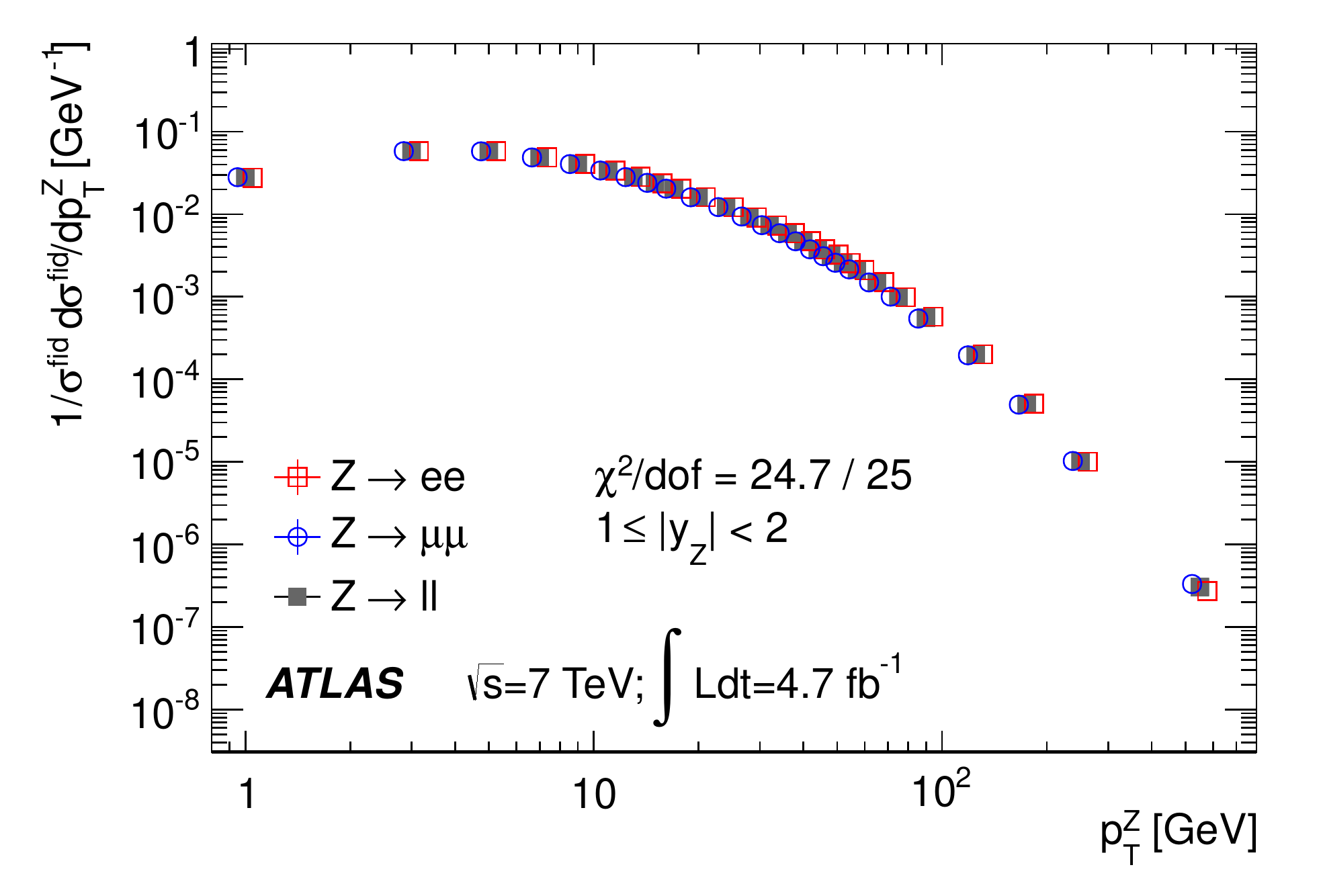}
  \includegraphics[width=0.8\textwidth]{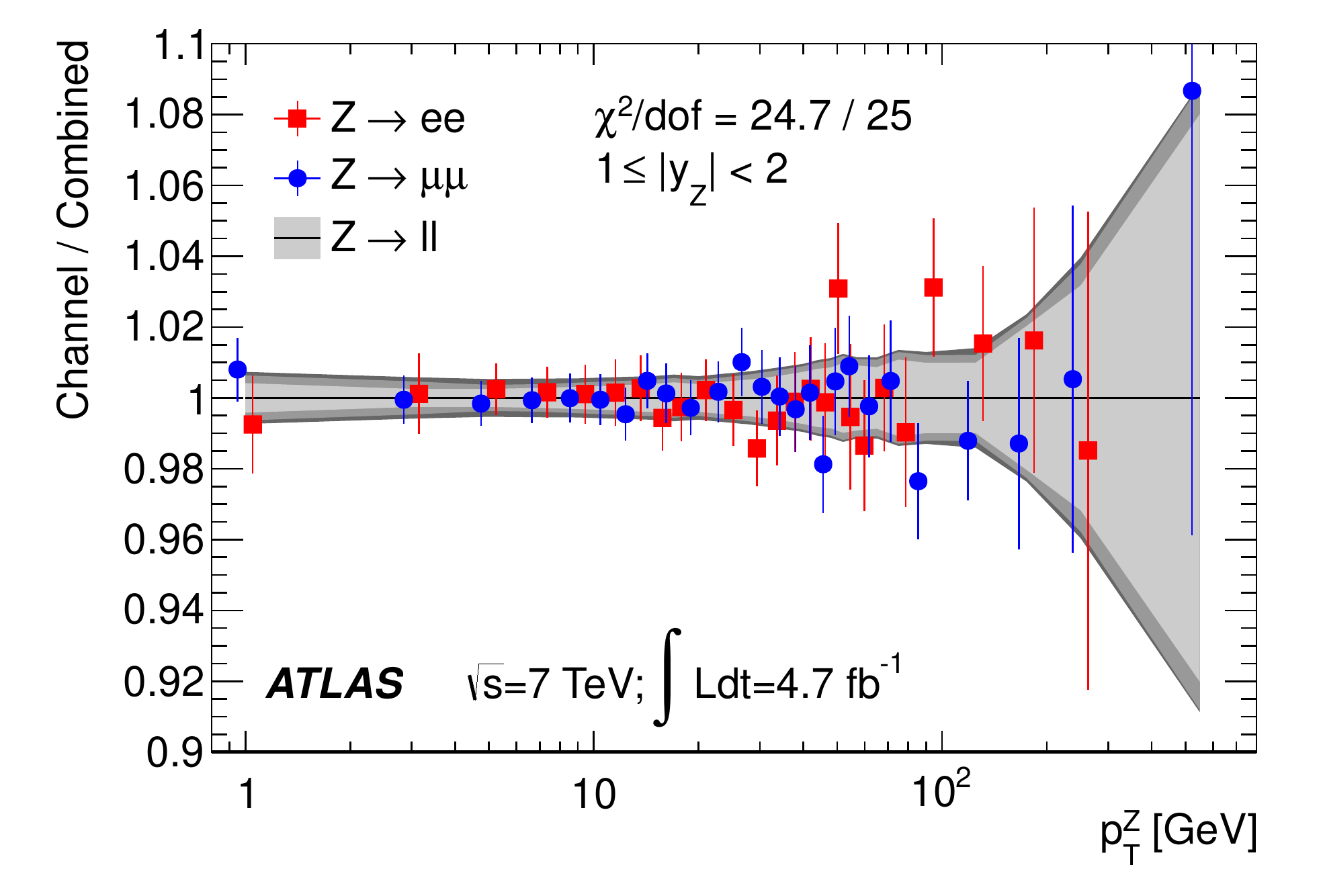}
  \caption{The measured normalized cross section
  $(1/\sigma^{\mathrm{fid}})(\der\sigma^{\mathrm{fid}}/\der\ptz)$
  for $1\leq|\yz|<2$, as a function of \ptz\ for the electron and muon channels and the
  combined result (top). Ratio of the electron and muon channels to the
  combined result (bottom). The uncertainty bands represent the statistical, total
  uncorrelated and total uncertainties, from light gray to dark gray respectively.}
  \label{fig:combinationRap2}
\end{figure}

\begin{figure}
  \centering
  \includegraphics[width=0.8\textwidth]{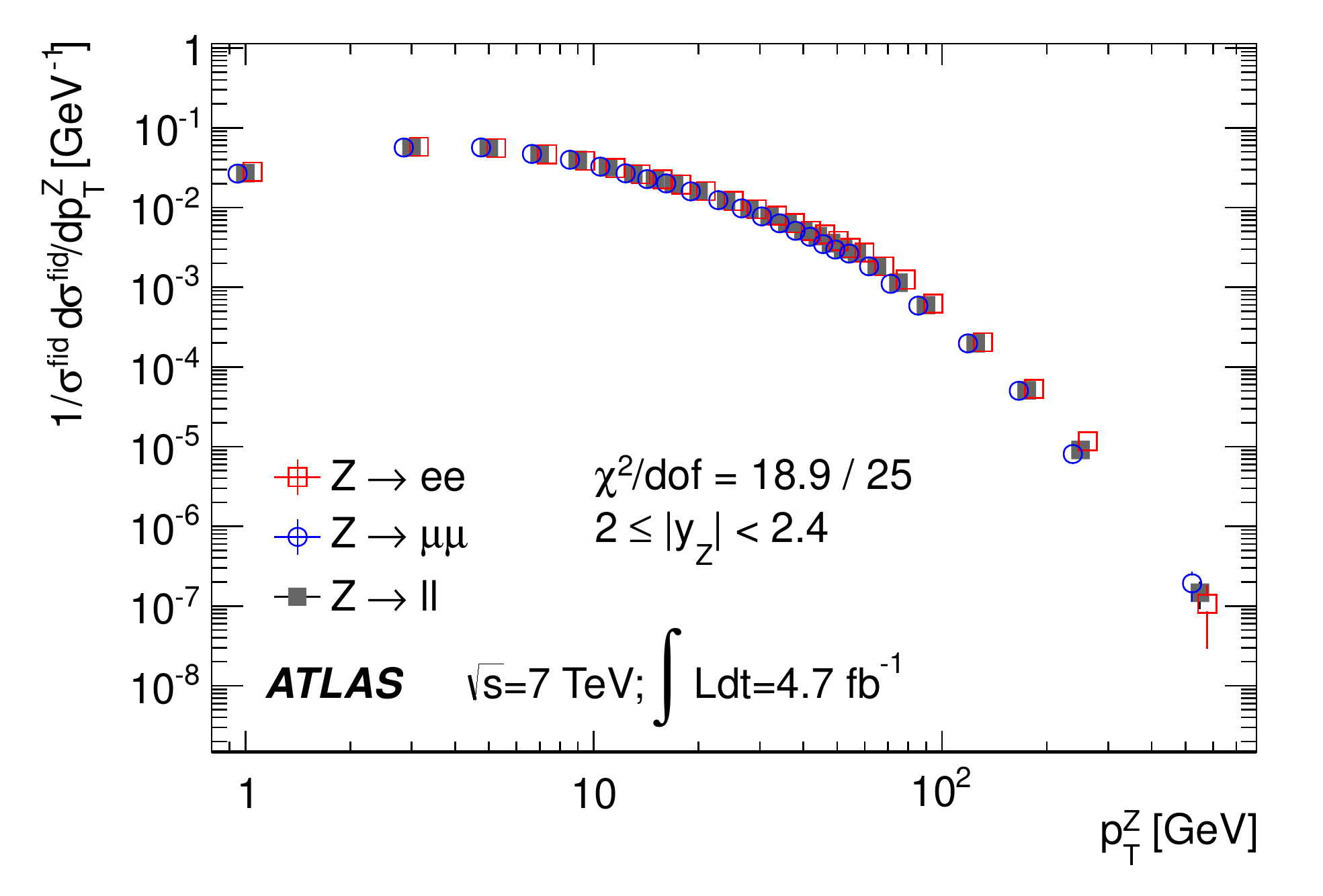}
  \includegraphics[width=0.8\textwidth]{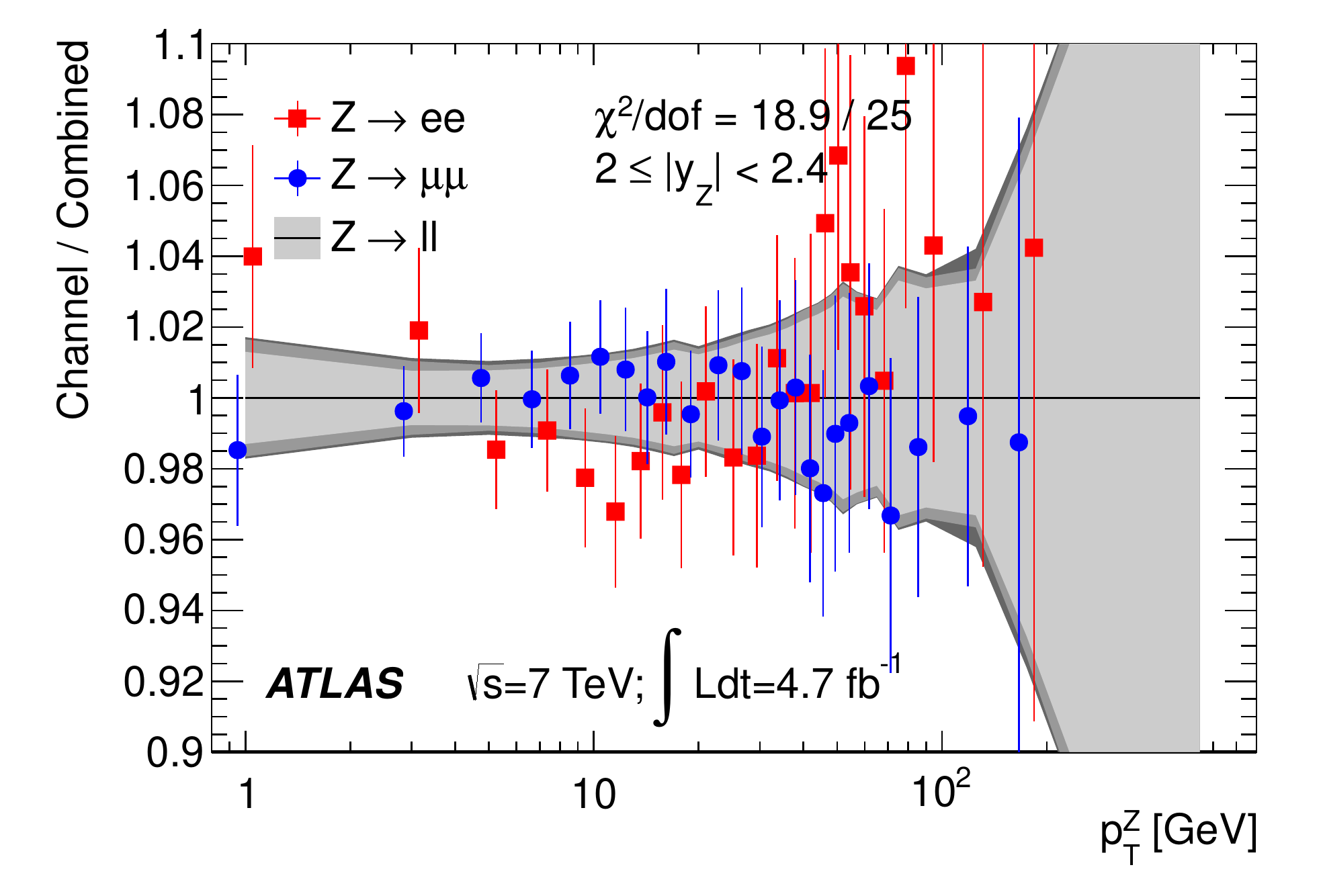}
  \caption{The measured normalized cross section
  $(1/\sigma^{\mathrm{fid}})(\der\sigma^{\mathrm{fid}}/\der\ptz)$
  for $2\leq|\yz|<2.4$, as a function of \ptz\ for the electron and muon channels and the
  combined result (top). Ratio of the electron and muon channels to the
  combined result (bottom). The uncertainty bands represent the statistical, total
  uncorrelated and total uncertainties, from light gray to dark gray respectively.}
  \label{fig:combinationRap3}
\end{figure}

\clearpage
\newpage

\section{Comparison to QCD predictions}
\label{sec:compaqcd}

In figure~\ref{fig:comp_theo}, the Born-level combined result is
compared to theoretical predictions at fixed order
from {\sc Fewz} and {\sc Dynnlo}, to {\sc ResBos} and to the NNLO+NNLL calculation of ref.~\cite{Banfi:2012du}. 
{\sc Fewz}, {\sc Dynnlo} and {\sc ResBos} use the CT10 PDFs,
while the NNLO+NNLL calculation of ref.~\cite{Banfi:2012du} uses the CTEQ6m PDFs~\cite{Pumplin:2002vw}.

The uncertainty on the predictions, estimated from the PDF uncertainties and
renormalization and factorization scale variations, are in all cases much larger  
than the measurement uncertainties. 
The disagreement between the data and the 
{\sc Fewz} and {\sc Dynnlo} predictions is larger than the
data uncertainties, reaching 10\% 
around 50~\GeV\ and diverging at low \ptz\ as expected from the absence of
resummation effects in these calculations. {\sc Fewz} and {\sc Dynnlo}
agree with each other when using QCD renormalization and
factorization scales, $\mu_{\rm R}$ and $\mu_{\rm F}$, defined as
$\mu_{\rm R}=\mu_{\rm F}=m_{Z}$ and leading-order electroweak perturbative
accuracy. 
The influence of the QCD scale choice is studied with {\sc
Dynnlo} by using the alternative dynamic scale $E_{\rm T}^Z$, defined
as the sum in quadrature of $m_{Z}$ and \ptz. The resulting \ptz\ shape is in better agreement with the
data for $\ptz>30$~GeV, but the normalization remains low by 10\% in
this region. 
NLO electroweak corrections to $Z$+jet production~\cite{Denner:2011vu} are applied 
to the dynamic-scale {\sc Dynnlo} prediction and lead to a decrease of
the cross section of 10\% in the highest \ptz\ bin.

The {\sc ResBos-GNW} prediction agrees with the data within 5--7\%; the
prediction uncertainties are defined from PDF, 
renormalization scale and factorization scale variations. The {\sc
ResBos}-BLNY prediction, to which the previous ATLAS
measurements~\cite{Aad:2011gj,Aad:2011fp,Aad:2013ps} were compared, is
included for reference. The NNLO+NNLL calculation following ref.~\cite{Banfi:2012du} matches the data
within 10--12\%. The uncertainties on this prediction are defined from resummation, renormalization and
factorization scale variations; PDF uncertainties are neglected. In
both cases, the prediction uncertainties are almost sufficient to
cover the difference with the data. 

\begin{figure}
  \centering
  \includegraphics[width=0.49\textwidth]{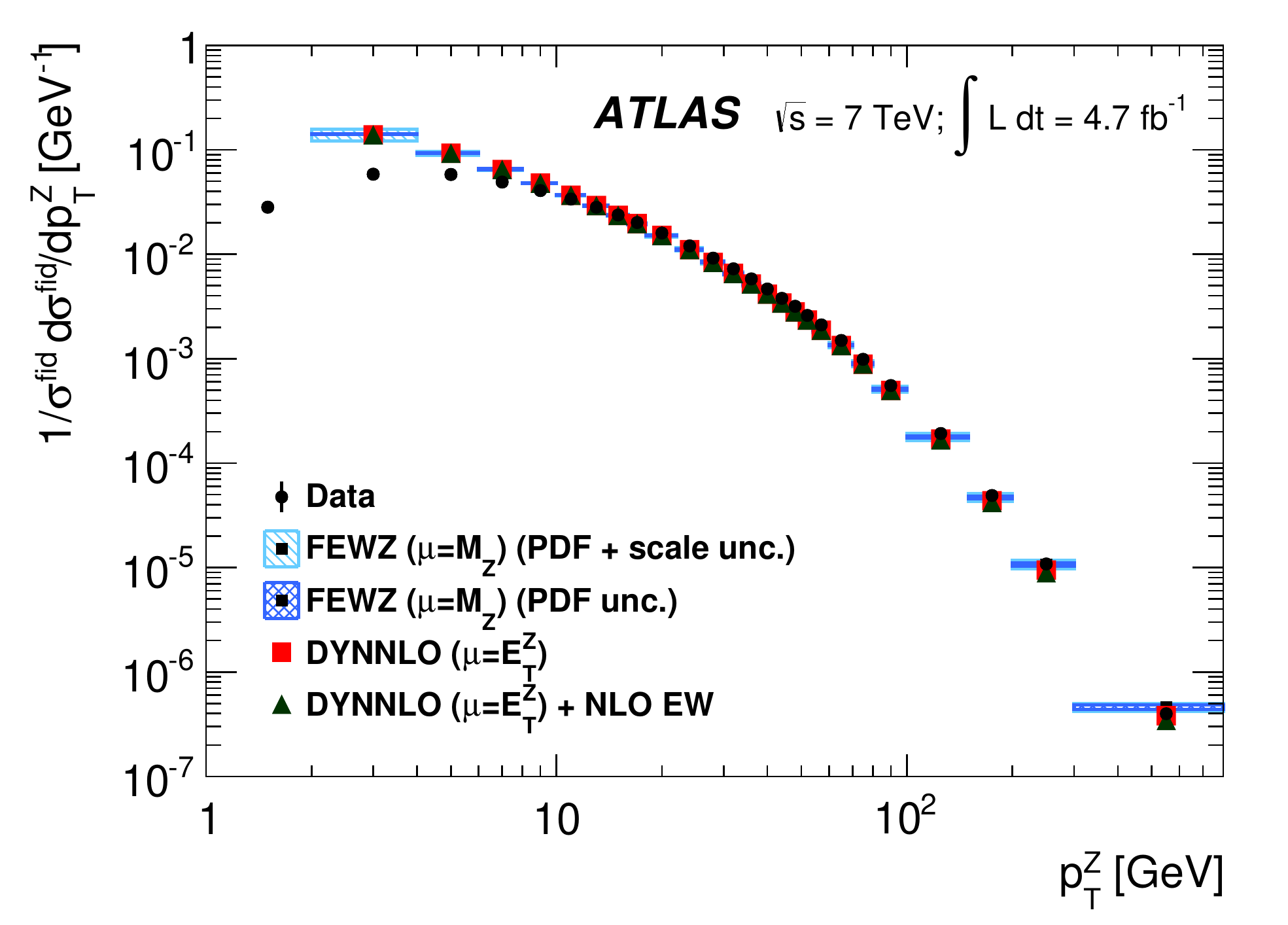}
  \includegraphics[width=0.49\textwidth]{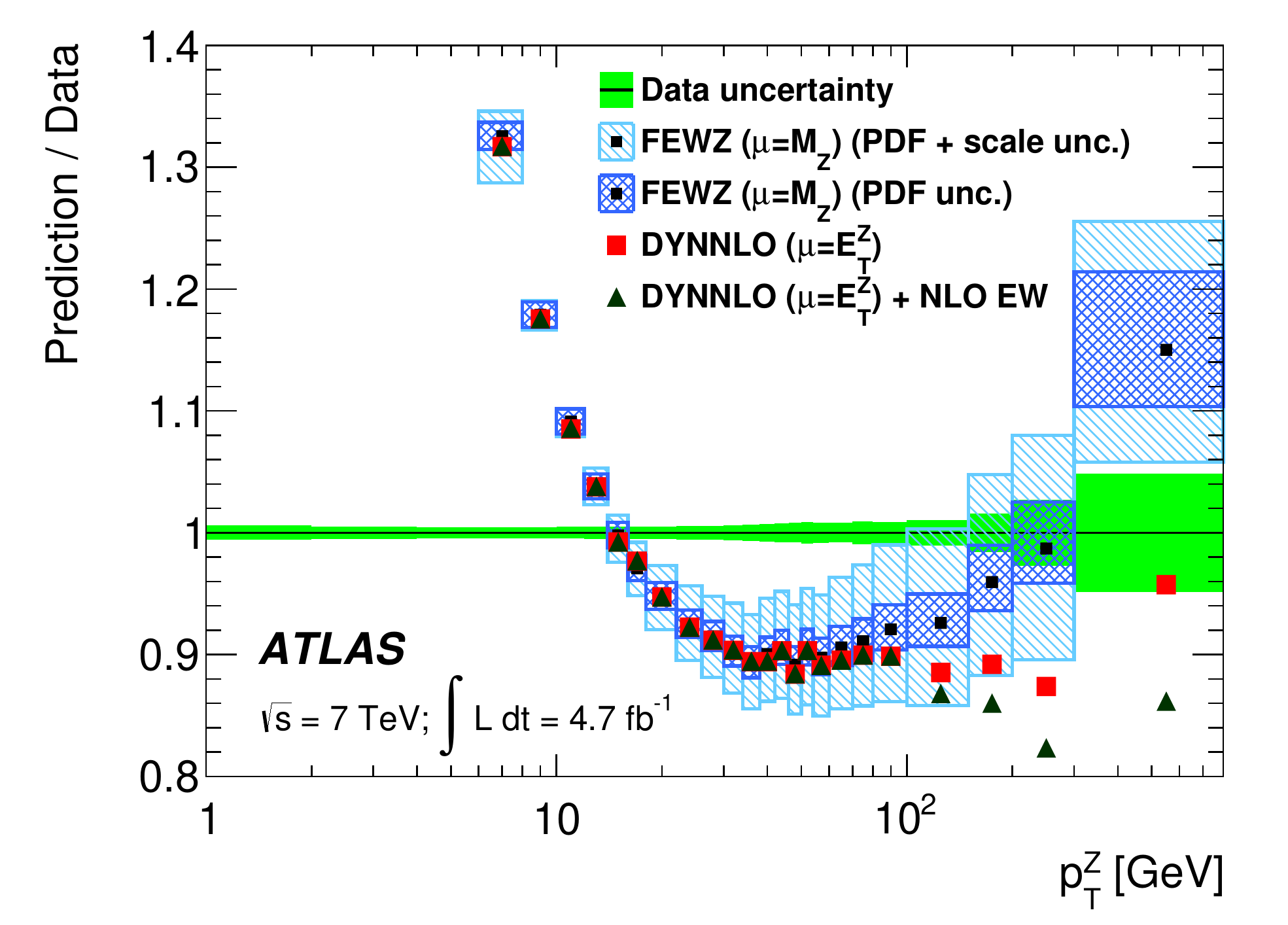}
  \includegraphics[width=0.49\textwidth]{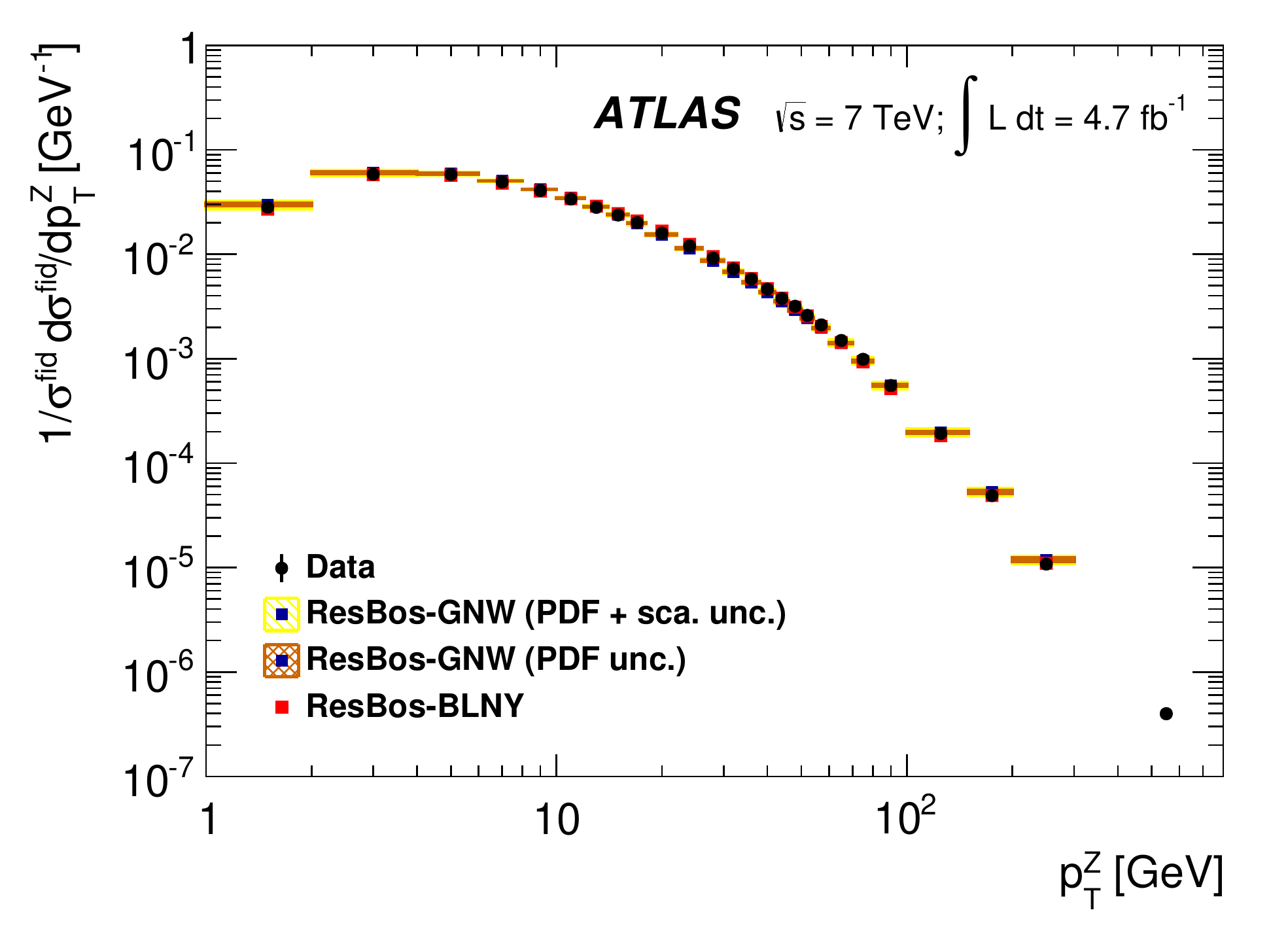}
  \includegraphics[width=0.49\textwidth]{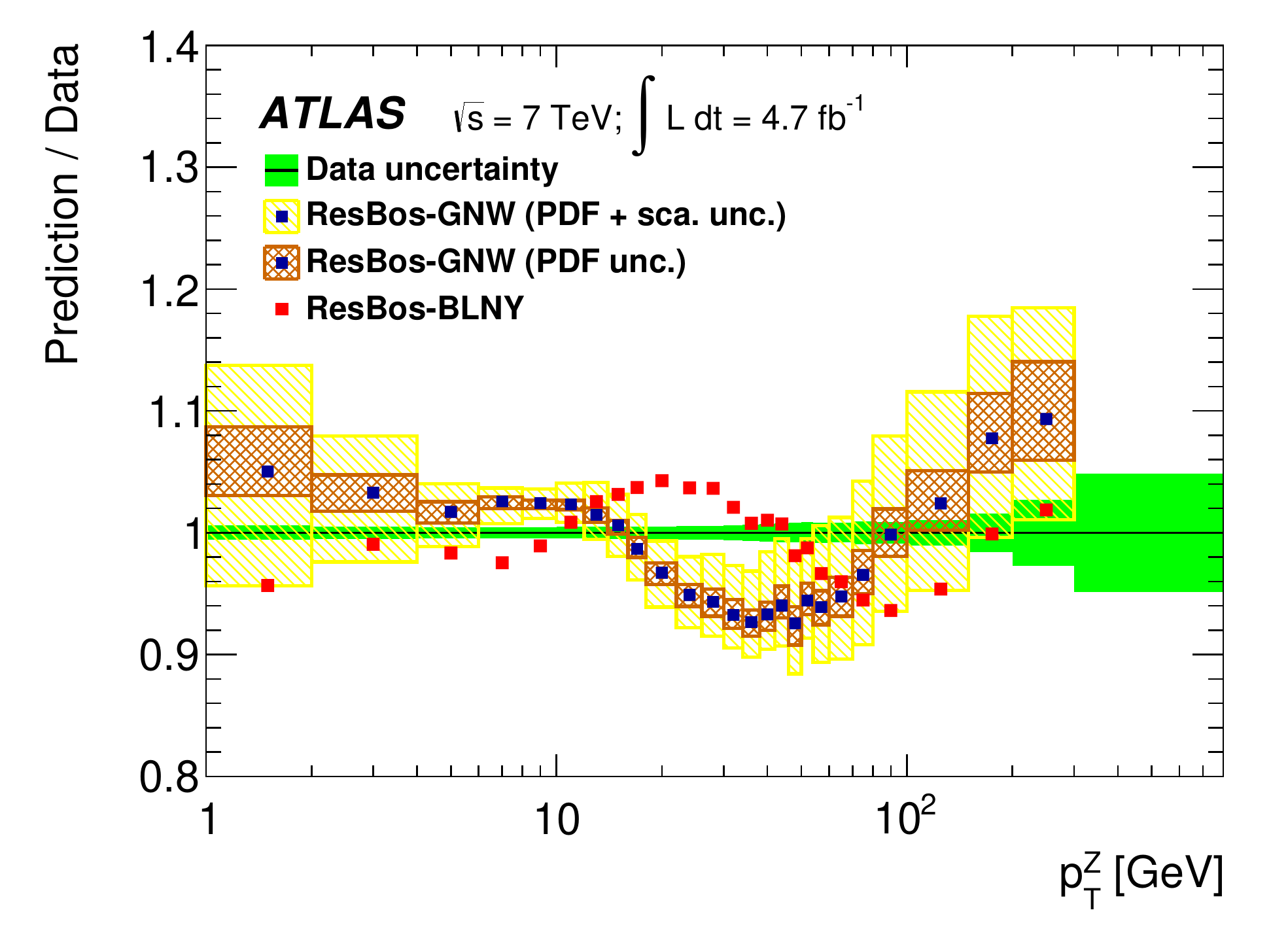}
  \includegraphics[width=0.49\textwidth]{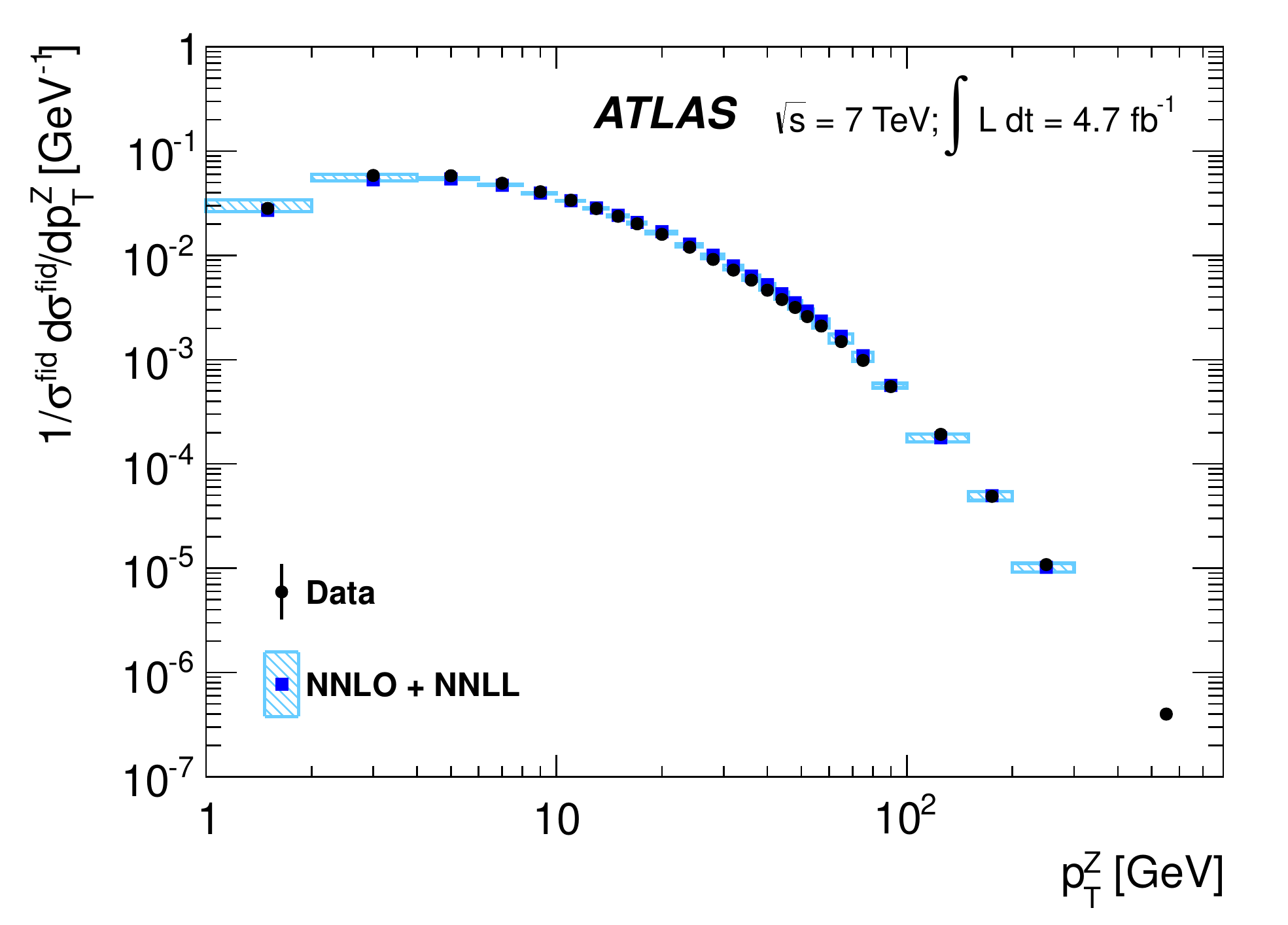}
  \includegraphics[width=0.49\textwidth]{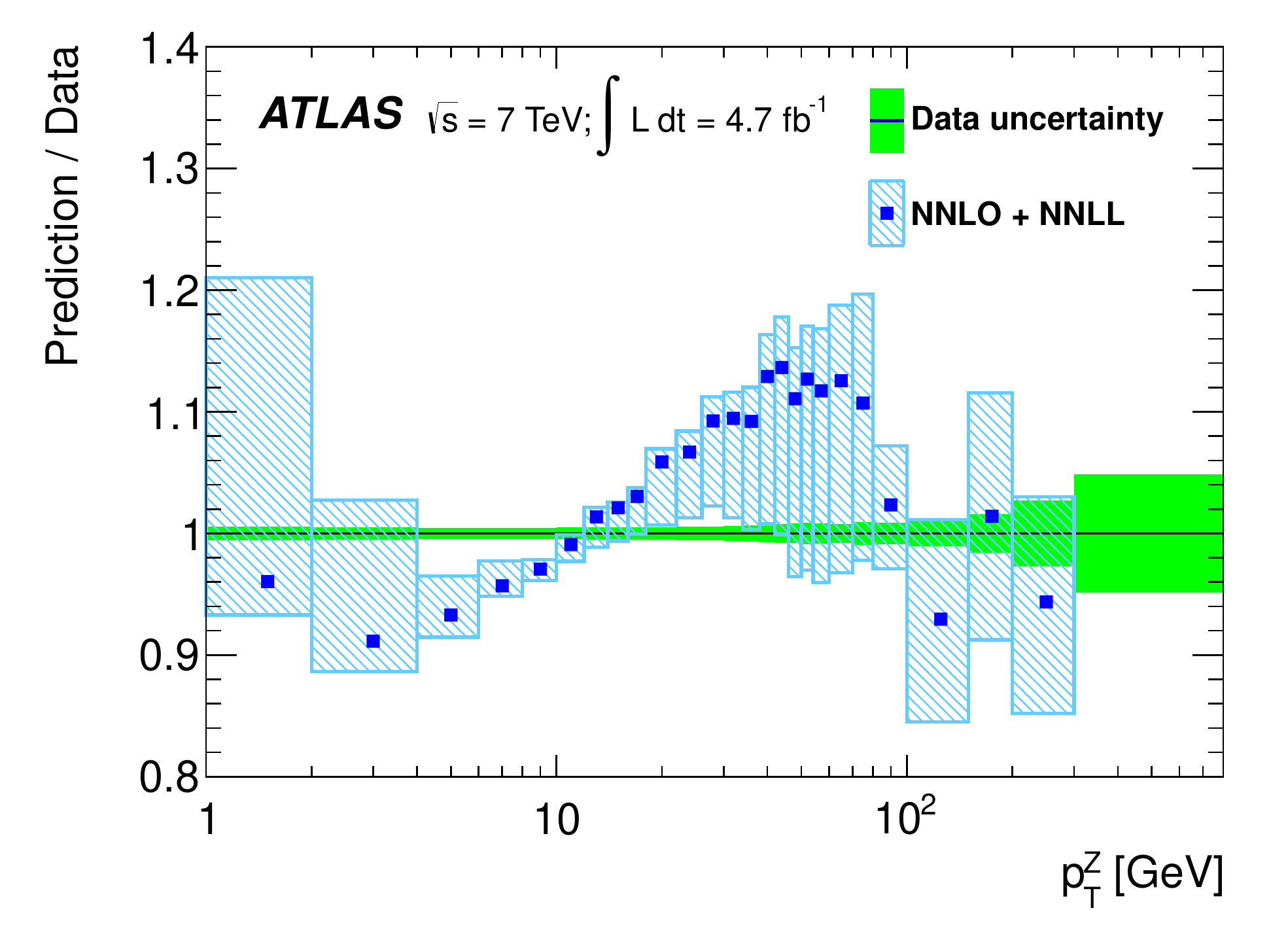}
  \caption{Left : comparison of the \ptz\ distributions predicted by different computations: {\sc Fewz} and {\sc Dynnlo}
   (top),  {\sc ResBos} (middle) and the NNLO+NNLL calculation of ref.~\cite{Banfi:2012du} (bottom) with the
    Born-level combined measurement, inclusively in $\yz$. Right :
    ratios between these predictions and the combined measurement.}
  \label{fig:comp_theo}
\end{figure}

Figure~\ref{fig:comp_gen} shows the ratio of the \ptz\ distributions
predicted by different generators to the combined measurements performed inclusively in  $Z$ rapidity,
and in the three exclusive $Z$ rapidity bins described above.
The {\sc Pythia} and {\sc Powheg} generators agree with the data to within ~5\% in the $2<\ptz<60$~\GeV\ range, 
and to within 20\% over the full range. 
\mcatnlo\ shows a similar level of agreement with the data for $\ptz<30$~GeV
 but develops a discrepancy up to around 40\% at the end of the spectrum. {\sc
Sherpa} and {\sc Alpgen} agree with the data to within about 5\% for
$5<\ptz<200$~\GeV, but tend to overestimate the distribution near the
end of the spectrum.

\begin{figure}
  \centering
  \includegraphics[width=0.49\textwidth]{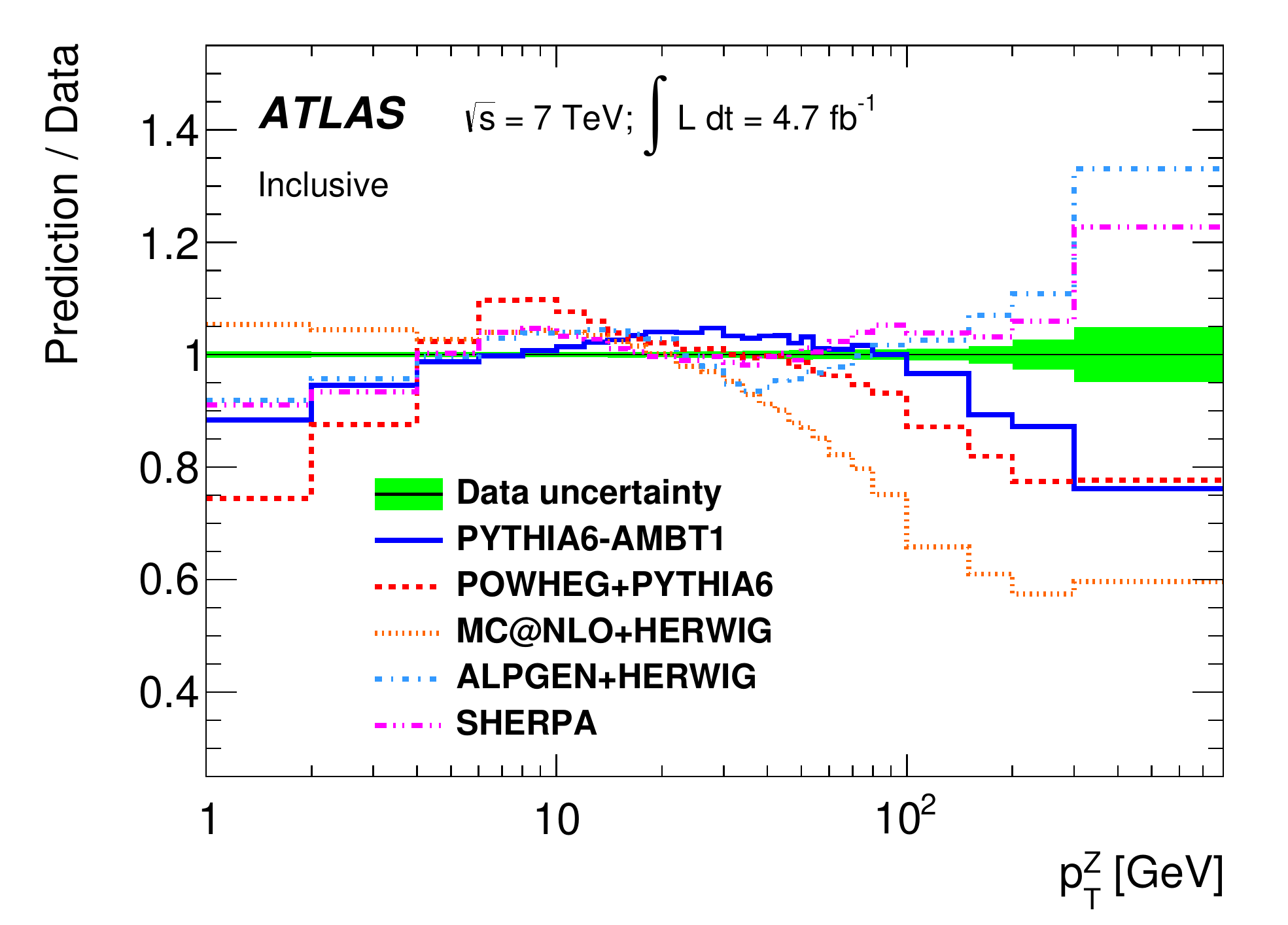}
  \includegraphics[width=0.49\textwidth]{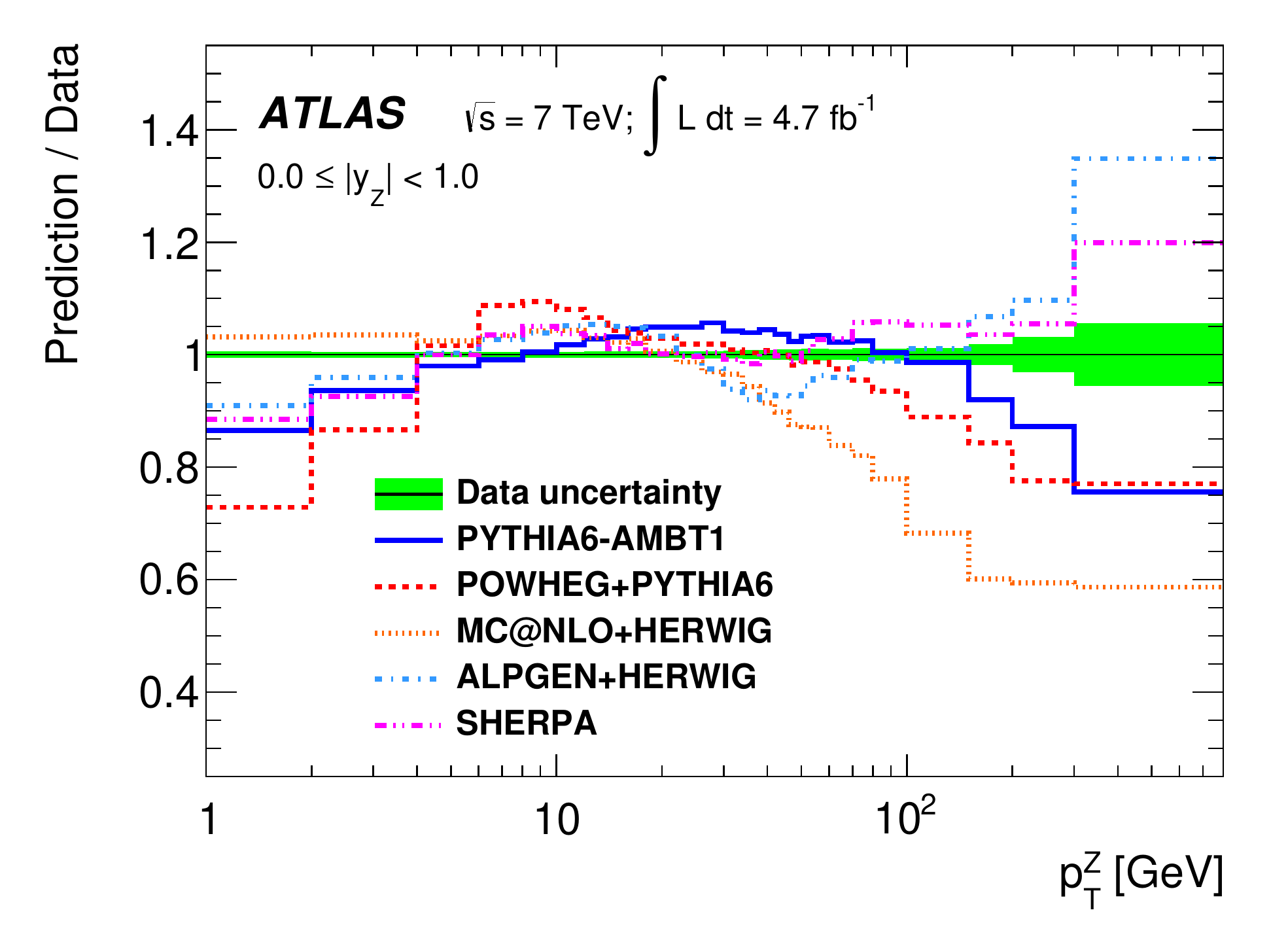}
  \includegraphics[width=0.49\textwidth]{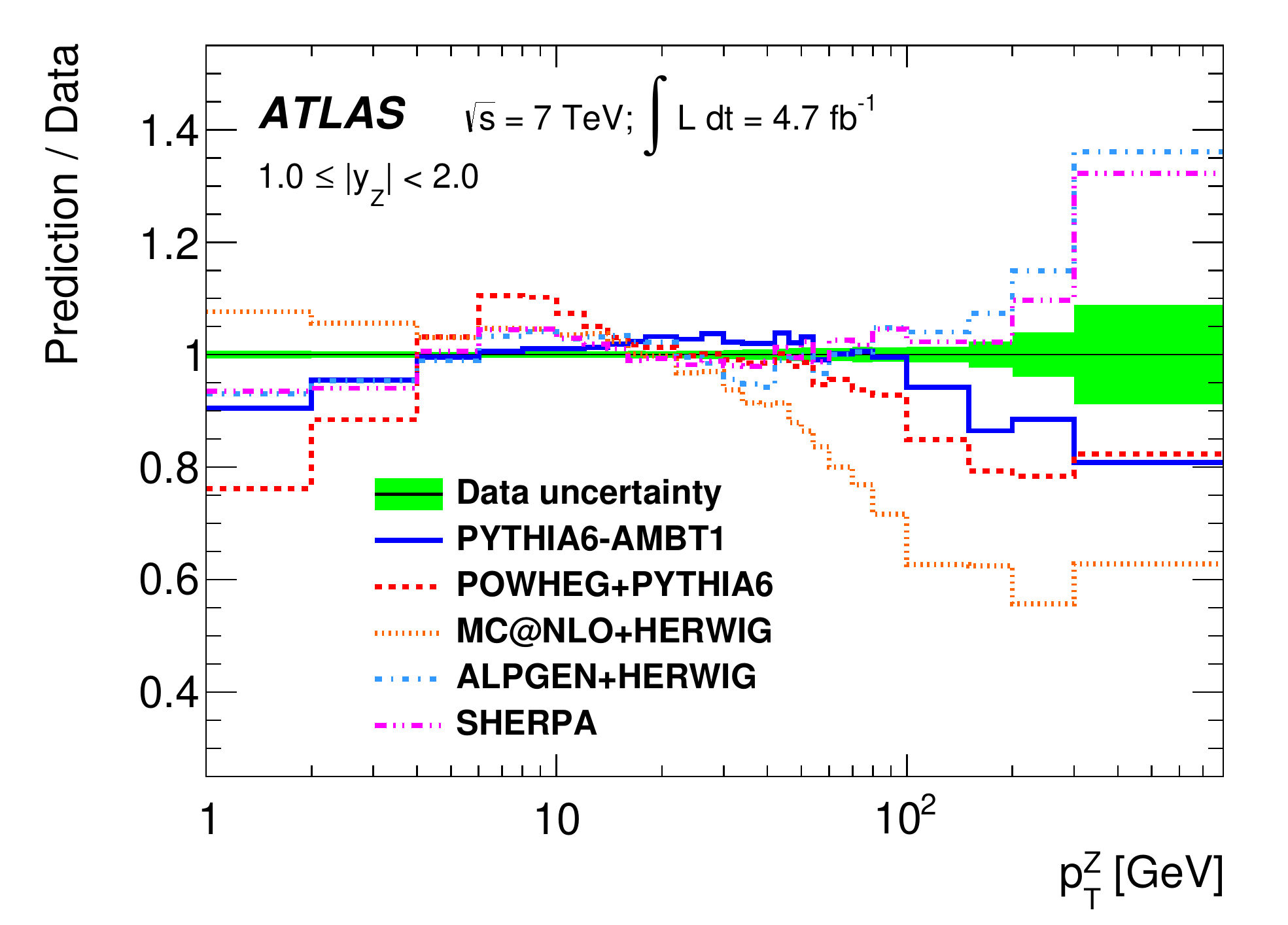}
  \includegraphics[width=0.49\textwidth]{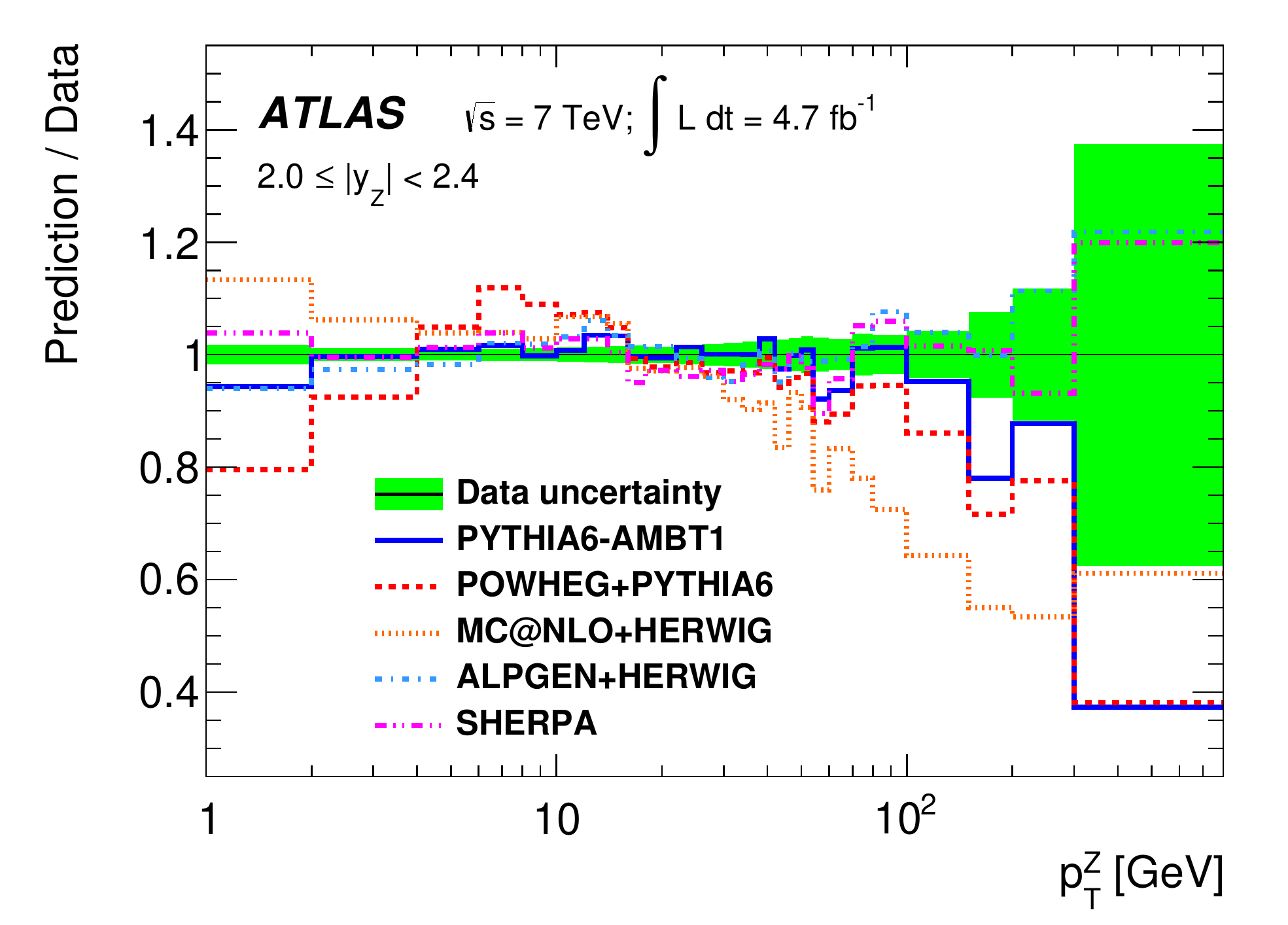}
  \caption{Ratio of the \ptz\ distribution predicted by different MC generators to
the Born-level combined measurement, for the inclusive measurement and
  for $0\leq|\yz|<1$, $1\leq|\yz|<2$ and $2\leq|\yz|<2.4$.}
  \label{fig:comp_gen}
\end{figure}

\section{Tuning of {\sc Pythia8} and {\sc Powheg}+{\sc Pythia8}}
\label{sec:tuning}

The parton shower tunes presented below are performed to determine the
sensitivity of the measured \ptz\ cross sections presented here to
parton shower model parameters in state-of-the-art MC generators, and
to constrain the models by trying to achieve precise predictions of
vector boson production. The ATLAS $\phi^\star_\eta$
measurement~\cite{Aad:2013ps} is also exploited as it is highly
correlated to \ptz\ and  is hence sensitive to the same model
components.   

The  {\sc Pythia8} generator with the $p_{\rm {T}}$-ordered, interleaved parton
shower is chosen for these studies. {\sc Pythia8} is used in standalone mode and in a
configuration interfaced to {\sc Powheg}. To minimize dependence on QED
final-state corrections, the tunes use the dressed-level measurement
results. The study is restricted to the low \ptz\ range, where parton
shower effects dominate. The tunes are performed for \ptz$<26$~GeV,
which is found to be most sensitive to the model parameters described below, and
$\phi^{\star}_\eta<0.29$, which covers a similar transverse momentum 
range. The measurement inclusive in rapidity is used for the tuning, and the 
compatibility of the tuned predictions with the data in the separate
rapidity bins is then evaluated. 

For {\sc Pythia8}, the parton shower model components under
consideration include the strong coupling constant used for the  parton shower 
evolution $\alpha_{\rm {S}}^{\rm {ISR}}(m_Z)$, and the parton shower lower cut-off $p_{\rm{}T0}$ in the
non-perturbative regime, implemented as a smooth damping factor $p_{\rm{}T}^2
/ (p_{\rm{}T0}^2 + p_{\rm{}T}^2)$. To populate the region below $p_{\rm{}T0}$,
the partons initiating the hard scattering process are assumed to have
a primordial transverse momentum $k_{\rm {T}}$ following
a Gaussian distribution with tunable width. The {\sc Pythia8} parton shower also
includes QED emissions, but the corresponding cut-off values and coupling
strength are left to the program defaults. The steerable parameters
not used in the tuning are set to the values defined  by the
tune 4C~\cite{Corke:2010yf}. 

{\sc Powheg} calculates the hardest (highest \pt) QCD radiation provided that it is above  
a transverse momentum threshold $p_{\rm{}T,min}^2$, which is a steerable parameter in the
program. Below $p_{\rm{}T,min}^2$, {\sc Powheg} generates events
without extra radiation and the phase space is populated by
{\sc Pythia8}. Therefore, the upper  limit of the {\sc Pythia8} parton
shower should match the {\sc Powheg} cut-off
value. The tunes are performed using $p_{\rm{}T,min}^2=4$~GeV$^2$,
corresponding to $\ptz = 2$~GeV. In addition, in order to avoid
discontinuities in the matched spectrum, the $\alpha_{\rm {S}}(m_Z)$ value used to
calculate the QCD radiation in {\sc Powheg} should match
$\alpha_{\rm {S}}^{\rm {ISR}}(m_Z)$ in {\sc Pythia}; $\alpha_{\rm {S}}(m_Z)=0.118$ is used as in 
the CT10 PDFs. Correspondingly the running of $\alpha_{\rm {S}}$  in the parton shower calculation is set to NLO.
The tuning of {\sc Powheg+Pythia8} hence only varies the shower
cut-off and the primordial $k_{\rm{}T}$ in {\sc Pythia8}. The
other steerable parameters not used in the tuning are set to the values
defined by the 4C tune. 

The tunes are performed using the Professor~\cite{Buckley:2009bj} package, 
which interpolates the dependence of MC predictions on the model
parameters as originally proposed in ref.~\cite{Abreu:1996na}. Predictions
for the \ptz\ distribution are generated at randomly chosen parameter
settings (anchor points) in the ranges indicated in table~\ref{tab:tune}. 
A fourth-order polynomial is used to approximate the generator predictions
between the anchor points. The optimal parameter values are determined
using a $\chi^2$ minimization between the interpolated generator
response and the data.

\begin{table}
\begin{center}
    \begin{tabular}{lcc}
\toprule
       Parameter           &    Variation Range & Variation Range\\ 
                                 &     {\sc Pythia8}  tune &  {\sc Pythia8+Powheg} tune \\
\midrule
       Primordial $k_{\rm{}T}$     [GeV]       &  1.0--2.5        &    0.5--2.5    \\
       ISR $\alpha_{\rm {S}}^{\rm {ISR}}(m_Z)$                    & 0.120--0.140    &    0.118     \\
       ISR cut-off          [GeV]             &  0.5--2.5       & 0.5--3.0 \\ 
       ISR $\alpha_{\rm {S}}$ order                   &  LO             &      NLO   \\
       {\sc Pythia8} base tune               & tune 4C          & tune 4C \\
\midrule
       {\sc Powheg} cut-off  [GeV$^2$]      & -                              & 4.0\\
 \bottomrule
     \end{tabular}
\end{center}
\caption{Parameter ranges and model switches used in the tuning of {\sc Pythia8} and {\sc Pythia8+Powheg} described in
section~\ref{sec:tuning}.}
\label{tab:tune}
\end{table}

The sensitivity of the generator parameters to the \ptz\
and $\phi^{\star}_{\eta}$ measurements is probed by performing
tunes of {\sc Pythia8} and {\sc Powheg+Pythia8} to each measurement separately. 
As shown in table~\ref{tab:ptphitunes} both measurements have
comparable sensitivity and yield compatible tuned parameter values.
As a further check of the compatibility between the \ptz\ and $\phi^\star_\eta$ measurements,
the \ptz-tuned and $\phi^\star_\eta$-tuned predictions are compared to
the measured \ptz\ distribution. The tuning uncertainty is obtained
from variations of the eigenvector components of the parameters error matrix 
over a range covering $\Delta\chi^2 = \chi^2_{\rm min}$/dof.
Figure~\ref{fig:tune_compatibility} shows that the tuned predictions
agree with the measured cross sections within 2\% for $\ptz<50$~GeV,
and with each other within the tuned parameter uncertainties.

\begin{table}
\begin{center}
\resizebox{\textwidth}{!}{
    \begin{tabular}{lcccc}
     \toprule
                             & \multicolumn{2}{c}{\sc Pythia8}               &  \multicolumn{2}{c}{\sc Powheg+Pythia8} \\
                             & \ptz{}             & $\phi^\star_\eta$          & \ptz{}                  & $\phi^\star_\eta$\\
\midrule
 Primordial $k_{\rm{}T}$ [GeV] & $1.74 \pm 0.03$     & $1.73 \pm 0.03$      & $1.75 \pm 0.03$    & $1.75 \pm 0.04 $ \\
 ISR $\alpha_{\rm {S}}^{\rm {ISR}}(m_Z)$         & $0.1233 \pm 0.0003$  & $0.1238 \pm
 0.0002$  & 0.118 (fixed)                  & 0.118 (fixed)               \\
 ISR cut-off [GeV]           & $0.66 \pm 0.14 $    & $0.58 \pm 0.07$      & $2.06 \pm 0.12$    & $1.88 \pm 0.12$  \\ 
\midrule
 $\chi^2_{\rm{}min}$/dof              & 23.9/19            & 59.9/45             & 18.5/20           & 68.2/46          \\
\bottomrule
     \end{tabular}}
\end{center}
\caption{Results of the {\sc Pythia8} and {\sc Powheg+Pythia8} tuning
to the \ptz\ and $\phi^\star_\eta$ data.} 
\label{tab:ptphitunes}
\end{table}

\begin{figure}
    \centering
    \includegraphics[width=0.8\textwidth]{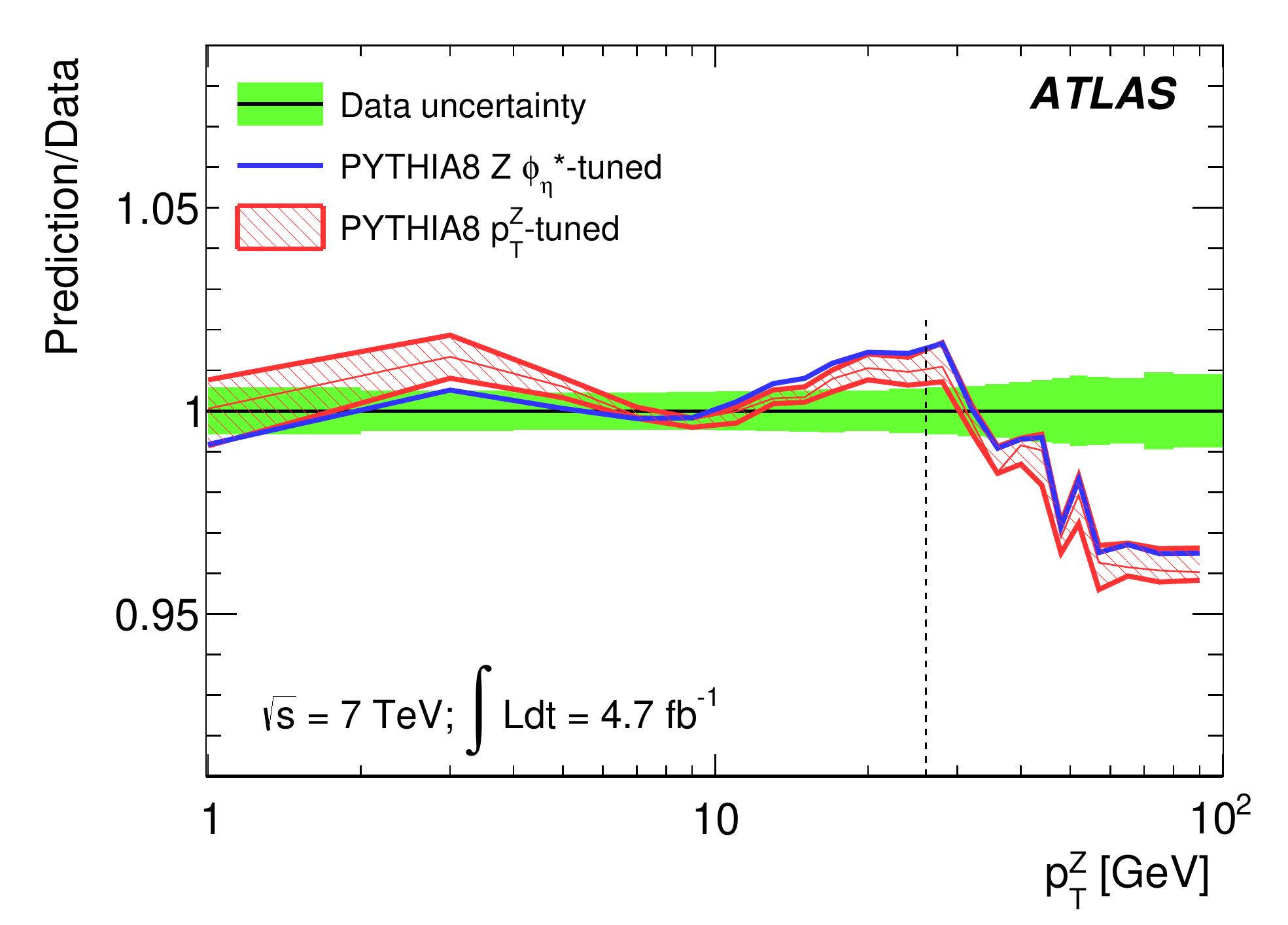}
    \includegraphics[width=0.8\textwidth]{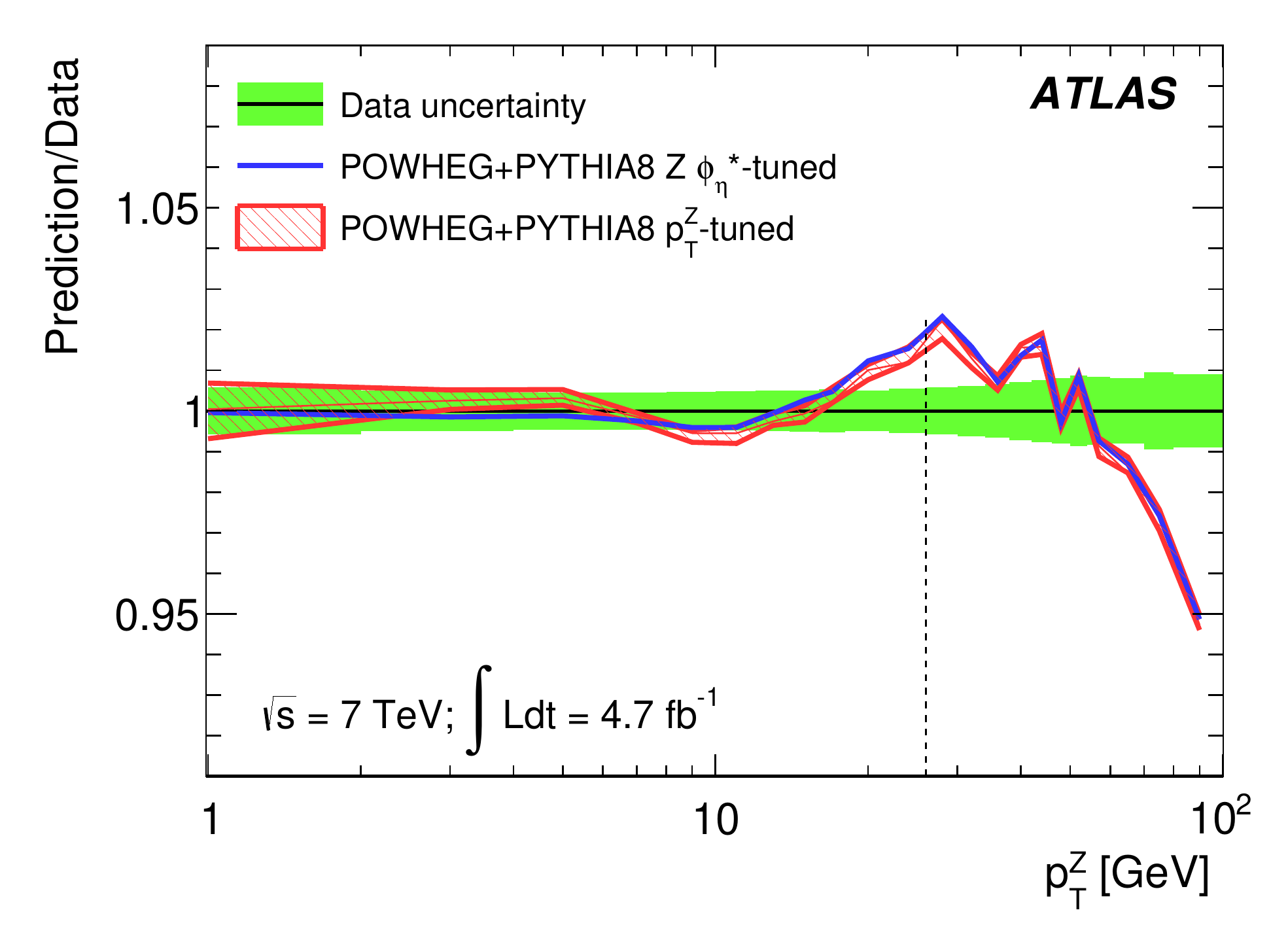}
   \caption{Comparison of the {\sc Pythia8} (top) and {\sc
   Powheg+Pythia8} (bottom) tuned predictions based on the $\phi^\star_\eta$ and \ptz\
   measurements with the data, for dressed kinematics. The vertical dashed lines show the upper limit of the
   tuning range.}   \label{fig:tune_compatibility}
\end{figure}

Since the \ptz\ and $\phi^\star_\eta$~observables provide similar
sensitivity to the parton shower parameters and to avoid correlations between
these measurements, the final tune
optimally combines the most precise independent single measurements, namely
the muon channel \ptz\ measurement, and the electron channel
$\phi^{\star}_{\eta}$ measurement. The same tuning range is
used. Table~\ref{tab:AZ} shows the tune results and
figure~\ref{fig:tuneAZ} shows the comparison of the tuned predictions
to the data. The final tunes are referred to as AZ and AZNLO for {\sc
Pythia8} and {\sc Powheg+Pythia8} respectively. The tuned predictions
agree with the measurement to better than 2\% in the range used for
the tuning, and below $\ptz=50$~\GeV. The primordial $k_{\rm{}T}$ and
ISR cut-off parameters are essentially constrained by 
the data in the region $\ptz<12$~GeV and not affected by the choice of upper bound for the
tuning range. In contrast, $\alpha_{\rm {S}}^{\rm {ISR}}(m_Z)$ is tightly
constrained for a given choice of range but its tuned value varies by
2\% when increasing the upper bound to 50~\GeV.
At higher transverse momentum, discrepancies of around 15\% for {\sc Pythia8} and 20\% 
for {\sc Powheg+Pythia8} remain, indicating the limited accuracy of the
NLO signal matrix element and suggesting the need for contributions from higher parton multiplicity.

\begin{table}
\begin{center}
    \begin{tabular}{lccc}
\toprule
                              & {\sc Pythia8}        & {\sc Powheg+Pythia8} & Base tune \\
\midrule
Tune Name                     & AZ                   & AZNLO             &  4C    \\
\midrule                                                                             
Primordial $k_{\rm{}T}$ [GeV]   & $1.71 \pm 0.03$      & $1.75 \pm 0.03$   & 2.0    \\
ISR $\alpha_{\rm {S}}^{\rm {ISR}}(m_Z)$           & $0.1237 \pm 0.0002$  & 0.118 (fixed)             & 0.137  \\
ISR cut-off  [GeV]            & $0.59 \pm 0.08$	     & $1.92 \pm 0.12$   & 2.0    \\
\midrule                                                                             
$\chi^2_{\rm{}min}$/dof                &       45.4/32        & 46.0/33           & -     \\
\bottomrule
     \end{tabular}
\end{center}
\caption{Final {\sc Pythia8} and {\sc Powheg+Pythia8} tuning results,
and comparison to the {\sc Pythia8} base tune.} 
\label{tab:AZ}
\end{table}

\begin{figure}
    \centering
    \includegraphics[width=0.49\textwidth]{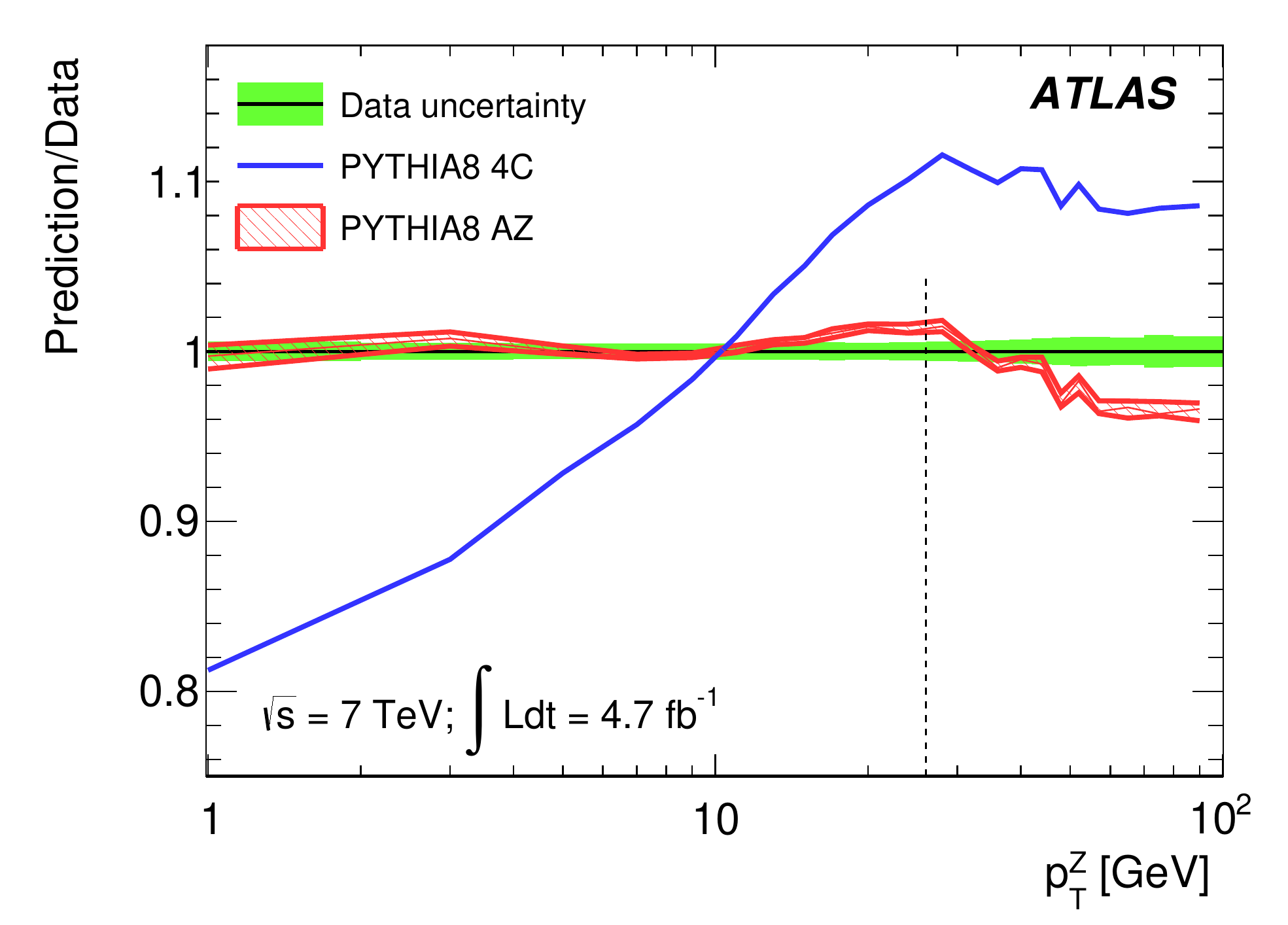}
    \includegraphics[width=0.49\textwidth]{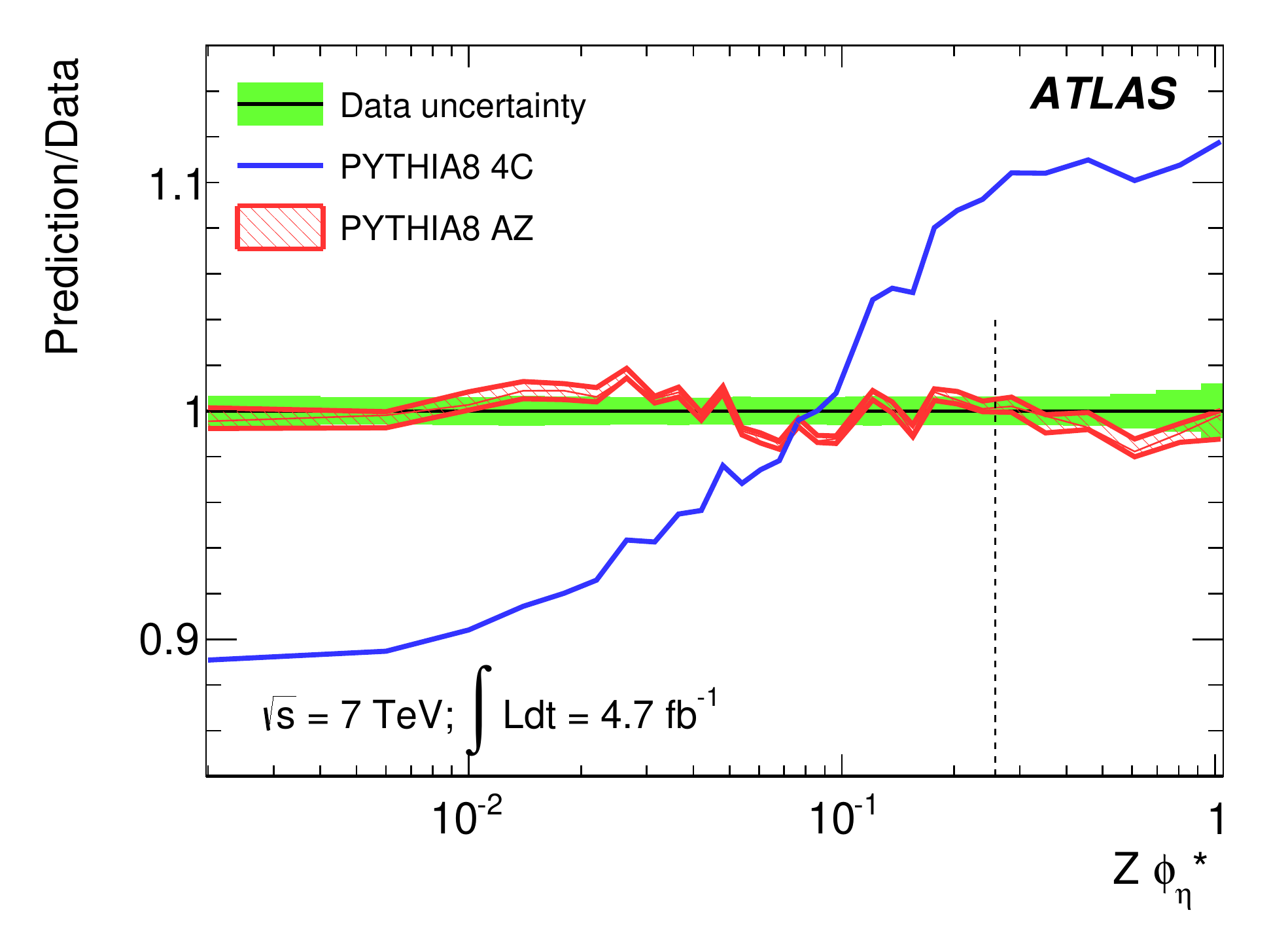}
    \includegraphics[width=0.49\textwidth]{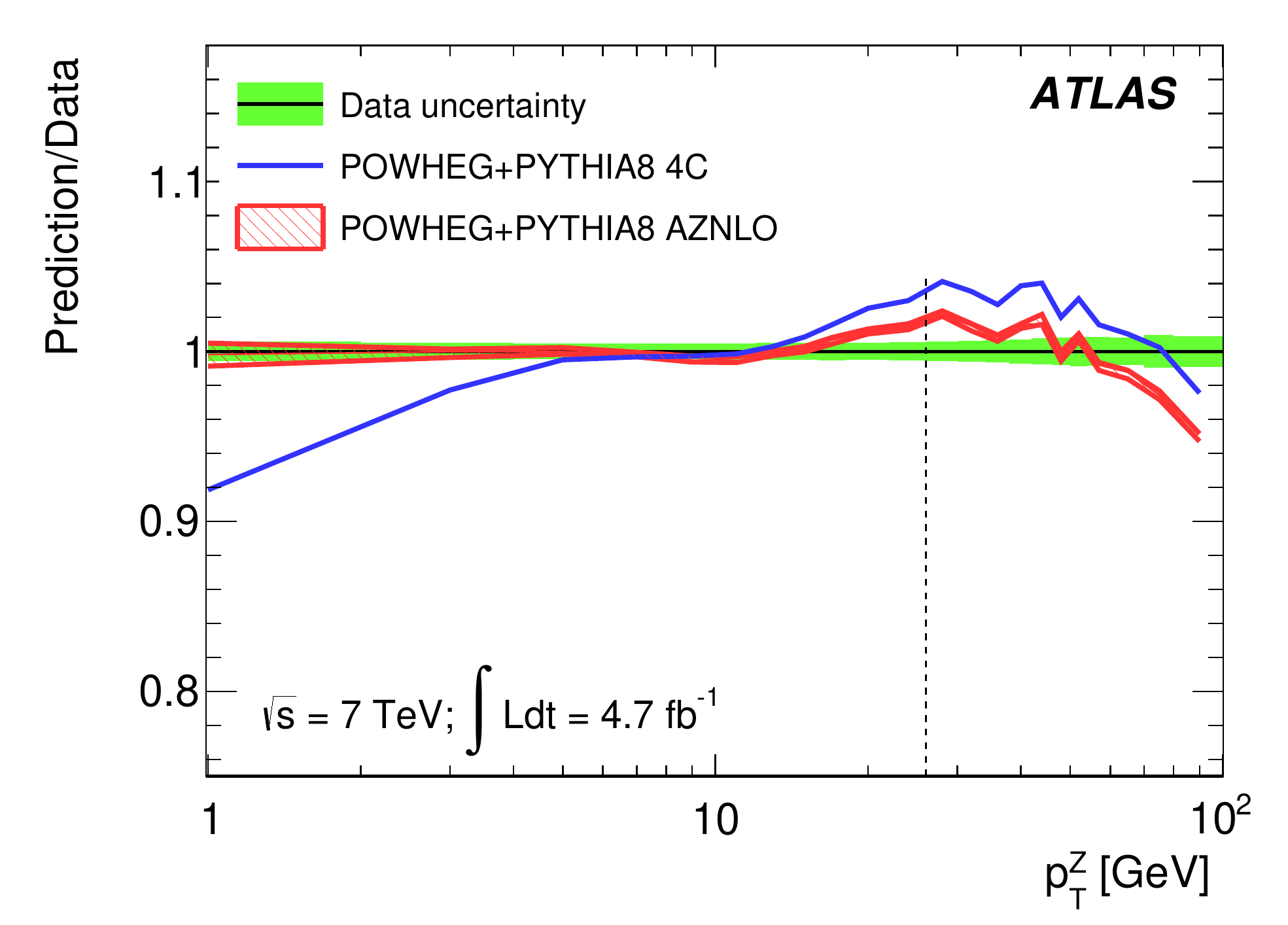}
    \includegraphics[width=0.49\textwidth]{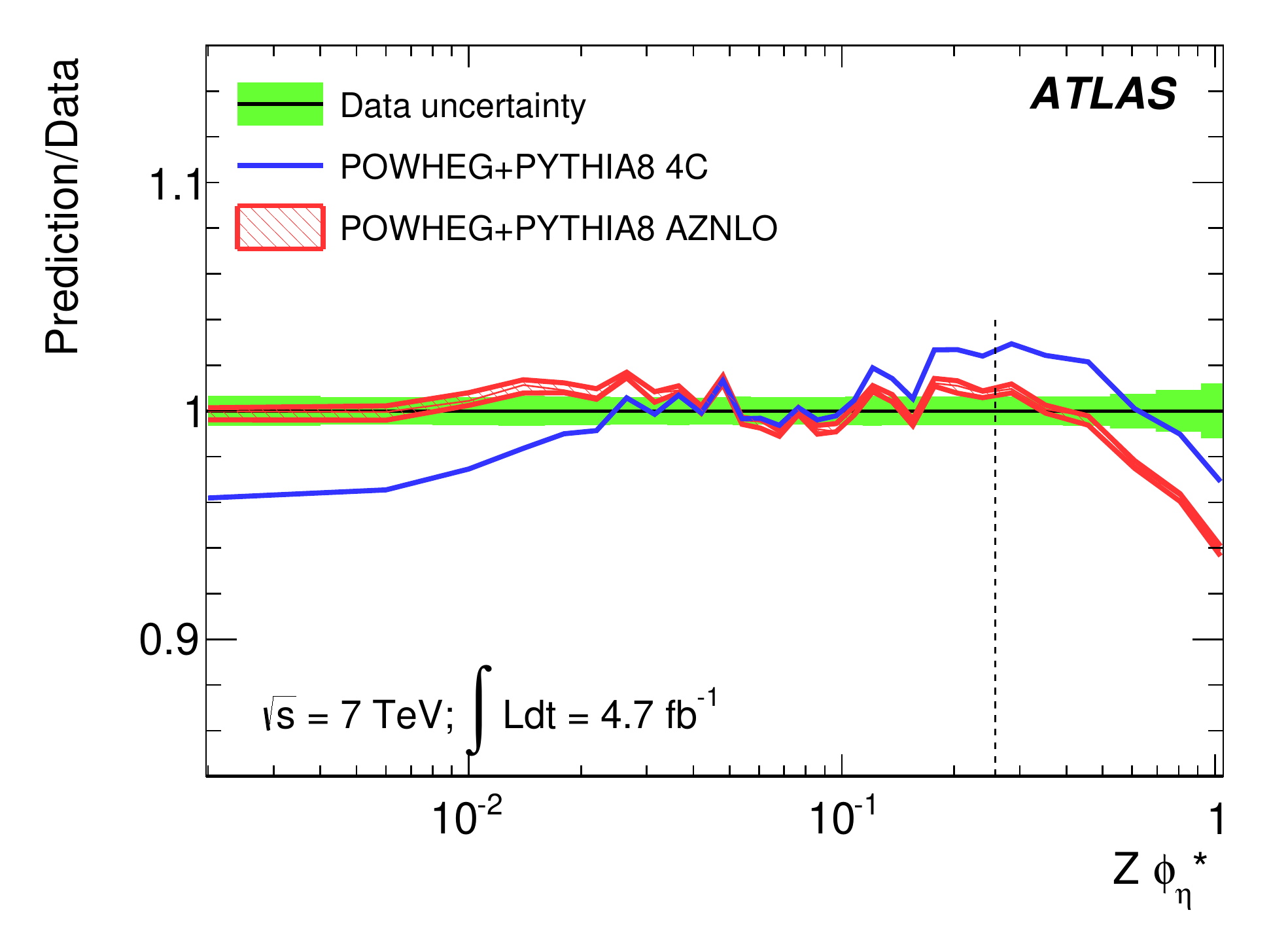}
  \caption{Comparison of tuned predictions to
    the \ptz\ and $\phi^\star_\eta$  differential cross sections, for dressed kinematics and in the full rapidity
    range. Comparison of the {\sc Pythia8} generator 
  with the 4C and AZ tunes to the muon-channel \ptz\ data and electron-channel $\phi^\star_\eta$ data
  (top). Comparison of the {\sc Powheg+Pythia8} set-up with the 4C and
    AZNLO tunes to the same data (bottom).
   The vertical dashed lines show the upper limit of the tuning range.} 
  \label{fig:tuneAZ}
\end{figure}

Tuned predictions based on the parameter values given in table~\ref{tab:AZ} are produced
in the different $Z$ rapidity bins and compared to the measured cross
sections with the aim of assessing how accurately the tune based on the inclusive 
measurement reproduces the data in each $Z$ rapidity bin.
The results are shown in
figure~\ref{fig:tunevsrapidity}. A satisfactory description across
rapidity is obtained in the case of {\sc Pythia8}; in the case of {\sc
  Powheg+Pythia8}, the prediction at low \ptz\  undershoots the data
for $0\leq|\yz|<1$, and overshoots the data for $2\leq|\yz|<2.4$. The
inclusive tune thus appears as a compromise between the different
$|\yz|$ regions in this case.  

\begin{figure}
    \centering
    \includegraphics[width=0.49\textwidth]{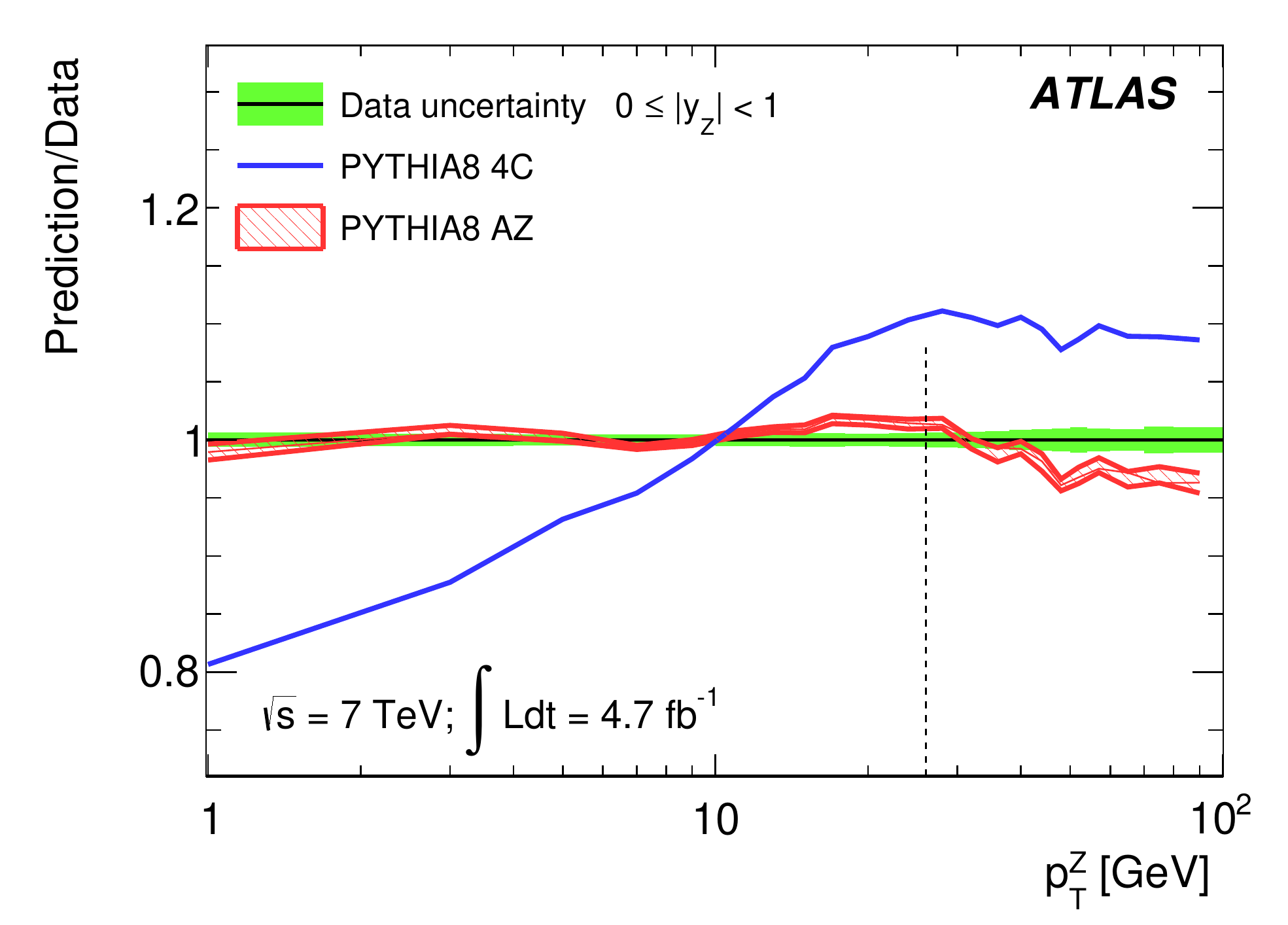}
    \includegraphics[width=0.49\textwidth]{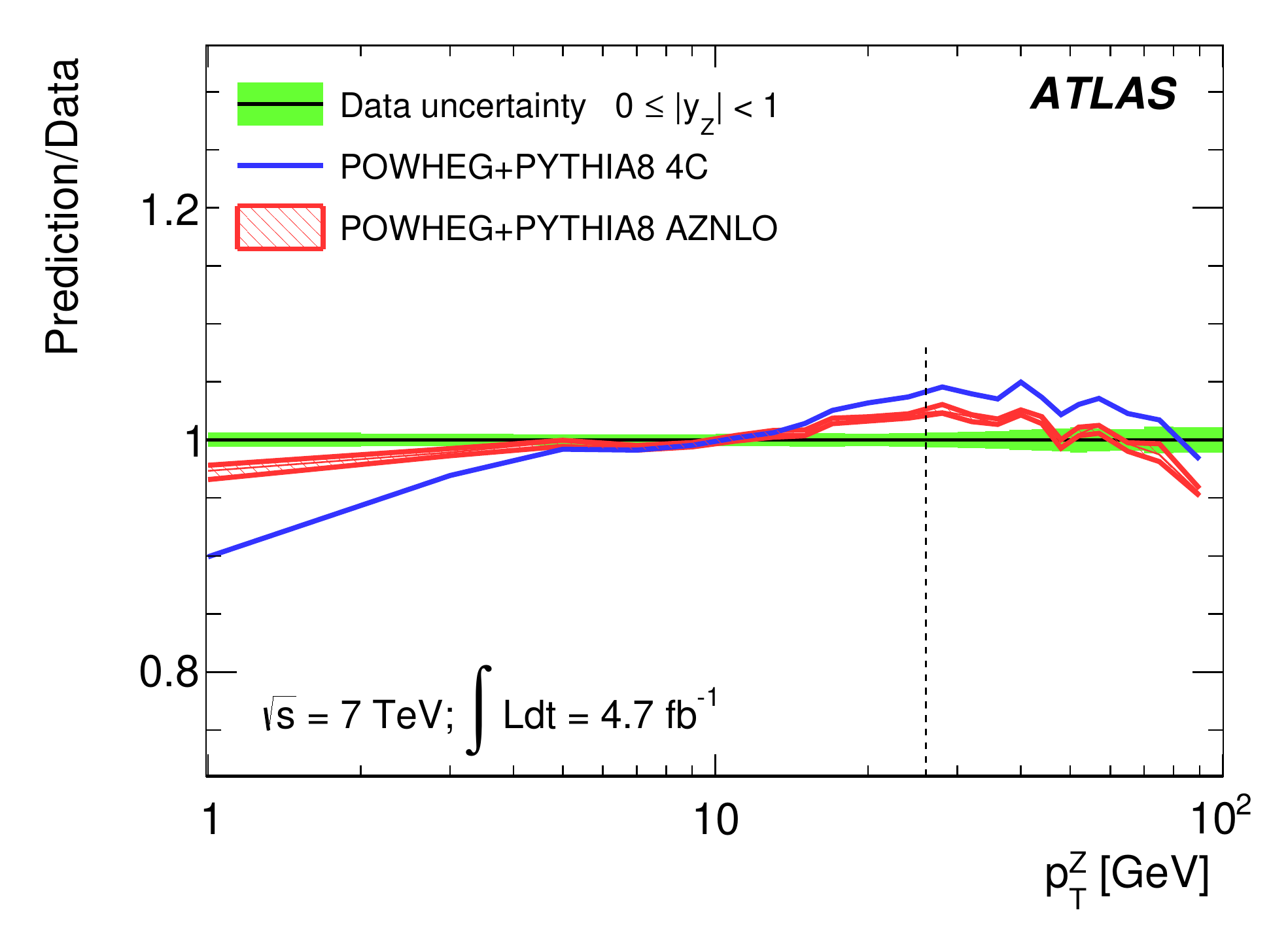}
    \includegraphics[width=0.49\textwidth]{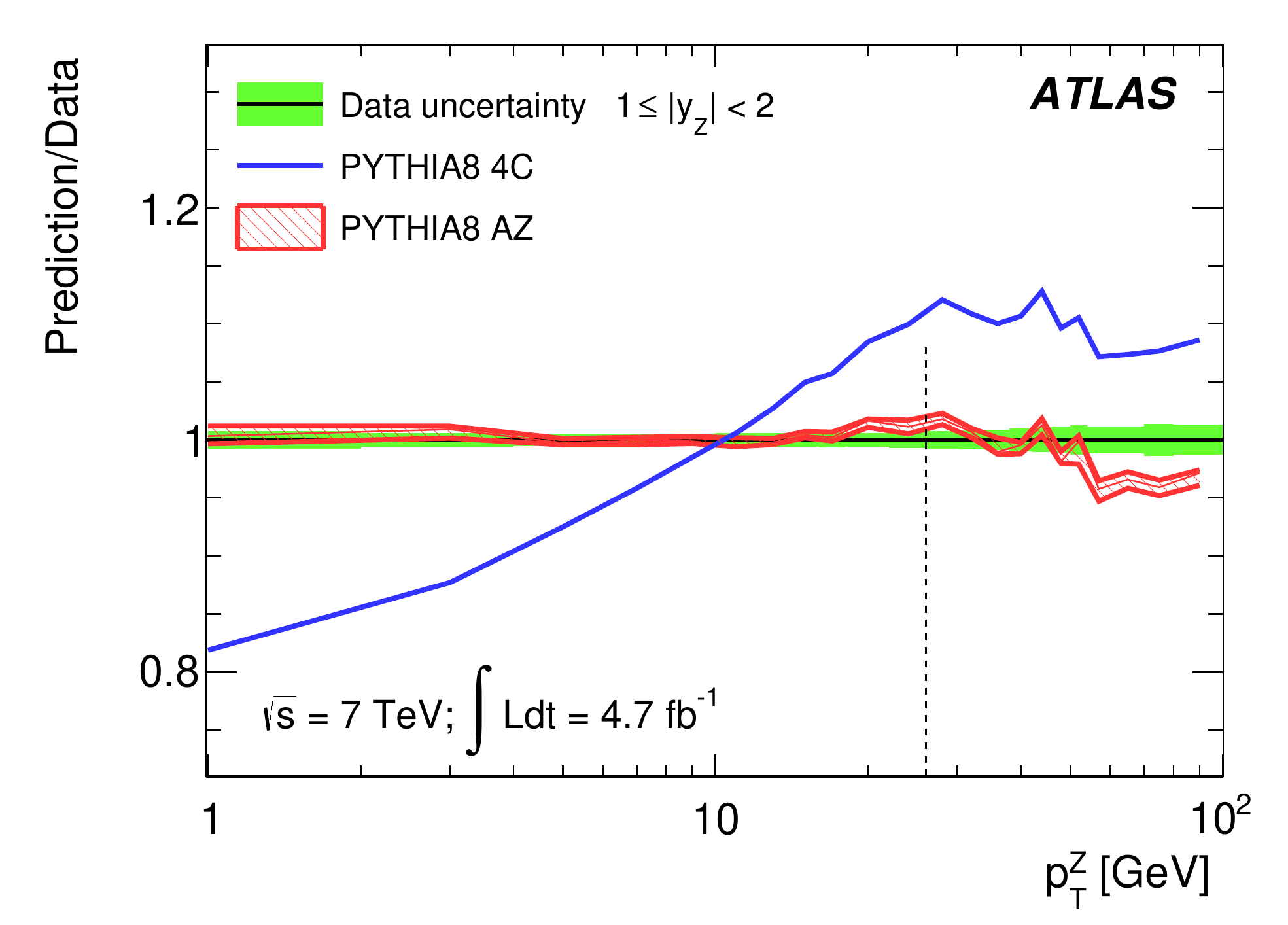}
    \includegraphics[width=0.49\textwidth]{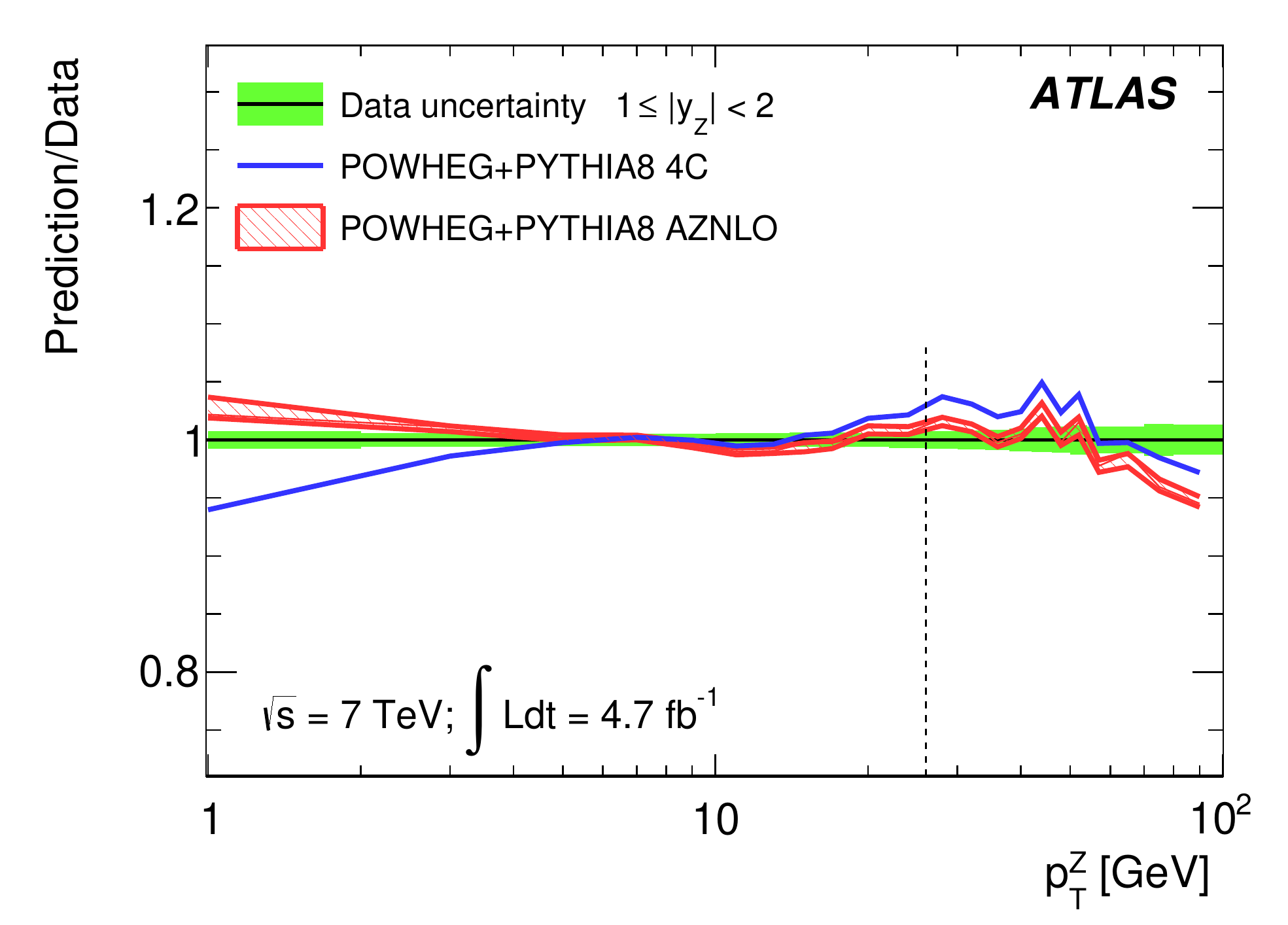}
    \includegraphics[width=0.49\textwidth]{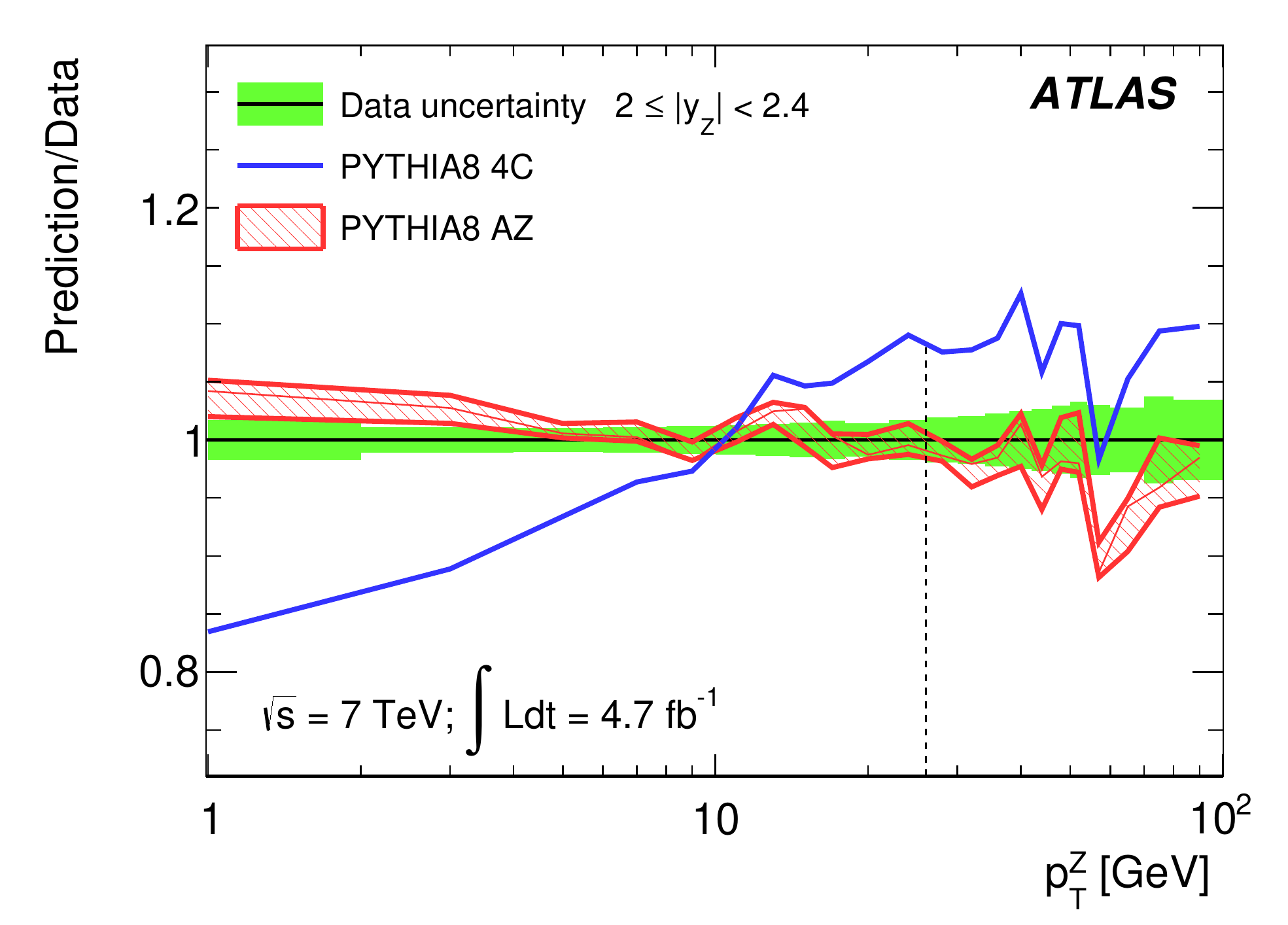}
    \includegraphics[width=0.49\textwidth]{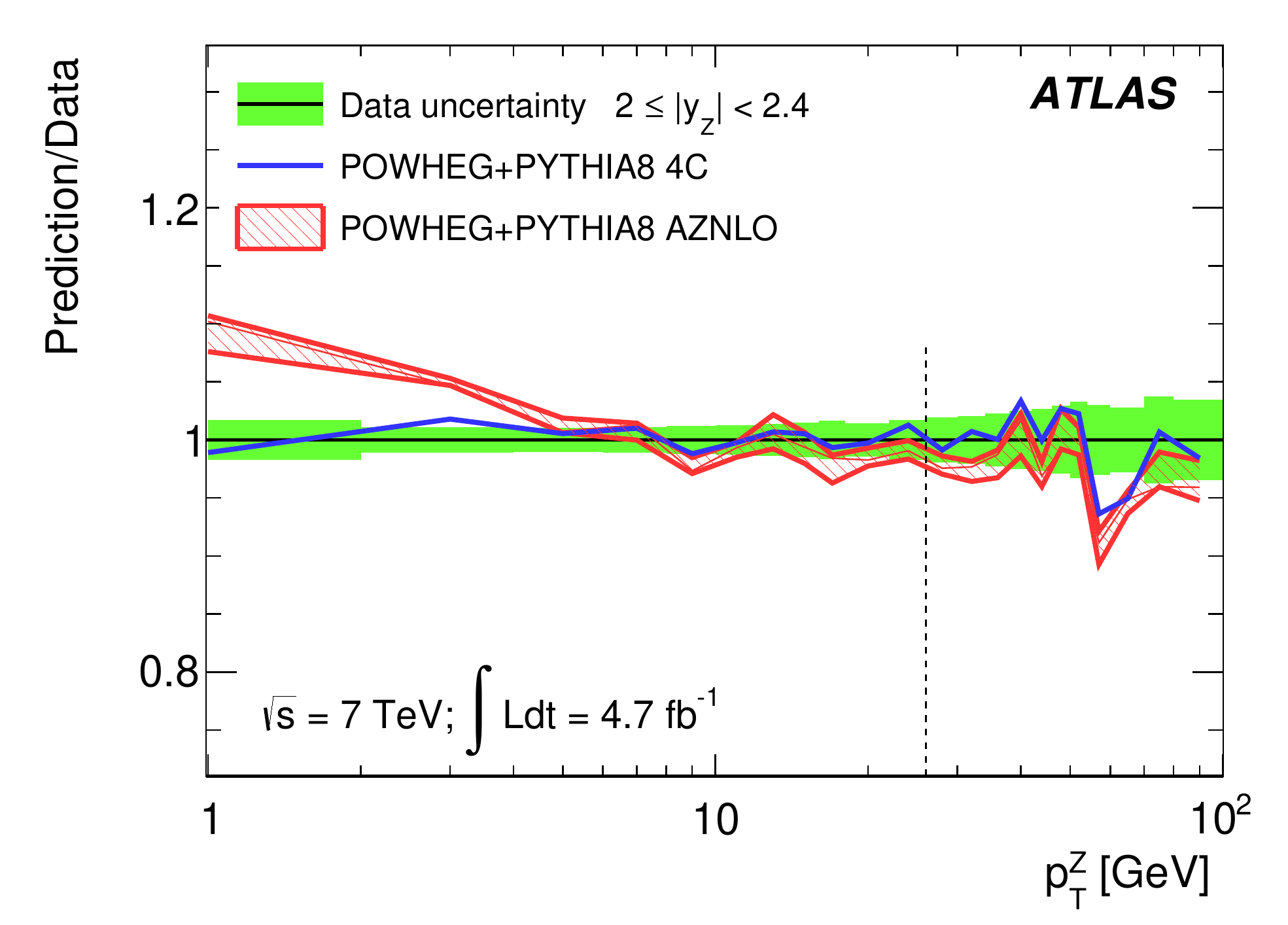}
    \caption{Tuned predictions based on
    table~\ref{tab:AZ} for {\sc Pythia8} (left) and {\sc
      Powheg+Pythia8} (right) for $0\leq |y_Z|<1$ (top), $1\leq |y_Z|<2$ (middle)
    and $2\leq |y_Z|<2.4$ (bottom),
    compared to the corresponding combined measurements, for dressed
    kinematics. The vertical dashed lines show the
    upper limit of the tuning range.} 
    \label{fig:tunevsrapidity}
\end{figure}

The sensitivity of the parton shower tune to other model components
provided in {\sc Pythia8}, such as multiple parton interactions (MPI),
which affect the event overall hadronic activity, was studied by 
varying the corresponding parameters. No effect on the parton shower
parameters is found. In order to compensate for the changes in energy and
particle flow induced by the modifications in the parton shower and primordial $k_{\rm T}$
parameters, the \pt\ threshold for the QCD $2\rightarrow 2$ scattering processes used
in MPI is changed from 2.085~GeV (tune 4C) to 2.18~GeV for {\sc Pythia8}, following the underlying event 
data measured in Drell--Yan events by ATLAS~\cite{Aad:2014jgf}.
For the {\sc Powheg+Pythia8} configuration, 
 the interleaving of MPI in the  parton shower model in {\sc Pythia8} is adapted to also properly take into account the 
  {\sc Powheg} emissions. It is tuned in the same way as 
 {\sc Pythia8} in standalone mode and an optimized value of 2.00 for the  low-\pt\ regularization of MPI is found.
Finally, comparing tunes based on the native {\sc Pythia8} QED final-state corrections with {\sc Pythia8} results using 
{\sc Photos} for QED final-state radiation, the results are found insensitive to the differences in the
QED FSR implementations. 

A consistent description of the \ptz\ and
$\phi^\star_\eta$ data is reached with a single tune. Both
observables are also found to provide similar sensitivity to the
parameters of interest. The inclusive tune provides an accurate
description of the different rapidity bins in the case of {\sc
Pythia8}, while the agreement versus $|\yz|$ is slightly worse in the case of
{\sc Powheg+Pythia8}.

\section{Conclusion}
\label{conclusion}

A measurement of the \Zg\ transverse momentum spectrum in the
\Zgee\ and \Zgmm\ channels with the ATLAS detector is presented, using 4.7 fb$^{-1}$
of LHC proton--proton collision data at a centre-of-mass energy of
$\rts=7$~TeV. Normalized differential cross sections as a function of 
\ptz\ are measured for the range $0<\ptz<800$~GeV and the individual
channel results are combined. The measurement is performed inclusively in rapidity, and
for $0\leq|\yz|<1$, $1\leq|\yz|<2$ and $2\leq|\yz|<2.4$. The large data sample allows
a fine binning in \ptz\ with a typical uncertainty on the combined
result better than 1\% for $\ptz<100$~GeV, rising to~5\% towards the
end of the spectrum.

The cross-section measurements are compared to pQCD and resummed
predictions. While {\sc Fewz} and {\sc Dynnlo} do not include
resummation and the observed disagreement at low \ptz\ is expected,
{\sc ResBos} and the NNLO+NNLL prediction of ref.~\cite{Banfi:2012du} marginally agree with the
data given the large uncertainties on these predictions.
 The data 
are also compared to predictions from {\sc Pythia6-AUET2B}, {\sc
  Powheg+Pythia6-AUET2B}, {\sc MC@NLO}, {\sc Alpgen} and {\sc Sherpa}.  
The \pythia\ and {\sc Powheg} generators  agree with the data to within ~5\% in
the range $2<\ptz<60$~GeV and to within 20\% over the full range,
whereas  \mcatnlo\ shows a deficit of about 40\% at the end of
the measured spectrum. {\sc Alpgen} and {\sc Sherpa} provide good
agreement over a larger range, $\ptz<200$~GeV, but overestimate the
high end of the distribution. These patterns are compatible with what was
observed for the $\phi^\star_\eta$~measurement. Electroweak 
corrections and the choice of a dynamic QCD scale were found to have a
significant impact on the predictions at high \ptz; incorporating
these improvements provides a better description of the measured
distribution in the high \ptz\ region.

The \ptz\ and $\phi^\star_\eta$ measurements were used to tune the {\sc
  Pythia8} and {\sc Powheg+Pythia8} generators. Both measurements can
be interpreted consistently in terms of the \Zg\ bosons transverse
momentum distribution and provide similar sensitivity to parton shower
model parameters. The tuned predictions are in agreement with the data
within 2\% for \ptz\ below 50~GeV. The  best description is
provided by {\sc   Pythia8}, which is also able to describe the
different rapidity intervals with a single tune. 

\section*{Acknowledgements}

We thank CERN for the very successful operation of the LHC, as well as the
support staff from our institutions without whom ATLAS could not be
operated efficiently.

We acknowledge the support of ANPCyT, Argentina; YerPhI, Armenia; ARC,
Australia; BMWF and FWF, Austria; ANAS, Azerbaijan; SSTC, Belarus; CNPq and FAPESP,
Brazil; NSERC, NRC and CFI, Canada; CERN; CONICYT, Chile; CAS, MOST and NSFC,
China; COLCIENCIAS, Colombia; MSMT CR, MPO CR and VSC CR, Czech Republic;
DNRF, DNSRC and Lundbeck Foundation, Denmark; EPLANET, ERC and NSRF, European Union;
IN2P3-CNRS, CEA-DSM/IRFU, France; GNSF, Georgia; BMBF, DFG, HGF, MPG and AvH
Foundation, Germany; GSRT and NSRF, Greece; ISF, MINERVA, GIF, I-CORE and Benoziyo Center,
Israel; INFN, Italy; MEXT and JSPS, Japan; CNRST, Morocco; FOM and NWO,
Netherlands; BRF and RCN, Norway; MNiSW and NCN, Poland; GRICES and FCT, Portugal; MNE/IFA, Romania; MES of Russia and ROSATOM, Russian Federation; JINR; MSTD,
Serbia; MSSR, Slovakia; ARRS and MIZ\v{S}, Slovenia; DST/NRF, South Africa;
MINECO, Spain; SRC and Wallenberg Foundation, Sweden; SER, SNSF and Cantons of
Bern and Geneva, Switzerland; NSC, Taiwan; TAEK, Turkey; STFC, the Royal
Society and Leverhulme Trust, United Kingdom; DOE and NSF, United States of
America.

The crucial computing support from all WLCG partners is acknowledged
gratefully, in particular from CERN and the ATLAS Tier-1 facilities at
TRIUMF (Canada), NDGF (Denmark, Norway, Sweden), CC-IN2P3 (France),
KIT/GridKA (Germany), INFN-CNAF (Italy), NL-T1 (Netherlands), PIC (Spain),
ASGC (Taiwan), RAL (UK) and BNL (USA) and in the Tier-2 facilities
worldwide.

\bibliographystyle{JHEP}
\bibliography{STDM-2012-23}
\newpage

\onecolumn
\clearpage
\begin{flushleft}
{\Large The ATLAS Collaboration}

\bigskip

G.~Aad$^{\rm 84}$,
B.~Abbott$^{\rm 112}$,
J.~Abdallah$^{\rm 152}$,
S.~Abdel~Khalek$^{\rm 116}$,
O.~Abdinov$^{\rm 11}$,
R.~Aben$^{\rm 106}$,
B.~Abi$^{\rm 113}$,
M.~Abolins$^{\rm 89}$,
O.S.~AbouZeid$^{\rm 159}$,
H.~Abramowicz$^{\rm 154}$
% ...
\bigskip
\\
% ...
$^{11}$ Institute of Physics, Azerbaijan Academy of Sciences, Baku, Azerbaijan\\
% ...
$^{84}$ CPPM, Aix-Marseille Universit{\'e} and CNRS/IN2P3, Marseille, France\\
% ...
$^{89}$ Department of Physics and Astronomy, Michigan State University, East Lansing MI, United States of America\\
% ...
$^{106}$ Nikhef National Institute for Subatomic Physics and University of Amsterdam, Amsterdam, Netherlands\\
% ...
$^{112}$ Homer L. Dodge Department of Physics and Astronomy, University of Oklahoma, Norman OK, United States of America\\
$^{113}$ Department of Physics, Oklahoma State University, Stillwater OK, United States of America\\
% ...
$^{116}$ LAL, Universit{\'e} Paris-Sud and CNRS/IN2P3, Orsay, France\\
% ...
$^{152}$ Institute of Physics, Academia Sinica, Taipei, Taiwan\\
% ...
$^{154}$ Raymond and Beverly Sackler School of Physics and Astronomy, Tel Aviv University, Tel Aviv, Israel\\
% ...
$^{159}$ Department of Physics, University of Toronto, Toronto ON, Canada\\
% ...
\end{flushleft}

\end{document}